\documentclass[australian,english,prl, singlespace, twocolumn]{revtex4-1}
\usepackage[T1]{fontenc}
\usepackage[latin9]{inputenc}
\usepackage{color}
\usepackage{verbatim}
\usepackage{amsmath}
\usepackage{amssymb}
\usepackage{graphicx}

\makeatletter
\@ifundefined{definecolor}
 {\usepackage{color}}{}
\makeatother

\makeatother

\usepackage{babel}
\begin{document}
\title{A macroscopic quantum three-box paradox: finding consistency with
weak macroscopic realism}
\author{C. Hatharasinghe, M. Thenabadu, P. D. Drummond and M. D. Reid$^{1}$}
\affiliation{$^{1}$Centre for Quantum Science and Technology Theory, Swinburne
University of Technology, Melbourne, Australia}
\begin{abstract}
The quantum three-box paradox considers a ball prepared in a superposition
of being in one of three Boxes. Bob makes measurements by opening
either Box 1 or Box 2. After performing some unitary operations (shuffling),
Alice can infer with certainty that the ball was detected by Bob,
regardless of which box he opened, if she detects the ball after opening
Box 3. The paradox is that the ball would have been found with certainty
in either box, if that box had been opened. Resolutions of the paradox
include that Bob\textquoteright s measurement cannot be made non-invasively,
or else that realism cannot be assumed at the quantum level. Here,
we strengthen the case for the former argument, by constructing macroscopic
versions of the paradox. Macroscopic realism implies that the ball
is in one of the boxes, prior to Bob or Alice opening any boxes. We
demonstrate consistency of the paradox with macroscopic realism, if
carefully defined (as weak macroscopic realism, wMR) to apply to the
system at the times prior to Alice or Bob opening any Boxes, but \emph{after}
the unitary operations associated with preparation or shuffling. By
solving for the dynamics of the unitary operations, and comparing
with mixed states, we demonstrate agreement between the predictions
of wMR and quantum mechanics: The paradox only manifests if Alice's
shuffling combines both local operations (on Box 3) and nonlocal operations,
on the other Boxes. Following previous work, the macroscopic paradox
is shown to correspond to a violation of a Leggett-Garg inequality,
which implies non-invasive measurability, if wMR holds. 
\end{abstract}
\maketitle

\section{Introduction}

The quantum three-box paradox \cite{three-box-paradox} concerns results
inferred for a system at a time intermediate between two times where
the system is in a preselected and postselected state \cite{abl}.
The paradox was introduced by Aharonov and Vaidmann and has attracted
much interest \cite{kastner,Maroney,three-box-revisited,aharonov-disappearing,e-george-exp,e-resch-exp,e-kol-exp,kirkpatrick-2007,kirkpatrick2003,leier-spekkens-prl,liefer-spekkens,causal-app,finkelstein,lv-reply}.
The paradox involves a ball prepared in a quantum superposition of
being in one of the three boxes. Bob makes a measurement by opening
either Box 1 or Box 2, to determine whether or not the ball is in
the box he opens. Alice then makes specific transformations on the
system, by ``shuffling'' the ball between the boxes. After these
operations, she knows that if she detects the ball in the Box 3, then
Bob must have detected the ball in the box he opened.  The paradox
is that the ball would have been found with certainty in either box,
if that box had been opened. The paradox can be considered a quantum
game, in which Alice can infer that Bob detected a particle with odds
not possible classically \cite{three-box-revisited,Maroney}.

The three-box paradox raised questions about what quantum effect is
involved, and how to understand the paradox \cite{kirkpatrick2003,kirkpatrick-2007,three-box-revisited,Maroney,leier-spekkens-prl,finkelstein,liefer-spekkens,lv-reply}.
One response is to argue that for a system described as a quantum
superposition, it cannot be assumed that the system is \emph{in} one
of those states until measured. Hence, it cannot be assumed that the
ball is in one of the boxes, prior to the box being opened, and that
this is the origin of the paradox. This approach defies the concept
of \emph{realism}. The argument is not so convincing for a macroscopic
version of the paradox, where the states of the system are macroscopically
distinct. \emph{Macroscopic realism} (MR) would posit that the ball
must be in one of the boxes, irrespective of Alice or Bob opening
a box. On the other hand, it could be argued that decoherence prevents
the creation of macroscopic superposition states, so that the paradox
could not be realised. This motivates the challenge of a macroscopic
version of the three-box paradox. The three-box paradox has been experimentally
verified, but at microscopic levels only \cite{e-george-exp,e-kol-exp,e-resch-exp}.

In this paper, we present a mesoscopic three-box paradox corresponding
to $N$ quanta (the ball) placed in a superposition of being in one
of three modes (boxes). We also present a macroscopic paradox in
which the states of the system correspond to macroscopically distinct
coherent states of a single-mode field. Since macroscopic quantum
superposition states have been created  \cite{cat-states-super-cond,collapse-revival-super-circuit-1},
we anticipate these predictions could be tested. This motivates consideration
of other approaches to explain the paradox consistently with macroscopic
realism.

We show in this paper that the three-box paradox can be explained
consistently with macroscopic realism, \emph{if} the definition of
macroscopic realism is carefully refined, as \emph{weak macroscopic
realism} \cite{manushan-bell-cat-lg,ghz-cat,wigner-friend-macro}.
Following previous work \cite{wigner-friend-macro,ghz-cat,manushan-bell-cat-lg},
weak macroscopic realism (wMR) posits that the outcome of Bob's or
Alice's measurement in opening any one of the boxes at the time $t_{i}$
is predetermined: each box is either occupied by the ball or not,
and the ball is in one of the boxes. The outcome of Alice or Bob opening
box $K$ is represented by a variable $\lambda_{K}$ which takes the
value $-1$ or $+1$, if the ball is in the box $K$ or not. Weak
macroscopic realism posits \emph{specifically} that prior to Alice
or Bob opening the box $K$, this value is fixed, at the time $t_{i}$
\emph{once} any preparation and unitary transformations (shuffling)
involving the box $K$ is completed. We refer to any such transformations
involving Box $K$ as ``local'' to $K$.  The definition of wMR
also posits that the ball being in a particular box $K$ at the time
$t_{i}$ is \emph{not affected} by any further measurement or shuffling
procedure that might occur solely for the other boxes \cite{manushan-bell-cat-lg,ghz-cat,wigner-friend-macro}.
Since the boxes may be spatially separated, we refer to such operations
as ``nonlocal''.

It is important to note that the definition of wMR is not concerned
with microscopic details about the state of the ball. Hence, wMR does
\emph{not} posit that the 'state' of the ball prior to this measurement
needs to be a quantum state. In fact, for macroscopic superposition
states, it has been argued previously that if the predetermination
given by $\lambda$ is valid, it is not possible to associate the
'state' of the system given by $\lambda$ with a quantum state \cite{manushan-bell-cat-lg,s-cat}.

In this paper, we verify consistency with weak macroscopic realism
for the results of the paradox, showing that (in this wMR-model) the
paradox arises due to disturbance from Bob's measurement, \emph{but
a measurable paradoxical effect occurs in the final joint probabilities,
only when} Alice's further unitary transformations involve \emph{both}
the local box $K$ and the other nonlocal boxes. We refer to this
as a \emph{local-nonlocal} operation. This allows us to test the predictions
of weak macroscopic realism in a potential experiment. The dynamics
of Alice's transformations are modelled by specific interaction Hamiltonians,
and are illustrated by the $Q$ function.

Our results support earlier conclusions which emphasize the role of
Bob's measurement disturbance in explaining the paradox \cite{leier-spekkens-prl,Maroney,causal-app}.
Maroney pointed out that the conditions under which the paradox occurs
are the same as for the set-up to violate a Leggett-Garg inequality
\cite{Maroney}. Leggett-Garg inequalities are derived from the assumptions
of \emph{macrorealism} \cite{legggarg}. Macrorealism posits macroscopic
realism (MR) (that a system with two or more macroscopically-distinct
states available to it must be \emph{in} one of those states) and
also macroscopic noninvasive measurability (NIM)$-$ that it is possible
to measure which of two macroscopically-distinct states the system
is in, with minimal disturbance to the macroscopic dynamics of the
system. Violation of the  inequalities can therefore arise from a
failure of NIM, and be consistent with MR. By illustrating violation
of a new class of Leggett-Garg inequality, Maroney argues that the
quantum feature of the paradox is measurement disturbance that cannot
be explained classically. In this paper, we show that the negation
of macrorealism is possible for the macroscopic and mesoscopic versions
of the three-box paradox, along the lines proposed by Maroney, thereby
illustrating the quantum nature of the proposed experiments. Similarly,
Blasiak and Borsuk identify causal structures for the three-box paradox,
showing that a realist viewpoint necessitates measurement disturbance
in order to maintain consistency with the assumption of realism \cite{causal-app}.

Our model extends previous work, since a parameter $\alpha$ or $N$
is introduced, which defines the size of the separation of the states
of the system, corresponding to the ball being in the Box or not.
The disturbance to the system due to Bob's measurement can be evaluated,
by comparing the states before and after. We find that the $Q$ function
for the two states becomes identical, as the system becomes macroscopic
($\alpha\rightarrow\infty$). This supports Leggett and Garg's macrorealism
premise that a noninvasive measurement should exist, for a sufficiently
large system. However, when we evaluate the predictions for Alice
detecting a Ball in Box 3, we find that the difference between the
predictions, depending on whether Bob makes a measurement or not,
remains macroscopically measurable, even as $\alpha\rightarrow\infty$.
The paradoxical results arise from microscopic differences existing
at a prior time, similar to a quantum revival.

These results also support the work of Thenabadu et al \cite{manushan-bellcats-noon}
and Thenabadu and Reid \cite{manushan-bell-cat-lg,manushan-cat-lg},
who gave predictions for violations of Leggett-Garg and Bell inequalities
  involving macroscopically distinct coherent states, $|\alpha\rangle$
and $|-\alpha\rangle$. These authors  showed how the violations
were consistent with wMR, demonstrating similarly that the Bell-nonlocal
effect only occurs when the unitary transformations that determine
the measurement settings are carried out at \emph{both} locations
(in a \emph{local-nonlocal} operation) \cite{ghz-cat}.  Similar
results are obtained in \cite{delayed-choice-cats,wigner-friend-macro,ghz-cat}.

The layout of the paper is as follows. In Section II, we summarise
the three-box paradox. In Section III, we present a mesoscopic paradox
where the three boxes are distinct modes, and the ball corresponds
to $N$ quanta.  A macroscopic (modified) version of the paradox
involving macroscopically distinct coherent states (cat states) is
presented in Section V.  In Sections IV and V, we give the definition
of wMR, and show consistency with wMR for both versions of the paradox.
The violation of the Leggett-Garg inequalities is demonstrated for
the mesoscopic and macroscopic three-box paradox, in Section VI.

\section{Three box paradox}

We first summarise the states, transformations, measurements involved
in the paradox \cite{three-box-paradox}. The system has three boxes
and one ball. The state of the ball being in box $k$ is denoted $|k\rangle$.
The system is prepared with the ball in box three i.e. in state $|3\rangle$.
A unitary transformation $U_{i}$ transforms the initial system into
the superposition state
\begin{equation}
|\psi_{1}\rangle=\frac{1}{\sqrt{3}}(|1\rangle+|2\rangle+|3\rangle)\label{eq:sup1}
\end{equation}
at time $t_{1}$, so that $U_{i}|3\rangle=|\psi_{1}\rangle$. Using
the basis set $\{|1\rangle,$ $|2\rangle$, $|3\rangle$\}, we find
\begin{eqnarray}
U_{i} & = & \frac{1}{\sqrt{6}}\left(\begin{array}{ccc}
\sqrt{3} & 1 & \sqrt{2}\\
\sqrt{3} & 1 & \sqrt{2}\\
0 & 2 & \sqrt{2}
\end{array}\right)\label{eq:Ui}
\end{eqnarray}
where the basis states $|k\rangle$ correspond to column matrices
$(a_{j1})$ with coefficients given as $a_{j1}=\delta_{jk}$. The
transformation can be carried out by first applying
\begin{equation}
U_{1i}=\frac{1}{\sqrt{3}}\left(\begin{array}{ccc}
\sqrt{3} & 0 & 0\\
0 & 1 & \sqrt{2}\\
0 & \sqrt{2} & 1
\end{array}\right)\label{eq:U1i}
\end{equation}
to create the superposition $\sqrt{\frac{2}{3}}|2\rangle+\frac{1}{\sqrt{3}}|3\rangle$,
and then applying
\begin{equation}
U_{2i}=\frac{1}{\sqrt{2}}\left(\begin{array}{ccc}
-1 & 1 & 0\\
1 & 1 & 0\\
0 & 0 & \sqrt{2}
\end{array}\right)\label{eq:U2i}
\end{equation}
to create the superposition $\frac{1}{\sqrt{2}}(|2\rangle+|1\rangle)$
from $|2\rangle$. Hence, $U_{i}=U_{2i}U_{1i}$.

Now, Bob can make a measurement to determine whether the system is
in $|1\rangle$ (the ball in box 1), or not. Assuming Bob makes an
ideal projective measurement, the state of the system after the measurement
according to quantum mechanics is $|1\rangle$ if he detects the ball
in box 1. Otherwise, the system is in the superposition state $\frac{1}{\sqrt{2}}(|2\rangle+|3\rangle)$.
Alternatively, Bob may make a measurement to determine if the system
is in state $|2\rangle$ (the ball is in box 2), or not. Assuming
an ideal projective measurement, the state of the system after the
measurement according to quantum mechanics is $|2\rangle$ if he detects
the ball in box 2. Otherwise, the system is in the superposition state
$\frac{1}{\sqrt{2}}(|1\rangle+|3\rangle)$.

After the interactions given by Bob's measurements, at time $t_{2}$,
Alice makes further measurements, to postselect for the state
\begin{equation}
|\psi_{f}\rangle=\frac{1}{\sqrt{3}}(|1\rangle+|2\rangle-|3\rangle)\label{eq:supf}
\end{equation}
which is orthogonal to both $\frac{1}{\sqrt{2}}(|2\rangle+|3\rangle)$
and $\frac{1}{\sqrt{2}}(|1\rangle+|3\rangle)$. The measurement is
realised as a transformation $U_{f}$, so that $|\psi_{f}\rangle$
maps to $|3\rangle$, at the time $t_{3}$. Alice then performs the
postselection by determining whether the ball is in box 3 at the time
$t_{3}$. $U_{f}$ has the property that its inverse satisfies
\begin{equation}
U_{f}^{-1}|3\rangle=|\psi_{f}\rangle\label{eq:Uf-1}
\end{equation}
To find $U_{f}$, we follow the above procedure and first apply the
unitary transformation $U_{1f}$ that transforms $|3\rangle$ into
the superposition $\sqrt{\frac{2}{3}}|2\rangle-\frac{1}{\sqrt{3}}|3\rangle$.
Then we create superposition $\frac{1}{\sqrt{2}}(|2\rangle+|1\rangle)$
from $|2\rangle$, which defines $U_{2f}=U_{2i}$. We find
\begin{equation}
U_{1f}=\frac{1}{\sqrt{3}}\left(\begin{array}{ccc}
\sqrt{3} & 0 & 0\\
0 & 1 & \sqrt{2}\\
0 & \sqrt{2} & -1
\end{array}\right)\label{eq:U1f}
\end{equation}
Hence, $U_{f}^{-1}=U_{2f}U_{1f}$. Hence, the required transformation
is $U_{f}=U_{1f}^{-1}U_{2f}^{-1}$ which, noting the transformations
are unitary so that $U_{2f}^{-1}=U_{2f}$ and $U_{1f}^{-1}=U_{1f}$,
becomes  $U_{f}^{-1}|3\rangle=|\psi_{f}\rangle$. Hence, Alice's
transformation is
\begin{eqnarray}
U_{f} & = & \frac{1}{\sqrt{6}}\left(\begin{array}{ccc}
-\sqrt{3} & \sqrt{3} & 0\\
1 & 1 & 2\\
\sqrt{2} & \sqrt{2} & -\sqrt{2}
\end{array}\right)\label{eq:Uf}
\end{eqnarray}

If after his measurements Bob determines the system to be in $|1\rangle$,
then the output after Alice's transformations is
\begin{eqnarray}
U_{f}|1\rangle & = & \frac{1}{\sqrt{6}}\left(\begin{array}{c}
-\sqrt{3}\\
1\\
\sqrt{2}
\end{array}\right)\label{eq:Uf-1-1}
\end{eqnarray}
If Bob measures that the system is not in $|1\rangle$, then the output
after Alice's operations is
\begin{eqnarray}
U_{f}\frac{1}{\sqrt{2}}\left(\begin{array}{c}
0\\
1\\
1
\end{array}\right) & = & \frac{1}{2\sqrt{3}}\left(\begin{array}{c}
\sqrt{3}\\
3\\
0
\end{array}\right)\label{eq:Uf-23}
\end{eqnarray}
If after his measurements Bob determined the system to be in $|2\rangle$,
then the final state is
\begin{eqnarray}
U_{f}|2\rangle & = & \frac{1}{\sqrt{6}}\left(\begin{array}{c}
\sqrt{3}\\
1\\
\sqrt{2}
\end{array}\right)\label{eq:Uf2}
\end{eqnarray}
If Bob measures that the system is not in $|2\rangle$, then the final
state is
\begin{eqnarray}
U_{f}\frac{1}{\sqrt{2}}\left(\begin{array}{c}
1\\
0\\
1
\end{array}\right) & = & \frac{1}{2\sqrt{3}}\left(\begin{array}{c}
-\sqrt{3}\\
3\\
0
\end{array}\right)\label{eq:Uf-13}
\end{eqnarray}
We see that if Bob determines that the system is not in state $|1\rangle$
(or $|2\rangle)$, then the probability of Alice determining that
the system is in state $|3\rangle$ at time $t_{3}$ is zero. Whenever
Alice measures that the ball is in Box 3 (i.e. when she confirms the
system at time $t_{3}$ is in the state $|3\rangle$), it is certain
that Bob found the ball in the box he measured. This leads to the
paradox.

The measured probabilities for the paradox can be summarised. We follow
the notation where $I_{k}$ represents the ball is in Box I at time
$t_{k}$. The probabilities for detection of a Ball if Bob opens the
Box 1 is $P_{B1}(1_{2})=1/3$. Similarly, $P_{B2}(2_{2})=1/3$. From
(\ref{eq:Uf-1-1}), if Bob detects the ball in Box 1, then $P_{B1}(3_{3}|1_{2})=1/3$.
If Bob opens Box 1, the joint probabilities are $P_{B1}(1_{2},3_{3})=P_{B1}(3_{3}|1_{2})P_{B1}(1_{2})=1/9$.
Also, if Bob opens Box 1, then $P_{B1}(3_{3})=1/9$. Hence, $P_{B1}(1_{2}|3_{3})=1$.
Similarly, if Bob opens Box 2, the probability of him detecting the
ball given Alice detects a Ball in Box 3 is $P_{B2}(2_{2}|3_{3})=1$.
We find 
\begin{equation}
P_{B1}(1_{2}|3_{3})=P_{B2}(2_{2}|3_{3})=1\label{eq:cond-1}
\end{equation}

If there is no measurement by Bob, then the final state at time $t_{3}$
after Alice's operations is 
\begin{eqnarray}
U_{f}|\psi_{1}\rangle & = & \frac{1}{3\sqrt{2}}\left(\begin{array}{c}
0\\
4\\
\sqrt{2}
\end{array}\right)\label{eq:no-m-1}
\end{eqnarray}
The probability of Alice detecting a ball if Bob makes no measurement
is $P_{N}(3_{3})=1/9$. We note that 
\begin{equation}
P_{N}(3_{3})=P_{B1}(3_{3})=P_{B2}(3_{3})=1/9\label{eq:imp-prob}
\end{equation}
Hence, the probability that Alice detects the ball in Box 3 is not
changed by Bob making a measurement. Maroney referred to such a measurement
as \emph{operationally nondisturbing} \cite{Maroney}, since Alice
cannot detect Bob's inteference, if she is restricted to opening Box
3. On the other hand, if Bob makes a measurement, then the final
state on average after Alice's transformations will be the mixture
\begin{eqnarray}
\rho_{mix,m} & = & \frac{1}{6}U_{f}|1\rangle\langle1|U_{f}^{\dagger}+\frac{2}{6}U_{f}(\frac{|2\rangle+|3\rangle}{\sqrt{2}})(\frac{\langle2|+\langle3|}{\sqrt{2}})U_{f}^{\dagger}\nonumber \\
 &  & +\frac{1}{6}U_{f}|2\rangle\langle2|U_{f}^{\dagger}+\frac{2}{6}U_{f}(\frac{|1\rangle+|3\rangle}{\sqrt{2}})(\frac{\langle1|+\langle3|}{\sqrt{2}})U_{f}^{\dagger}\nonumber \\
\label{eq:me-mix}
\end{eqnarray}
The relative probabilities for Alice detecting the Ball in Box 1,
2 or 3 if Bob makes a measurement are $1/3$, $5/9$ and $1/9$, compared
to $0,8/9,1/9$ given by (\ref{eq:no-m-1}), if Bob makes no measurement.
We see that overall, the probabilities are changed.\textcolor{red}{}

\section{Mesoscopic paradox}

A macroscopic version of the paradox can be constructed by considering
that the three states $|1\rangle$, $|2\rangle$ and $|3\rangle$
become macroscopically distinct. It is also necessary to identify
suitable unitary transformations. Macroscopic and mesoscopic versions
can be constructed in a number of ways. In this section, we first
consider a mesoscopic example which is a direct mapping of the original
three-box paradox, with the generalisation that the ``particle''
comprises $N$ quanta. In Section V, we consider a macroscopic example
involving coherent states of a single-mode field, which allows a greater
depth of study of the dynamics associated with the unitary operations.

\subsection{Number states}

We analyse a proposal that maps directly onto the original paradox
described in Section II, where the three boxes correspond to spatially
separated modes. We let
\begin{eqnarray}
|1\rangle & = & |N\rangle_{1}|0\rangle_{2}|0\rangle_{3}\nonumber \\
|2\rangle & = & |0\rangle_{1}|N\rangle_{2}|0\rangle_{3}\nonumber \\
|3\rangle & = & |0\rangle_{1}|0\rangle_{2}|N\rangle_{3}\label{eq:number-states}
\end{eqnarray}
where $|N\rangle$ is a number state. The state $|n\rangle_{i}$ of
the $i$th mode is denoted by the subscript $i=1,2,3$. The subscript
is omitted where the meaning is clear. The modes are prepared in the
superposition state
\begin{equation}
|\psi_{sup}\rangle\equiv|\psi_{1}\rangle=\frac{e^{i\varphi}}{\sqrt{3}}(|3\rangle+e^{i\varphi_{1}}(i|2\rangle-|1\rangle)\label{eq:sup-num}
\end{equation}
where $\varphi$ and $\varphi_{1}$ are phase shifts, and the modes
spatially separated. These states are tripartite extensions of NOON
states \cite{noon-dowling-1}.

\begin{figure}[t]

\begin{centering}
\includegraphics[width=0.5\columnwidth]{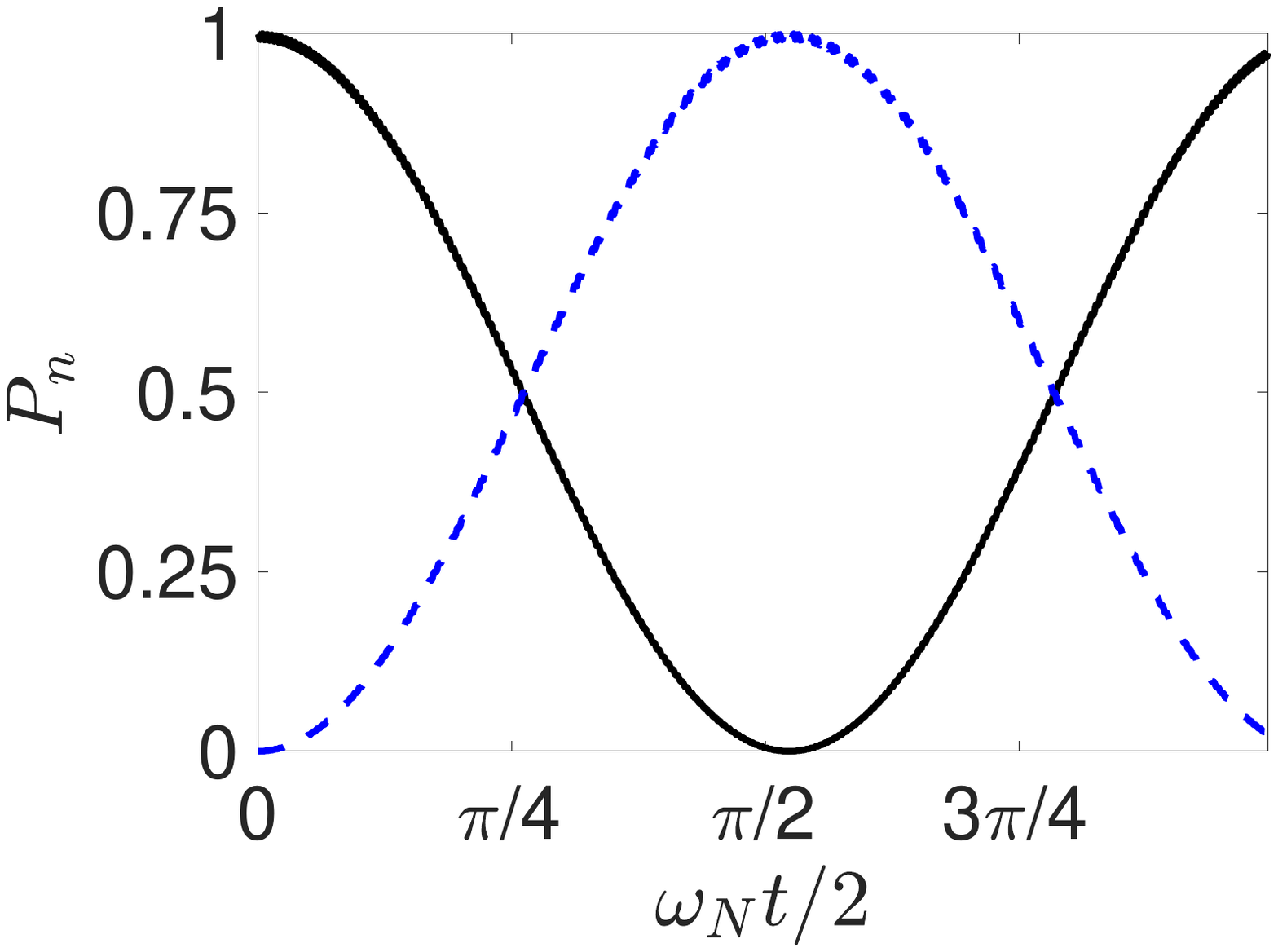}\includegraphics[width=0.5\columnwidth]{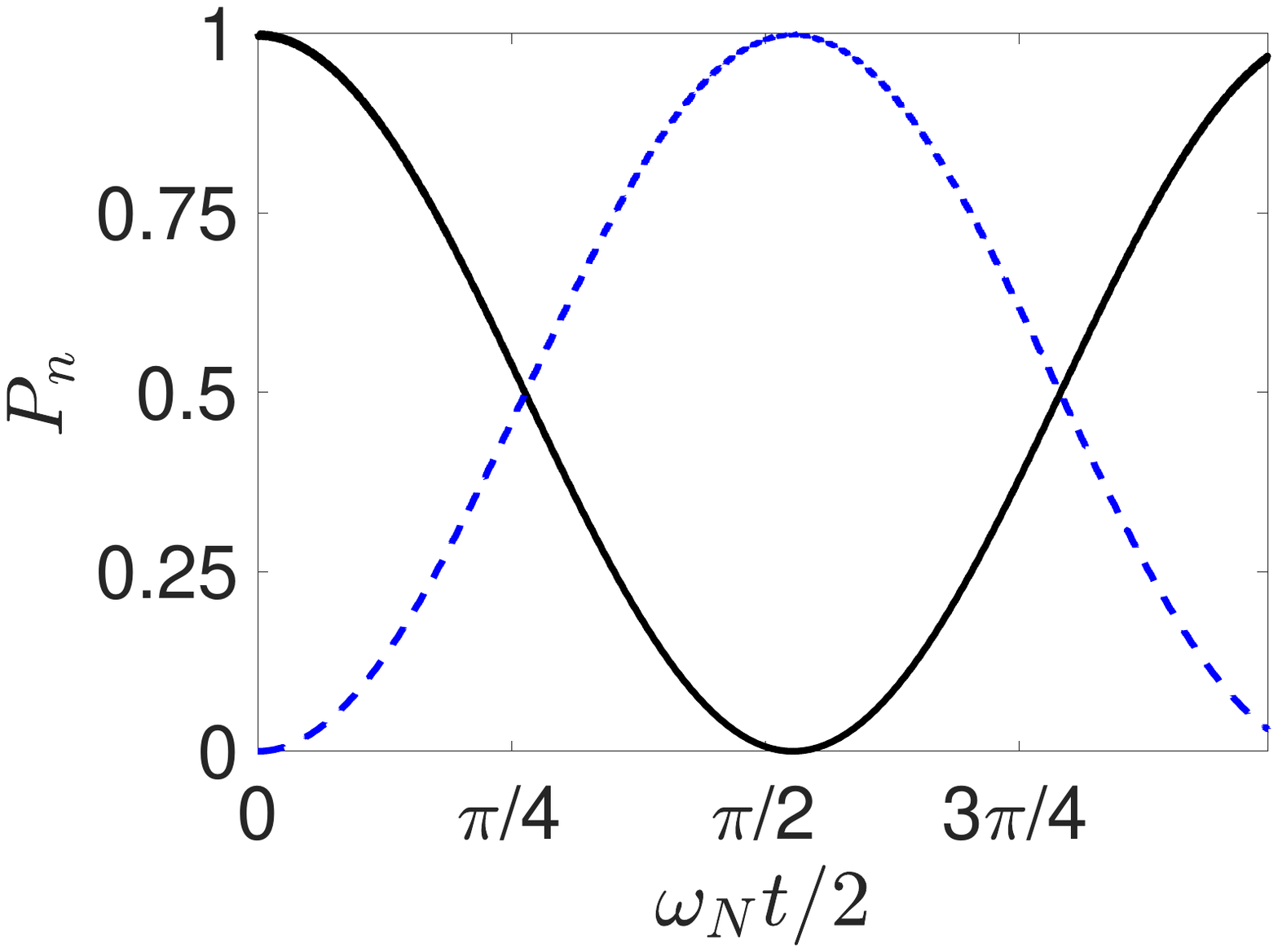}
\par\end{centering}
\caption{Realisation of a nonlinear beam splitter: Solutions are shown for
the Hamiltonian $H_{kl}$ after a time $t$ with initial state $|N\rangle_{k}|0\rangle_{l}$.
Top: $N=2$, $\kappa=1$, $g=30$ (left), and $N=5$, $\kappa=20$,
$g=333.33$ (right).  $P_{N}$ (black solid line) is the probability
for all $N$ bosons to be in mode $a_{k}$;  $P_{0}$ (blue dashed
line) is the probability for all $N$ bosons to be in mode $a_{l}$.
The parameters identify regimes optimal, or nearly optimal, for the
nonlinear beam splitter interaction, where $P_{N}+P_{0}\sim1$ and
$P_{N}\sim\cos^{2}\omega_{N}t$. \textcolor{red}{} \label{fig:n-solns}}
\end{figure}

The unitary transformations necessary for the three-box paradox are
achieved by an interaction $H_{kl}$ that transforms the state $|N\rangle_{k}|0\rangle_{l}$
into the superposition \cite{nonlinear-Ham-1,josHam-collett-steel-2-1,manushan-bellcats-noon,carr-two-well}\textcolor{red}{}
\begin{equation}
e^{i\varphi(\theta)}(\cos\theta|N\rangle_{k}|0\rangle_{l}-i\sin\theta|0\rangle_{k}|N\rangle_{l})\label{eq:hnumtrans}
\end{equation}
and also the state $|0\rangle_{k}|N\rangle_{l}$ into\textcolor{red}{
}
\begin{equation}
e^{i\varphi'(\theta)}(\sin\theta|N\rangle_{k}|0\rangle_{l}+i\cos\theta|0\rangle_{k}|N\rangle_{l})\label{eq:hnumtrans-1}
\end{equation}
For $N=1$, this is achieved by beam splitters or polarising beam
splitters. For $N\geq1$, we use the Josephson interaction that couples
modes $k$ and $l$, given as 
\begin{equation}
H_{kl}=\kappa(\hat{a}_{k}^{\dagger}\hat{a}_{l}+\hat{a}_{k}\hat{a}_{l}^{\dagger})+g\hat{a}_{k}^{\dagger2}\hat{a}_{k}^{2}+g\hat{a}_{l}^{\dagger2}\hat{a}_{l}^{2}\label{eq:hm-2}
\end{equation}
so that $U=e^{-iH_{kl}t/\hbar}$. Calculations have shown the result
(\ref{eq:hnumtrans}) to be realised to an excellent approximation,
for $N\lesssim100$, to the extent that Bell violations are predicted
for systems where the spin states $|\uparrow\rangle$ and $|\downarrow\rangle$
become the mesoscopically distinct states $|N\rangle_{k}|0\rangle_{l}$
and $|0\rangle_{k}|N\rangle_{l}$ \cite{manushan-bellcats-noon}.
Here, $\hat{a}_{k}$, $\hat{a}_{l}$ are the boson destruction operators
for two field modes $k$ and $l$, and $\kappa$ and $g$ are the
interaction constants. The $\theta$ is a function of the interaction
time $t$ and can be selected so that $0\leq\theta\leq2\pi$. We introduce
a scaled time \textcolor{black}{$\theta=\omega_{N}t$ where solutions
are given in \cite{carr-two-well}.} Here, we determine $\theta$
numerically, by solving for the time $T_{NOON}$ taken for the system
to evolve from $|N\rangle_{k}|0\rangle_{l}$ to (\ref{eq:hnumtrans})
where $\theta=\pi/4$. The solutions illustrating (\ref{eq:hnumtrans})
are shown in Figure \ref{fig:n-solns}. 

\begin{figure*}[t]
\begin{centering}
\includegraphics[width=0.6\columnwidth]{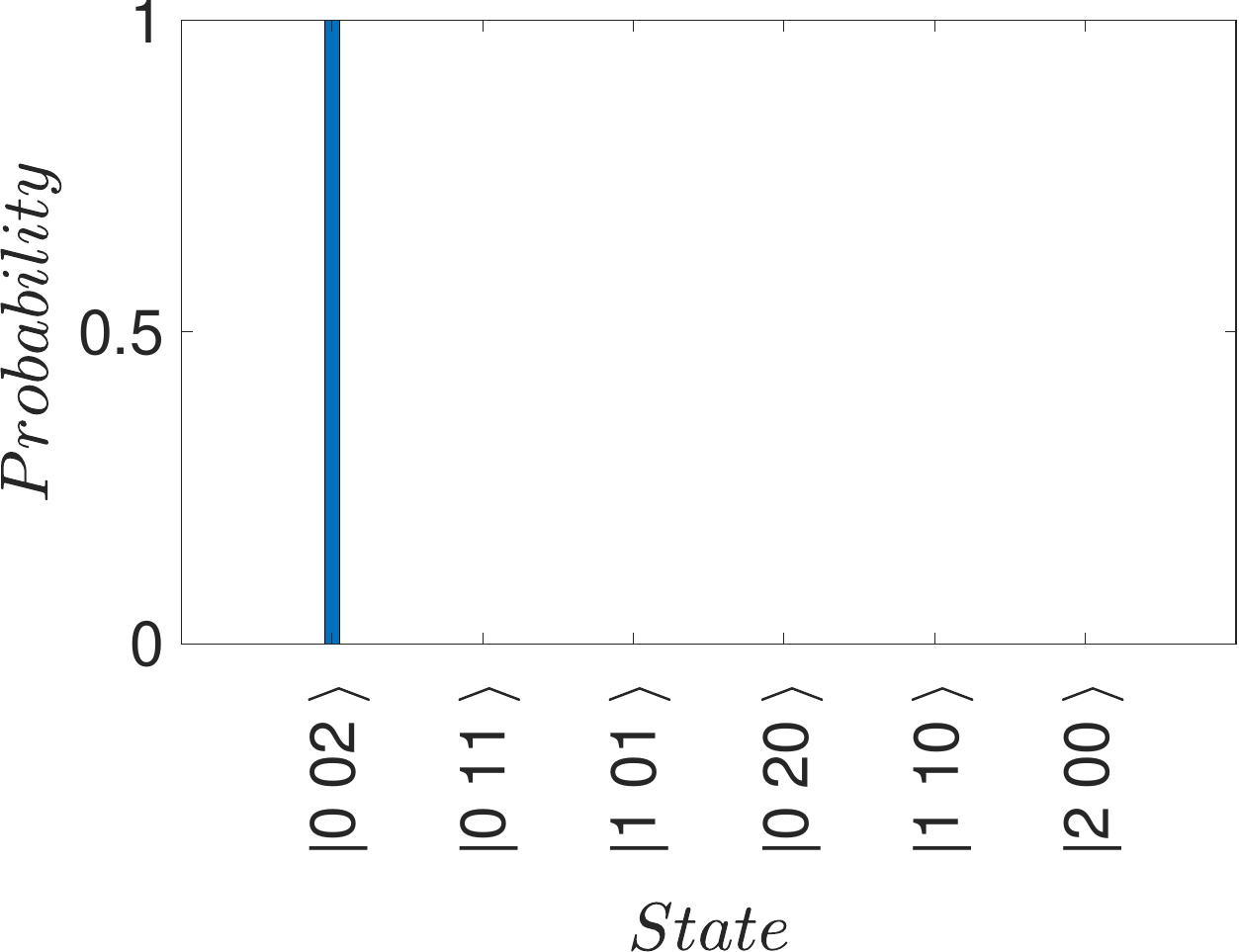}\hspace{0.03\columnwidth}\includegraphics[width=0.6\columnwidth]{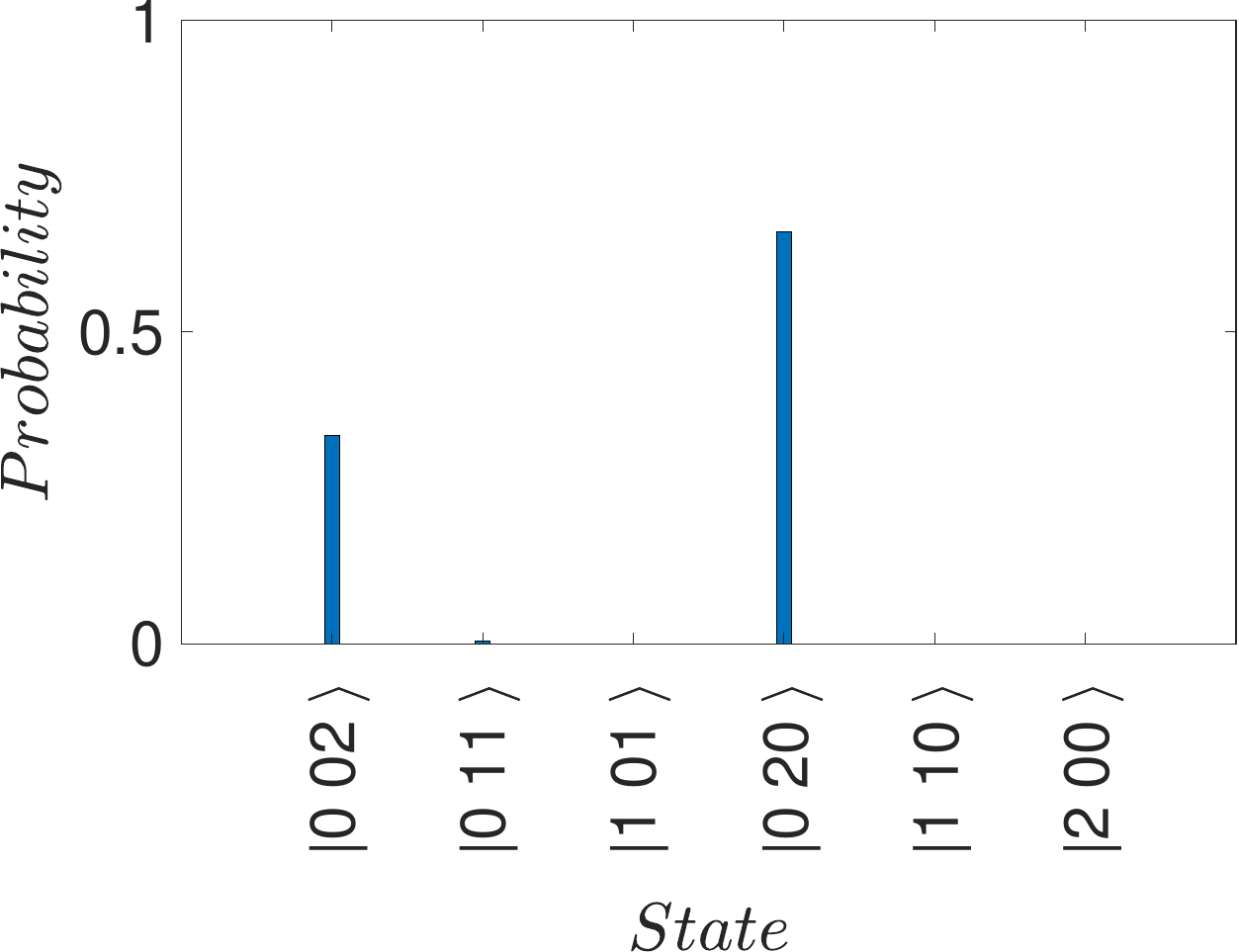}\hspace{0.03\columnwidth}\includegraphics[width=0.6\columnwidth]{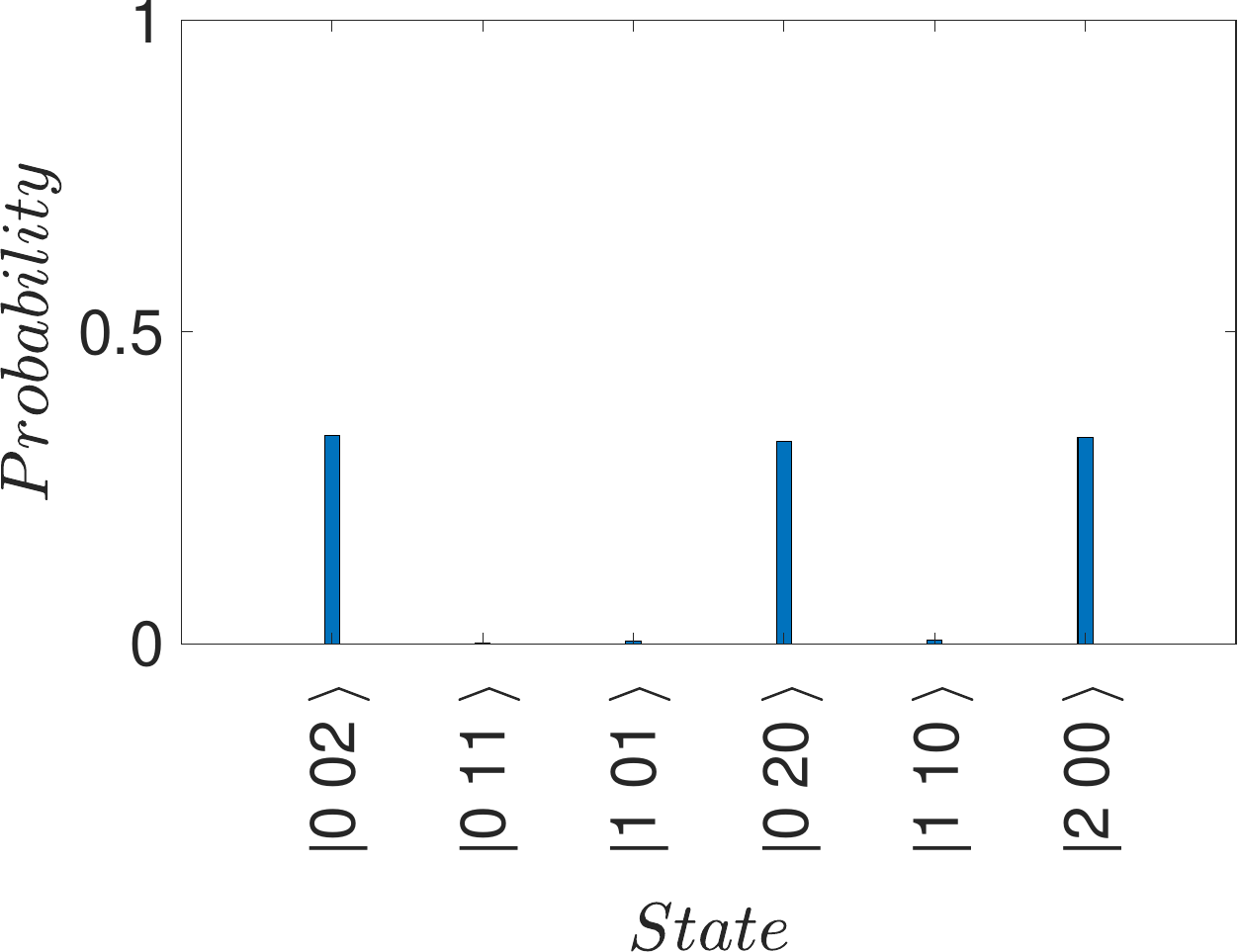}\textcolor{black}{}
\par\end{centering}
\caption{Creation of the superposition $|\psi_{sup}\rangle$ from $|3\rangle$,
for $N=2$. The values of $\kappa$ and $g$ are chosen as in Figure
\ref{fig:n-solns}. Each sequence shows the initial state $|3\rangle$
(far left), the intermediate state $U_{1i}|3\rangle$ (centre), and
the final state $U_{2i}U_{1i}|3\rangle$ (far right), where $U_{1i}=e^{-iH_{32}t_{1i}/\hbar}$
and $U_{2i}=e^{-iH_{21}t_{2i}/\hbar}$ for suitable choices of times
$t_{1i}$ and $t_{2i}$. Depicted is the probability the system is
in state $|k,l,m\rangle\equiv|k\rangle_{1}|l\rangle_{2}|m\rangle_{3}$
at the given time in the sequence. The probability that the system
is in a state different to $|1\rangle$, $|2\rangle$ or $|3\rangle$
is less than $3\times10^{-3}$.\textcolor{blue}{}\label{fig:n-solns-create|sup>}\textcolor{blue}{}}
\end{figure*}

The state $|\psi_{sup}\rangle$ is created from state $|3\rangle$
using the interaction $H_{kl}$ as follows. We define $U_{f}$ so
that
\begin{eqnarray}
U_{f}|\psi_{f}\rangle & = & |3\rangle\nonumber \\
U_{f}^{-1}|3\rangle & = & |\psi_{f}\rangle\label{eq:cond}
\end{eqnarray}
where $|\psi_{f}\rangle$ will be the postselected state. First, we
examine how to create the initial superposition state $|\psi_{sup}\rangle$
from $|3\rangle$ i.e. we find $U_{i}$ such that $U_{i}|3\rangle=|\psi_{sup}\rangle$.
The state
\begin{equation}
|\psi_{2i}\rangle=e^{i\varphi}(\frac{1}{\sqrt{3}}|3\rangle+i\sqrt{\frac{2}{3}}|2\rangle)\label{eq:first-sup}
\end{equation}
is first created from the initial state $|3\rangle$ by evolving with
$H_{32}$ for a suitable time $t_{1i}$, given by $\theta=\omega_{N}t_{1i}=\cos^{-1}(1/\sqrt{3})$
where $3\pi/2<\theta<2\pi$. We find $U_{1i}|3\rangle=e^{-iH_{32}t_{1i}/\hbar}|3\rangle$
where
\begin{equation}
U_{1i}=\frac{1}{\sqrt{3}}\left(\begin{array}{ccc}
\sqrt{3} & 0 & 0\\
0 & -ie^{i\varphi} & ie^{i\varphi}\sqrt{2}\\
0 & e^{i\varphi}\sqrt{2} & e^{i\varphi}
\end{array}\right)\label{eq:umat-7}
\end{equation}
Then the interaction $H_{21}$ for the time $t_{2i}$ given by $\theta=\omega_{N}t_{2}=7\pi/4$
creates $|2\rangle\rightarrow\frac{e^{i\varphi_{1}}}{\sqrt{2}}(|2\rangle+i|1\rangle)$,
to give\textcolor{blue}{}
\begin{equation}
|\psi_{sup}\rangle=\frac{e^{i\varphi}}{\sqrt{3}}(|3\rangle+e^{i\varphi_{1}}(i|2\rangle-|1\rangle)\label{eq:sup9}
\end{equation}
We find $U_{2i}|\psi_{2i}\rangle=e^{-iH_{21}t_{2i}/\hbar}|\psi_{2i}\rangle$
where
\begin{equation}
U_{2i}=\frac{1}{\sqrt{2}}\left(\begin{array}{ccc}
-ie^{i\varphi_{1}} & ie^{i\varphi_{1}} & 0\\
e^{i\varphi_{1}} & e^{i\varphi_{1}} & 0\\
0 & 0 & \sqrt{2}
\end{array}\right)\label{eq:m13}
\end{equation}
We find
\begin{eqnarray}
U_{i} & = & U_{2i}U_{1i}\nonumber \\
 & = & \frac{1}{\sqrt{6}}\left(\begin{array}{ccc}
-i\sqrt{3}e^{i\varphi_{1}} & e^{i\varphi_{1}}e^{i\varphi} & i\sqrt{2}e^{i\varphi_{1}}e^{i\varphi}\\
e^{i\varphi_{1}}\sqrt{3} & -ie^{i\varphi_{1}}e^{i\varphi} & ie^{i\varphi_{1}}\sqrt{2}e^{i\varphi}\\
0 & 2e^{i\varphi} & \sqrt{2}e^{i\varphi}
\end{array}\right)\label{eq:mat-rpod}
\end{eqnarray}
Hence
\begin{equation}
U_{i}^{\dagger}=\frac{1}{\sqrt{6}}\left(\begin{array}{ccc}
i\sqrt{3}e^{-i\varphi_{1}} & e^{-i\varphi_{1}}\sqrt{3} & 0\\
e^{-i\varphi_{1}}e^{-i\varphi} & ie^{-i\varphi}e^{-i\varphi_{1}} & 2e^{-i\varphi}\\
-\sqrt{2}e^{-i\varphi_{1}}e^{-i\varphi} & -i\sqrt{2}e^{-i\varphi_{1}}e^{-i\varphi} & \sqrt{2}e^{-i\varphi}
\end{array}\right)\label{eq:solnudag}
\end{equation}
The dynamics arising from the Hamiltonians $H_{32}$ and $H_{21}$
for actual values of $g$ and $\kappa$ does not in general constrain
the system to the states $|1\rangle$, $|2\rangle$ or $|3\rangle.$
Full solutions are depicted in Figure \ref{fig:n-solns-create|sup>}.
For the parameters given, the probability that the system is found
in a state different to $|1\rangle$, $|2\rangle$ or $|3\rangle$
is however negligible. 
\begin{figure*}[t]
\begin{centering}
\par\end{centering}
\begin{centering}
\includegraphics[width=0.6\columnwidth]{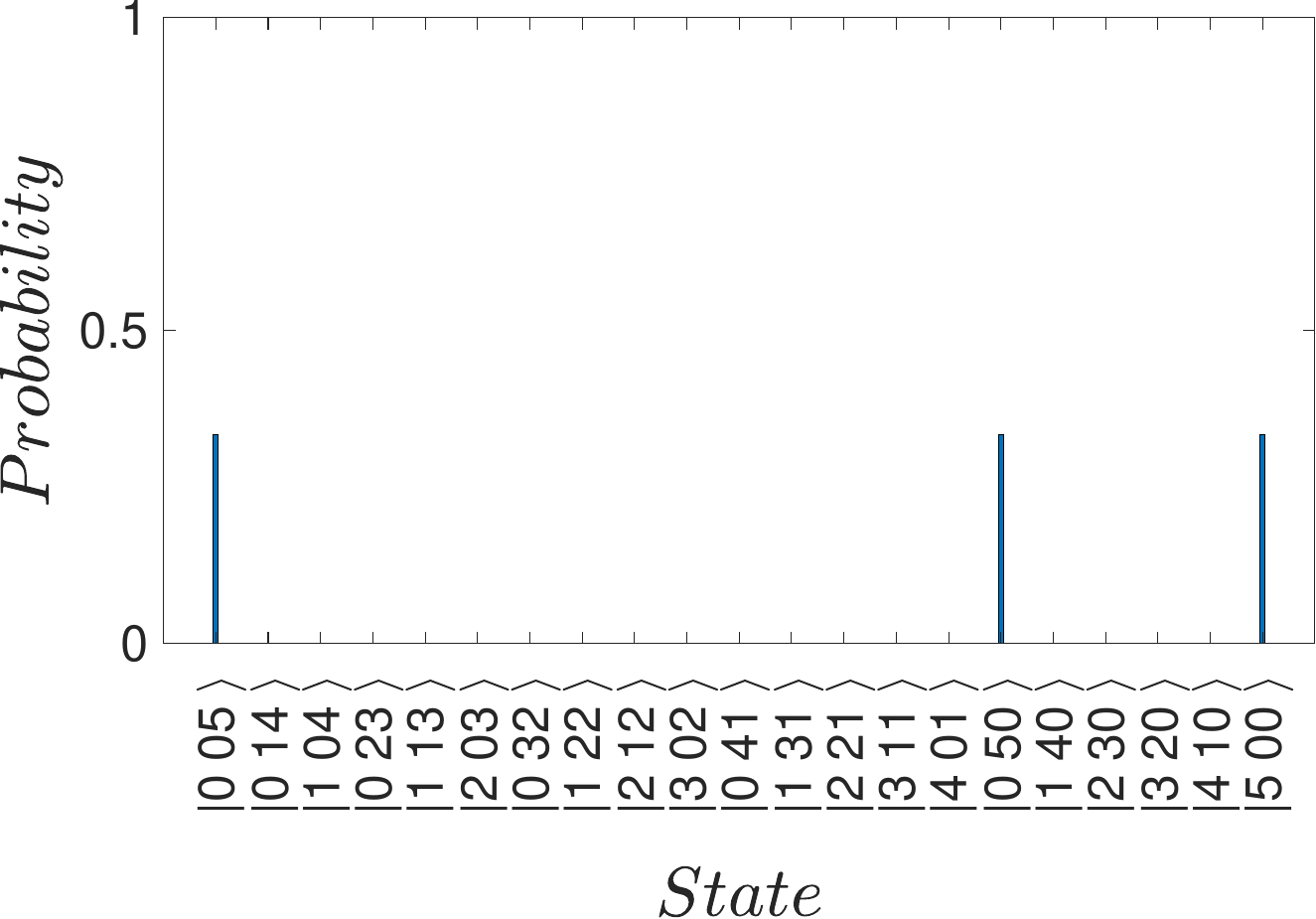}\hspace{0.03\columnwidth}\includegraphics[width=0.6\columnwidth]{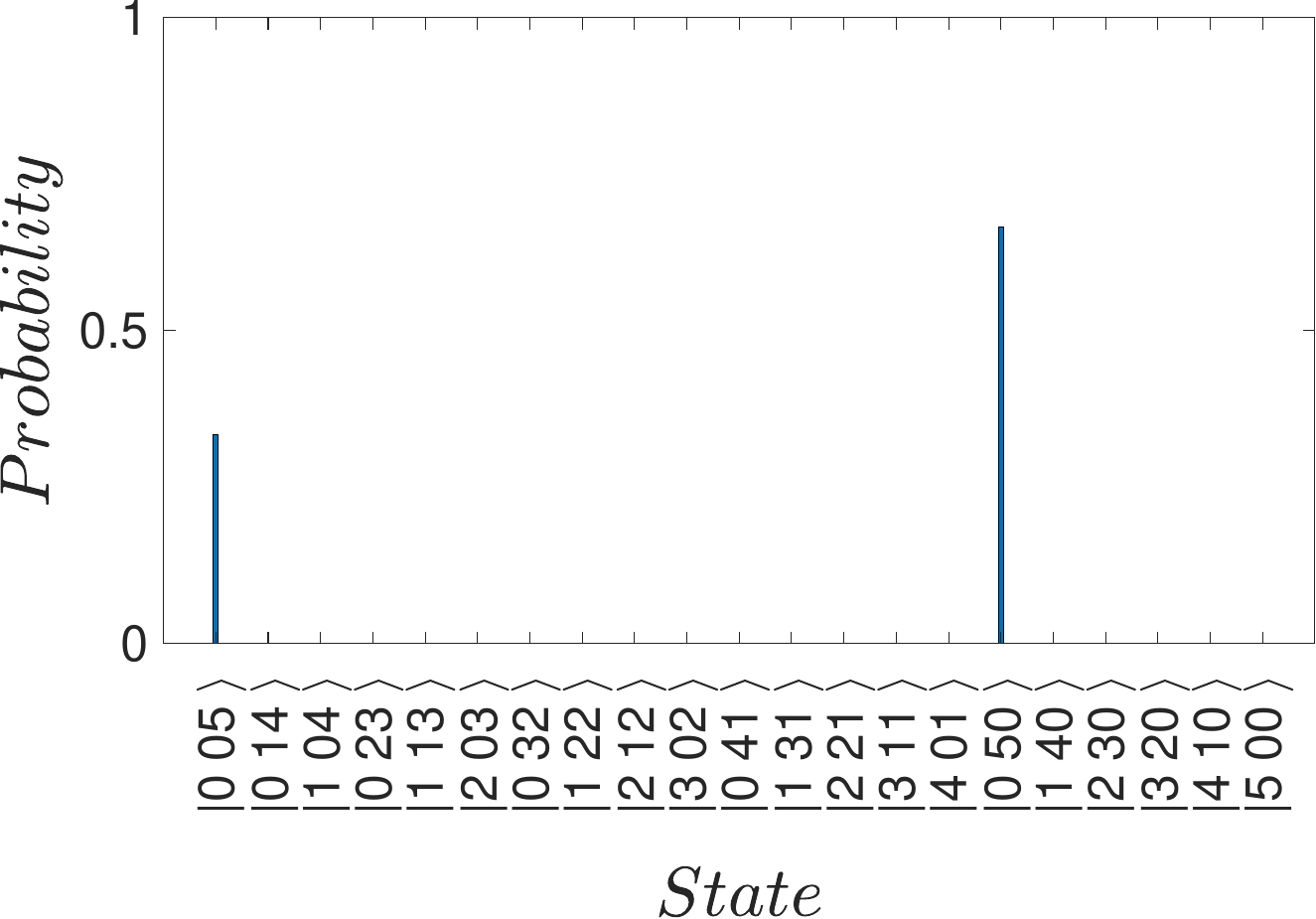}\hspace{0.03\columnwidth}\includegraphics[width=0.6\columnwidth]{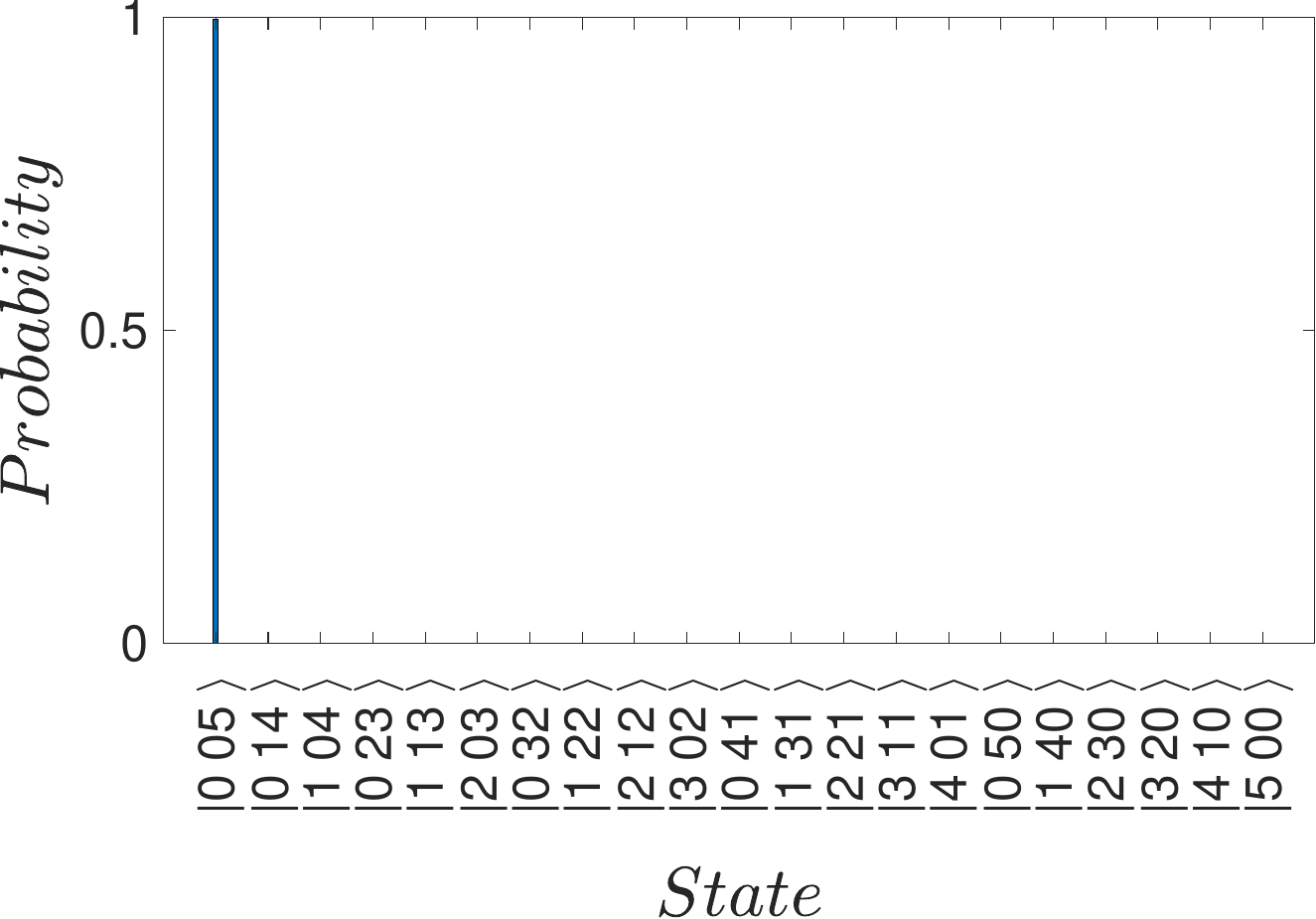}
\par\end{centering}
\begin{centering}
\par\end{centering}
\caption{\textcolor{red}{}Creation of the superposition $|3\rangle$ from
the postselected state $|\psi_{f}\rangle$ using the operations $U_{f}$,\textcolor{red}{\label{fig:n-solns-postselect}}
for  $N=5$. The values of $\kappa$ and $g$ are chosen as in Figure
\ref{fig:n-solns}. Each sequence shows the initial state $|\psi_{f}\rangle$
(far left), the intermediate state $U_{2f}^{-1}|\psi_{f}\rangle$
(centre), and the final state $U_{1f}^{-1}U_{2f}^{-1}|\psi_{f}\rangle$
(far right), where $U_{1f}\equiv U_{32}$ and $U_{2f}\equiv U_{21}$
as defined in the text. Depicted is the probability the system is
in state $|k,l,m\rangle\equiv|k\rangle_{1}|l\rangle_{2}|m\rangle_{3}$
at the given time in the sequence. The probability that the system
is in a state different to $|1\rangle$, $|2\rangle$ or $|3\rangle$
is less than $6\times10^{-3}$.\textcolor{blue}{}}
\end{figure*}
Now reversing, we see that the postselected state
\begin{equation}
|\psi_{f}\rangle=\frac{e^{i\varphi}}{\sqrt{3}}(-|3\rangle+e^{i\varphi_{1}}(i|2\rangle-|1\rangle))\label{eq:post-num}
\end{equation}
can be created from $|3\rangle$ using $U_{f}^{-1}$, by applying
$H_{32}$ with $\theta=\omega_{N}t$, so that $\cos\theta=-\frac{1}{\sqrt{3}}$,
$\sin\theta=-\sqrt{2/3}$, followed by $H_{21}$. We start with $|3\rangle$
and act with $U_{1f}=U_{32}=e^{-iH_{32}t/\hbar}$, for $t=\theta/\omega_{N}$,
so that where
\begin{equation}
U_{1f}=\frac{1}{\sqrt{3}}\left(\begin{array}{ccc}
\sqrt{3} & 0 & 0\\
0 & -ie^{i\varphi} & ie^{i\varphi}\sqrt{2}\\
0 & -e^{i\varphi}\sqrt{2} & -e^{i\varphi}
\end{array}\right)\label{eq:um1}
\end{equation}
Here,
\[
U_{1f}^{\dagger}=\frac{1}{\sqrt{3}}\left(\begin{array}{ccc}
\sqrt{3} & 0 & 0\\
0 & ie^{-i\varphi} & -e^{-i\varphi}\sqrt{2}\\
0 & -ie^{-i\varphi}\sqrt{2} & -e^{-i\varphi}
\end{array}\right)
\]
This creates state $|\psi_{1f}\rangle=\frac{e^{i\varphi}}{\sqrt{3}}(-|3\rangle+i\sqrt{2}|2\rangle)$.
Then we act on $|\psi_{1f}\rangle$ with $U_{2f}=U_{21}=e^{-iH_{21}t/\hbar}$
for $t$ such that\textcolor{red}{} $t=7\pi/4\omega_{N}$ so that
$U_{2f}=U_{2i}$. We see that 
\begin{eqnarray}
U_{f}^{-1} & = & U_{2f}U_{1f}\nonumber \\
 & = & \frac{1}{\sqrt{6}}\left(\begin{array}{ccc}
-i\sqrt{3}e^{i\varphi_{1}} & e^{i\varphi_{1}}e^{i\varphi} & -e^{i\varphi_{1}}e^{i\varphi}\sqrt{2}\\
\sqrt{3}e^{i\varphi_{1}} & -ie^{i\varphi_{1}}e^{i\varphi} & ie^{i\varphi}e^{i\varphi_{1}}\sqrt{2}\\
0 & -2e^{i\varphi} & -\sqrt{2}e^{i\varphi}
\end{array}\right)\nonumber \\
\label{eq:prod2-1}
\end{eqnarray}
This gives state $|\psi_{f}\rangle$.\textcolor{red}{{} }Hence, $U_{f}^{-1}|3\rangle=|\psi_{f}\rangle$
where $U_{f}^{-1}=U_{2f}U_{1f}$. Hence, $U_{f}=U_{1f}^{\dagger}U_{2f}^{\dagger}$.\textcolor{red}{{}
} Hence, Alice's measurements are $U_{2f}^{-1}$ followed by $U_{1f}^{-1}$.
This gives $U_{f}|\psi_{f}\rangle=|3\rangle$. We find\textcolor{green}{}
\begin{eqnarray}
U_{f} & = & \frac{1}{\sqrt{6}}\left(\begin{array}{ccc}
i\sqrt{3}e^{-i\varphi_{1}} & \sqrt{3}e^{-i\varphi_{1}} & 0\\
e^{-i\varphi_{1}}e^{-i\varphi} & ie^{-i\varphi_{1}}e^{-i\varphi} & -2e^{-i\varphi}\\
-e^{-i\varphi_{1}}e^{-i\varphi}\sqrt{2} & -ie^{-i\varphi}e^{-i\varphi_{1}}\sqrt{2} & -\sqrt{2}e^{-i\varphi}
\end{array}\right)\nonumber \\
\label{eq:alice-uf}
\end{eqnarray}
Full solutions are depicted in Figure \ref{fig:n-solns-postselect}.

\begin{figure*}[t]
\begin{centering}
\includegraphics[width=0.6\columnwidth]{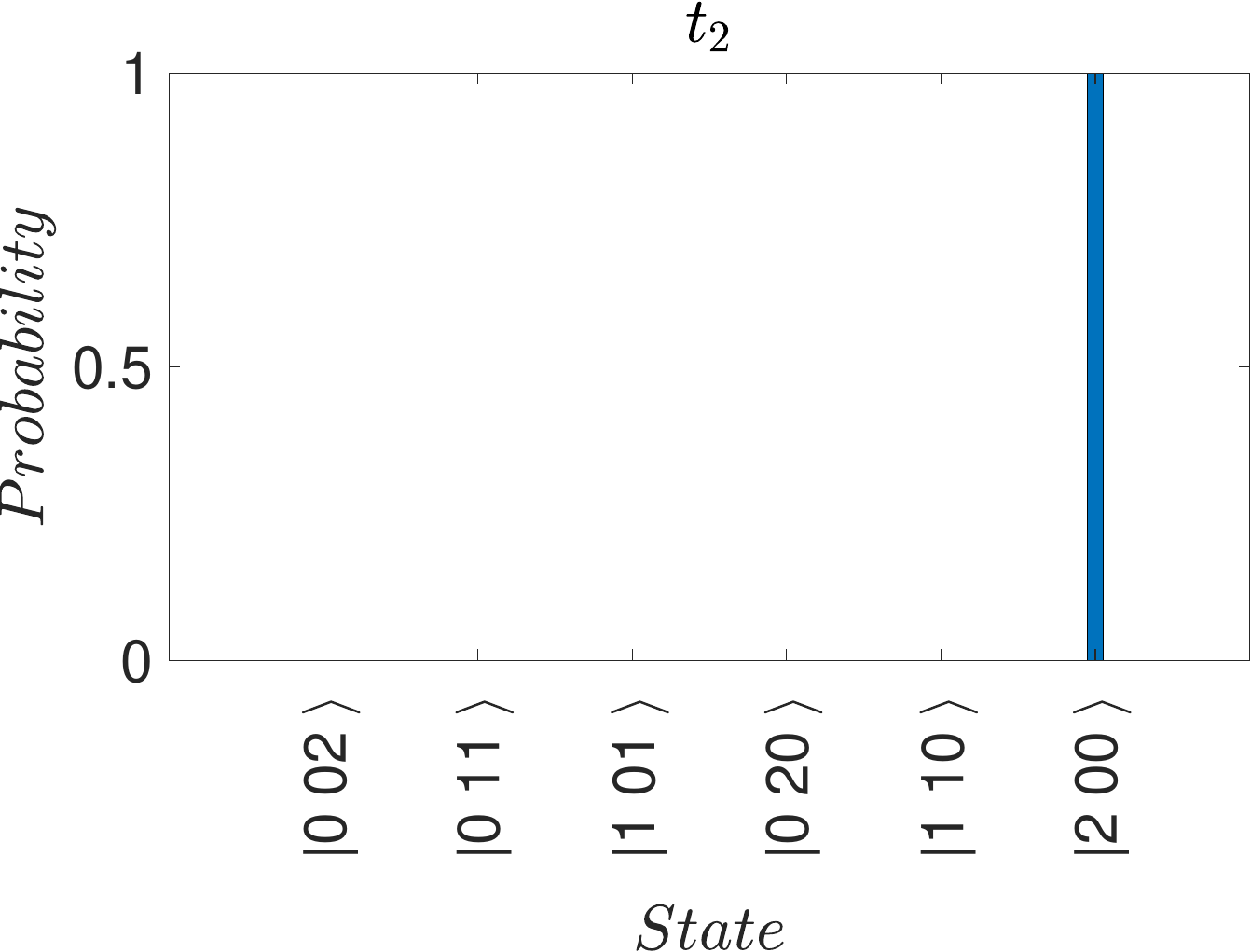}\hspace{0.03\columnwidth}\includegraphics[width=0.6\columnwidth]{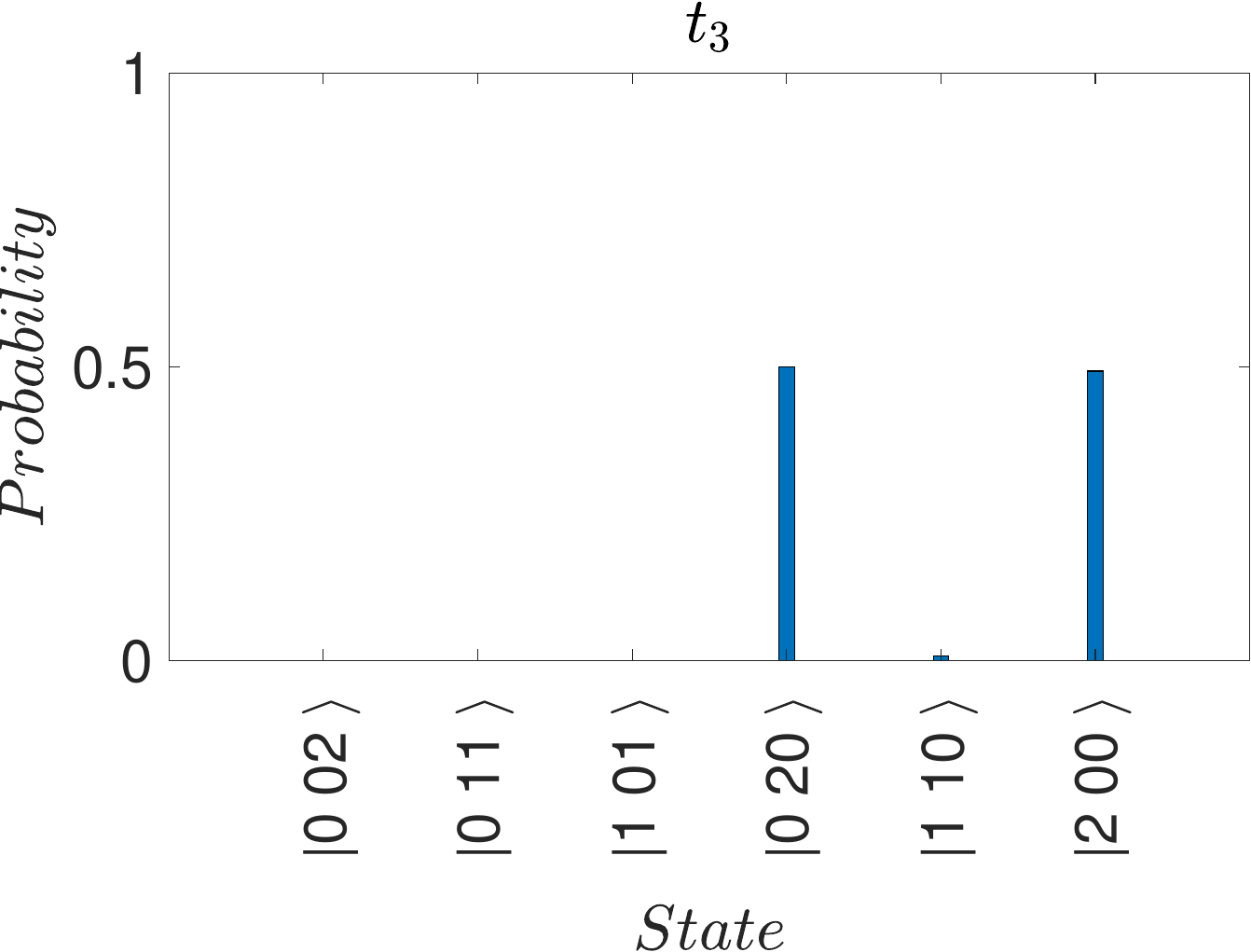}\hspace{0.03\columnwidth}\includegraphics[width=0.6\columnwidth]{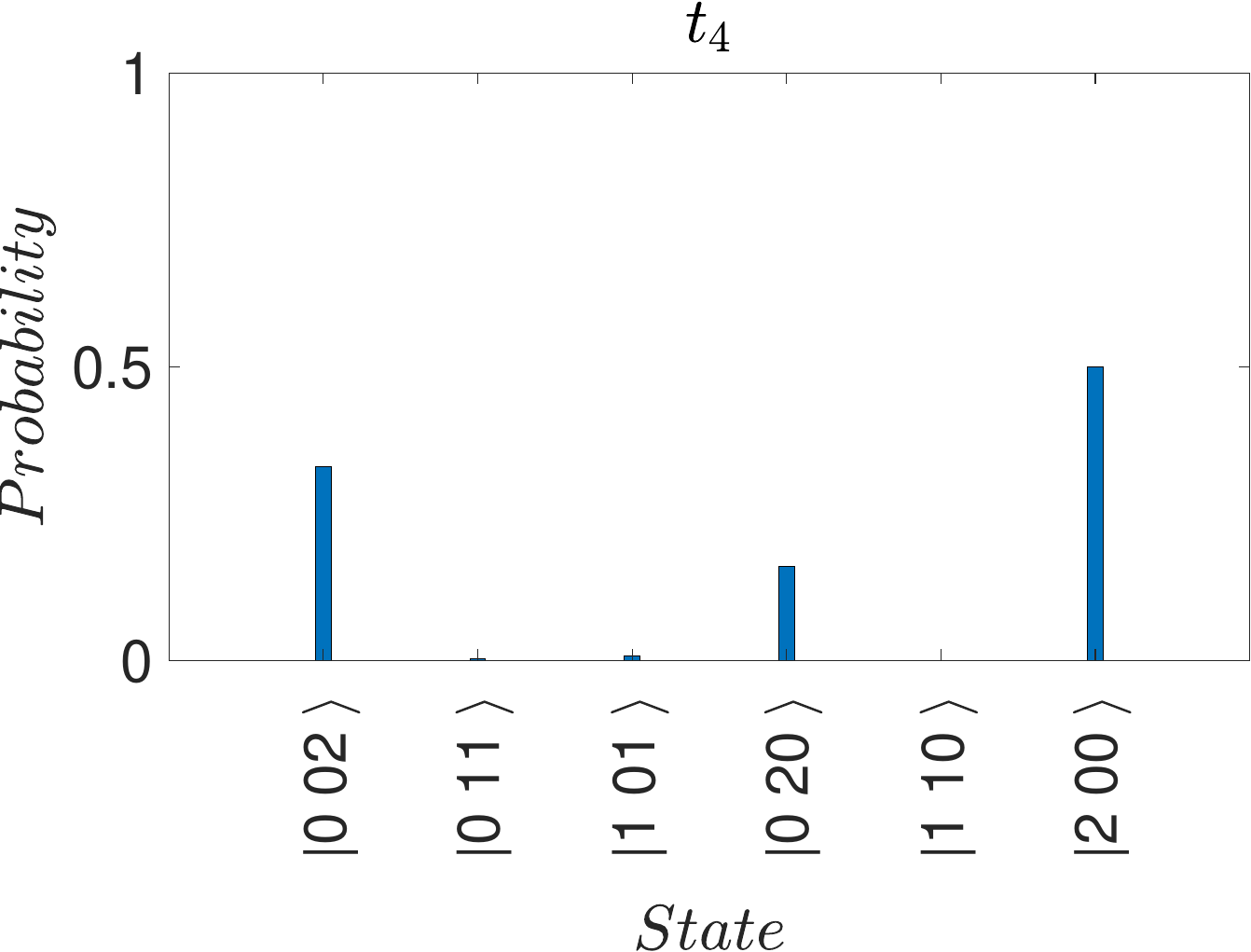}\medskip{}
\par\end{centering}
\begin{centering}
\includegraphics[width=0.6\columnwidth]{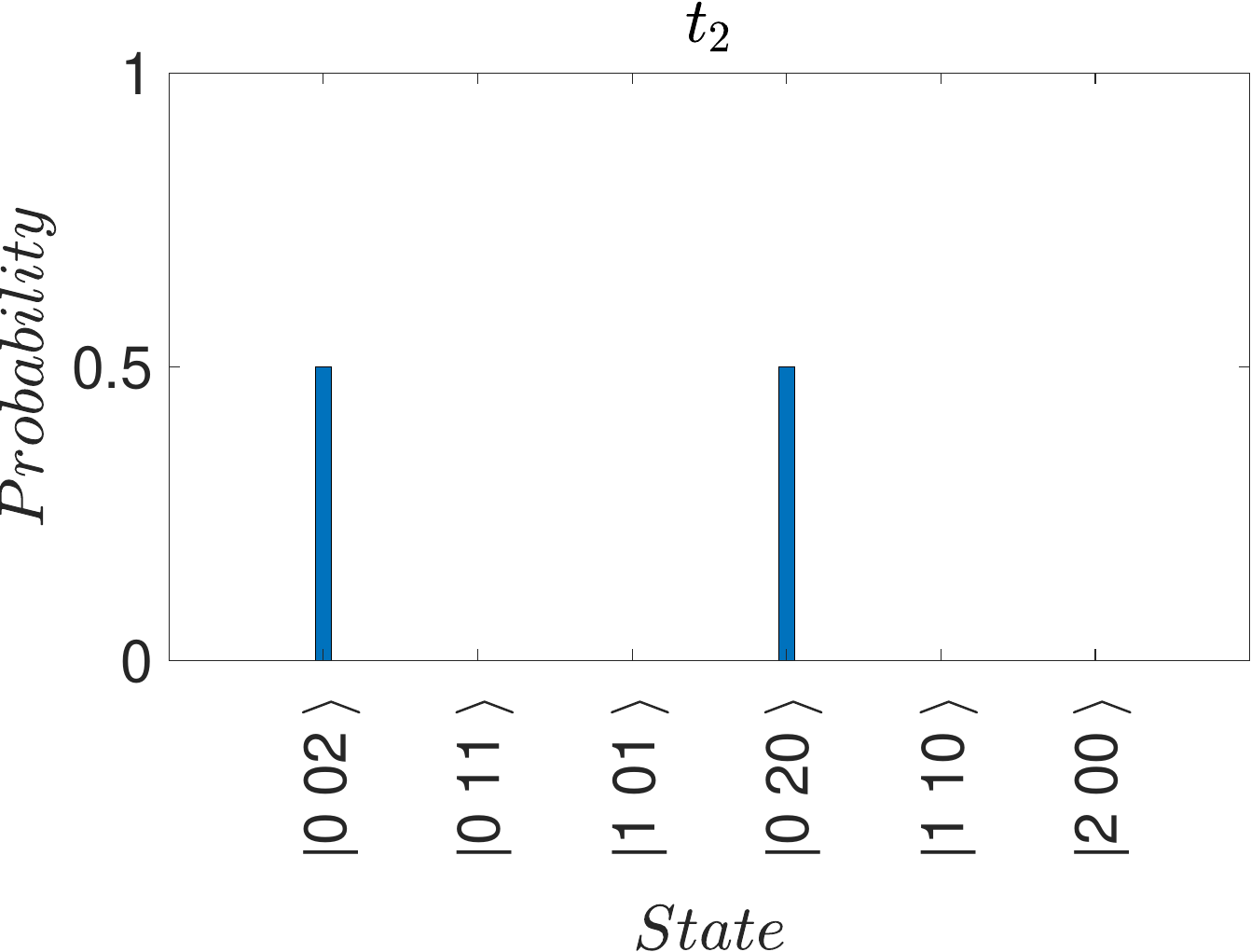}\hspace{0.03\columnwidth}\includegraphics[width=0.6\columnwidth]{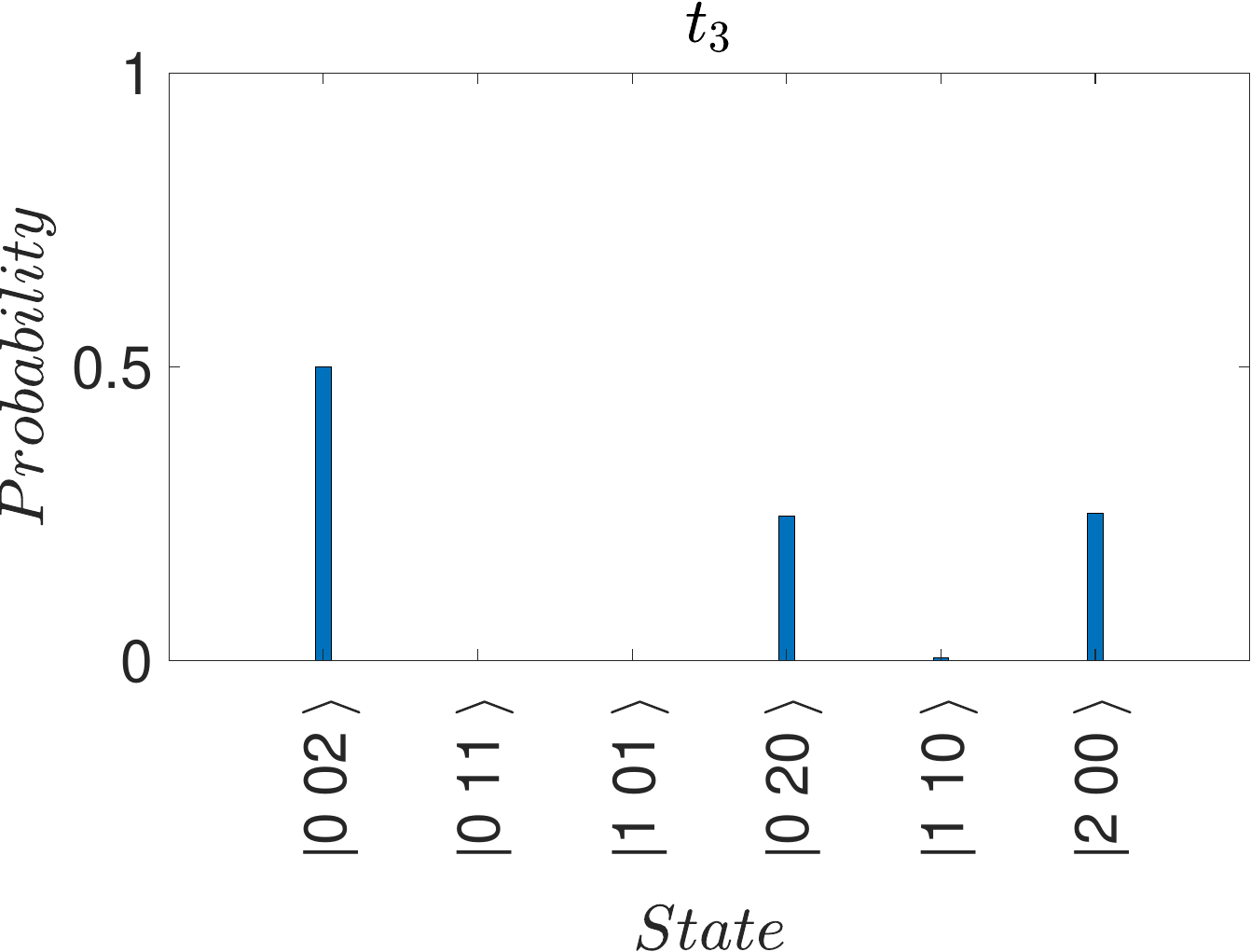}\hspace{0.03\columnwidth}\includegraphics[width=0.6\columnwidth]{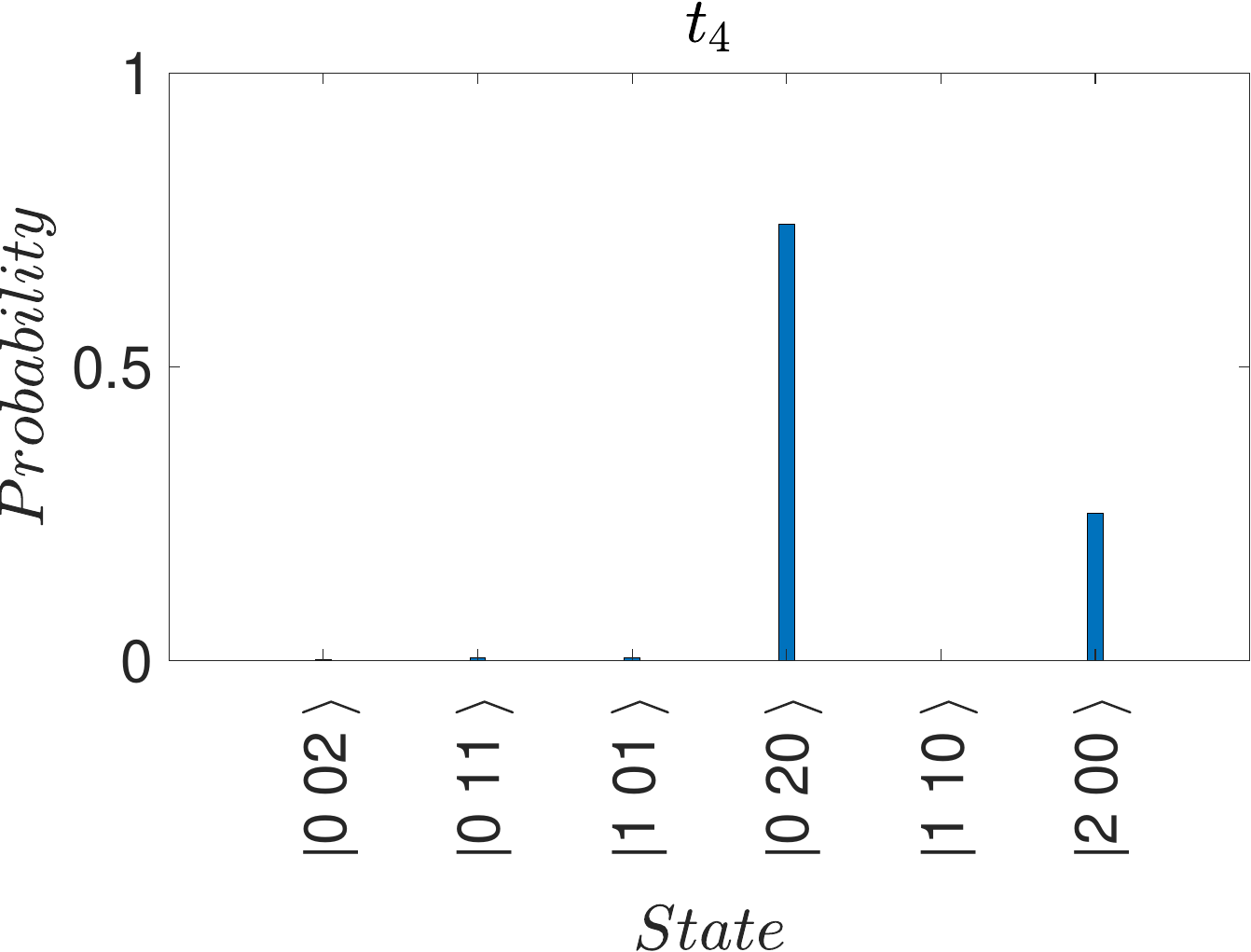}
\par\end{centering}
\caption{The dynamics corresponding to Alice's transformations $U_{f}$ if
(top row): Bob detects $N$ photons in Box 1 at time $t_{2}$ (top
row); or (lower row), if Bob detects the photons are not in Box 1
at time $t_{2}$. The histograms give the probabilities for detecting
$N$ photons in Box $k$. Depicted is the probability the system
is in state $|k,l,m\rangle\equiv|k\rangle_{1}|l\rangle_{2}|m\rangle_{3}$
at the given time in the sequence. Here, we show the initial state
after Bob's measurement, at time $t_{2}$ (far left), the state generated
at time $t_{3}$ after Alice's transformation $U_{2f}^{-1}$ (centre),
and the state generated at time $t_{4}$ after Alice's further transformation
$U_{1f}^{-1}$\textcolor{red}{{} }(far right). The final state after
Alice's total transformation $U_{f}$ is (\ref{eq:soln2}) to an excellent
approximation. The probability that the system is in any other state
apart from $|00N\rangle$, $|0N0\rangle$ or $|N00\rangle$ is less
than\textcolor{red}{{} }$8\times10^{-3}$.\textcolor{red}{{} }The solutions
are for $N=2$. \textcolor{red}{}\label{fig:n-solns-bob-measures-1-then-Uf/or not}.}
\end{figure*}

The paradox follows as for the original paradox. Bob determines whether
the system is in state $|1\rangle$ or not. Alternatively, he determines
whether the system is in state $|2\rangle$, or not. If after Bob's
measurements the system is in state $|1\rangle$, then after Alice's
measurements the system is in\textcolor{green}{}
\begin{equation}
U_{f}|1\rangle=\frac{e^{-i\varphi_{1}}}{\sqrt{6}}\left(\begin{array}{c}
i\sqrt{3}\\
e^{-i\varphi}\\
-e^{-i\varphi}\sqrt{2}
\end{array}\right)\label{eq:soln2}
\end{equation}
The solutions in Figure \ref{fig:n-solns-bob-measures-1-then-Uf/or not}
for the optimal choice of $\kappa$ and $g$ give agreement for $N=2$.
Similarly, if Bob determines that the system is in state $|2\rangle$,
then after Alice's measurements the system is in
\begin{equation}
U_{f}|2\rangle=\frac{e^{-i\varphi_{1}}}{\sqrt{6}}\left(\begin{array}{c}
\sqrt{3}\\
ie^{-i\varphi}\\
-ie^{-i\varphi}\sqrt{2}
\end{array}\right)\label{eq:soln3}
\end{equation}
If Bob determines the system is not in $|1\rangle$, then at time
$t_{2}$ the system is in $(|3\rangle+e^{i\varphi_{1}}i|2\rangle)/\sqrt{2}$.
The final state after Alice's transformations is
\begin{eqnarray}
U_{f}\frac{1}{\sqrt{2}}\left(\begin{array}{c}
0\\
ie^{i\varphi_{1}}\\
1
\end{array}\right) & = & \frac{1}{2\sqrt{3}}\left(\begin{array}{c}
\sqrt{3}\\
-3e^{-i\varphi}\\
0
\end{array}\right)\label{eq:soln5}
\end{eqnarray}
This is depicted in Figure \ref{fig:n-solns-bob-measures-1-then-Uf/or not}.
If Bob determines the system is not in $|2\rangle$, then at time
$t_{2}$ the system is in $(|3\rangle-e^{i\varphi_{1}}|1\rangle)/\sqrt{2}$.
The final state after Alice's transformations is
\begin{eqnarray}
U_{f}\frac{1}{\sqrt{2}}\left(\begin{array}{c}
-e^{i\varphi_{1}}\\
0\\
1
\end{array}\right) & = & \frac{1}{2\sqrt{3}}\left(\begin{array}{c}
-i\sqrt{3}\\
-3e^{-i\varphi}\\
0
\end{array}\right)\label{eq:soln6}
\end{eqnarray}
The paradox occurs because Alice finds that there is zero probability
of finding the system in the state $|3\rangle$ in both cases.

The calculations for the marginal and joint probabilities follows
along the same lines as in Section II for the original paradox. We
note that if there is no measurement by Bob, then the final state
at time $t_{3}$ is 
\begin{eqnarray}
U_{f}|\psi_{sup}\rangle & = & \frac{1}{3\sqrt{2}}\left(\begin{array}{c}
0\\
-4\\
\sqrt{2}
\end{array}\right)\label{eq:number-no-meas-by-bob}
\end{eqnarray}
This implies $P_{N}(3_{3})=1/9$. As for the original paradox, this
agrees with the value of $P_{B1}(3^{3})=1/9$, calculated from the
above results, where Bob opens Box 1. Similarly, $P_{B2}(3^{3})=1/9$.
Hence, 
\[
P_{B1}(1_{2}|3_{3})=1,P_{B2}(2_{2}|3_{3})=1
\]
If Bob makes a measurement, the system reduces to the mixture corresponding
to the outcomes obtained by Bob. Overall, if he opens Box 1, the state
is 
\begin{eqnarray}
\rho_{mix,1}(t_{3}) & = & \frac{1}{3}U_{f}|1\rangle\langle1|U_{f}^{\dagger}\nonumber \\
 &  & +\frac{1}{3}U_{f}(|3\rangle+e^{i\varphi_{1}}i|2\rangle)(\langle3|-ie^{-i\varphi_{1}}\langle2|)U_{f}^{\dagger}\nonumber \\
\label{eq:q-n-bobmeasures1}
\end{eqnarray}
Similarly, if he opens Box 2, the state is
\begin{eqnarray}
\rho_{mix,2}(t_{3}) & = & \frac{1}{3}U_{f}|2\rangle\langle2|U_{f}^{\dagger}\nonumber \\
 &  & +\frac{1}{3}U_{f}(|3\rangle-e^{i\varphi_{1}}|2\rangle)(\langle3|-e^{i\varphi_{1}}\langle2|)U_{f}^{\dagger}\nonumber \\
\label{eq:q-n-bobmeasures2}
\end{eqnarray}

We emphasize that the solution (\ref{eq:hnumtrans}) for the evolution
given by $H$ is approximate. Actual solutions are given in the figures,
and are sufficient to confirm the three-box paradox for moderate $N$,
illustrated by $N=2$ and $N=5$. 

\section{Consistency with weak macroscopic realism: three mode example}

By definition, macroscopic realism posits that the system described
by the macroscopic superposition states $|\psi_{k}\rangle$ at time
$t_{k}$ can be described by a set of values $\lambda_{i}^{(k)}$
which determines the outcome of $\hat{n}_{i}$ for each mode \cite{manushan-bell-cat-lg}.
The variable $\lambda_{i}^{(k)}$ assumes the values $+1$ or $-1$:
$1$ indicates the outcome to be $N$; $-1$ indicates the outcome
to be $0$.\textcolor{black}{}
\begin{figure*}[t]
\begin{centering}
\textcolor{black}{\includegraphics[width=0.5\columnwidth]{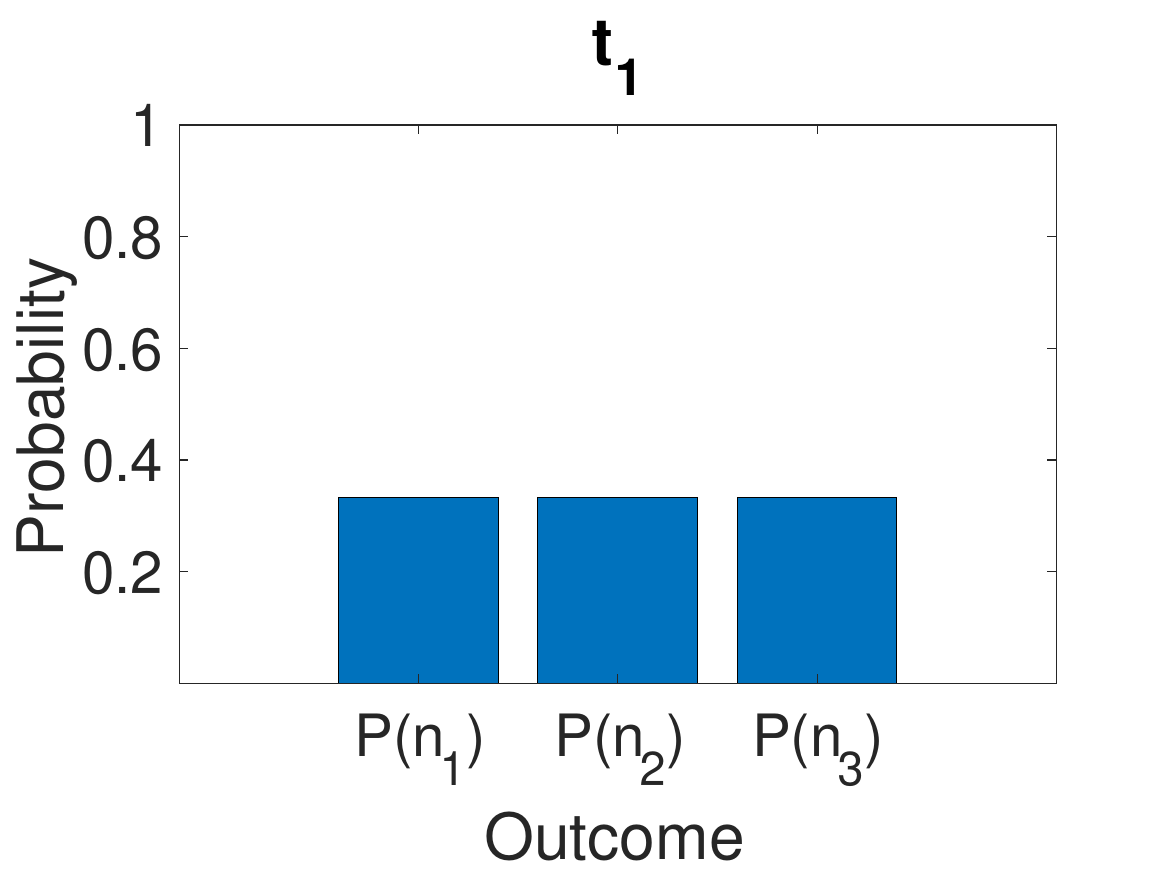}\includegraphics[width=0.5\columnwidth]{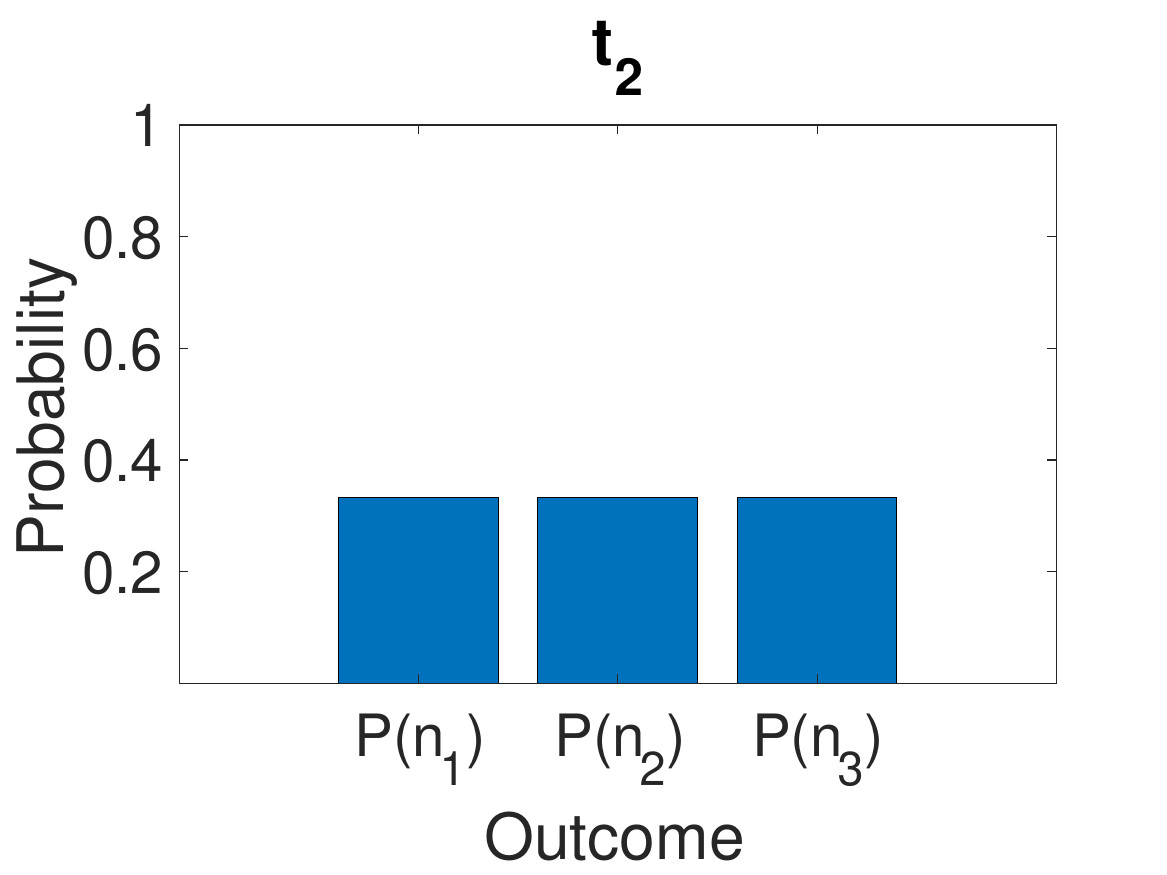}\includegraphics[width=0.5\columnwidth]{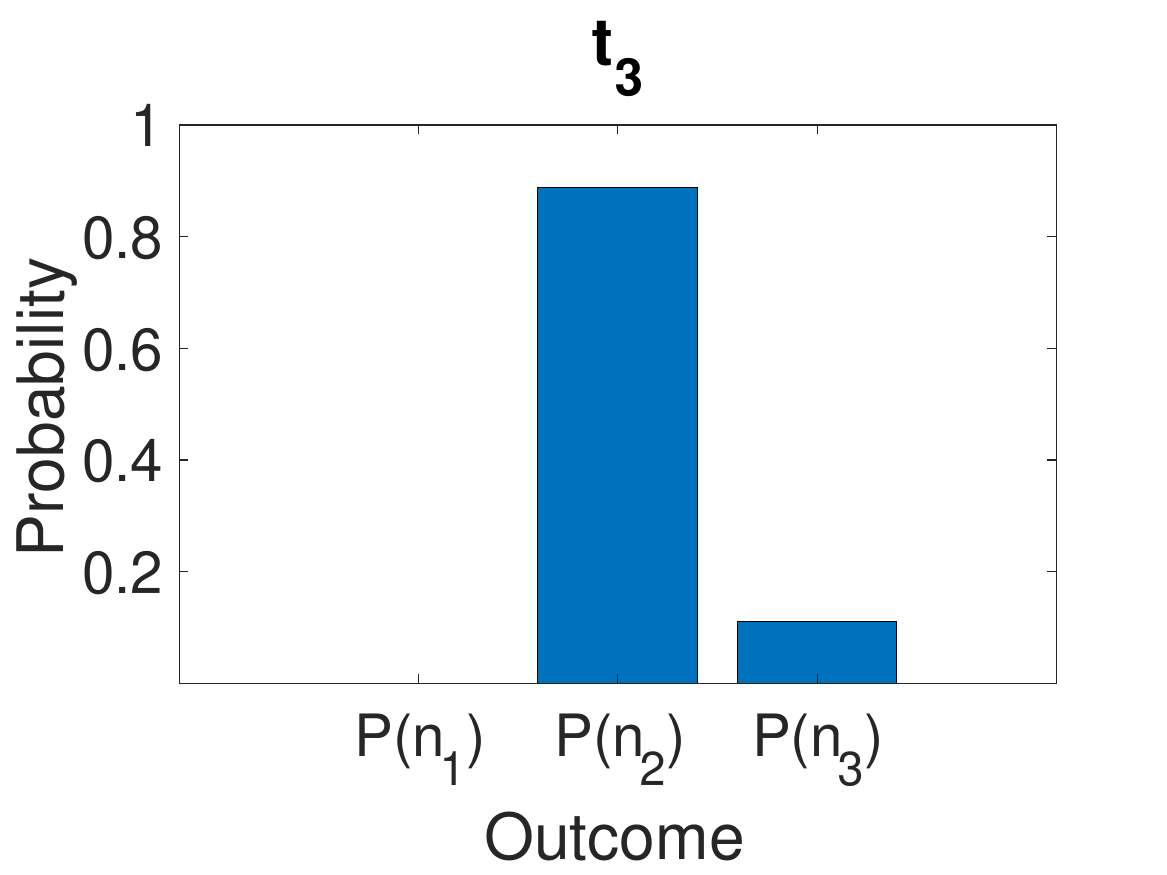}}
\par\end{centering}
\medskip{}

\begin{centering}
\textcolor{black}{\includegraphics[width=0.5\columnwidth]{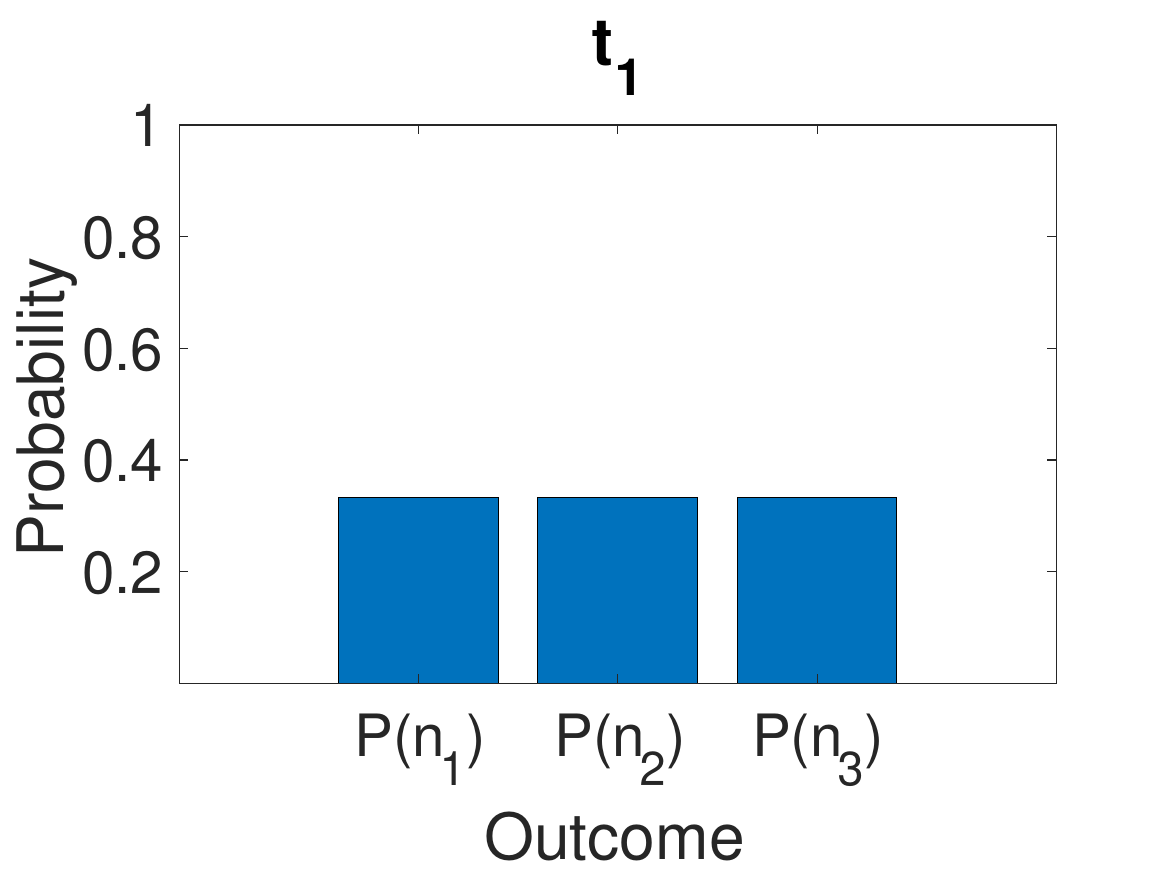}\includegraphics[width=0.5\columnwidth]{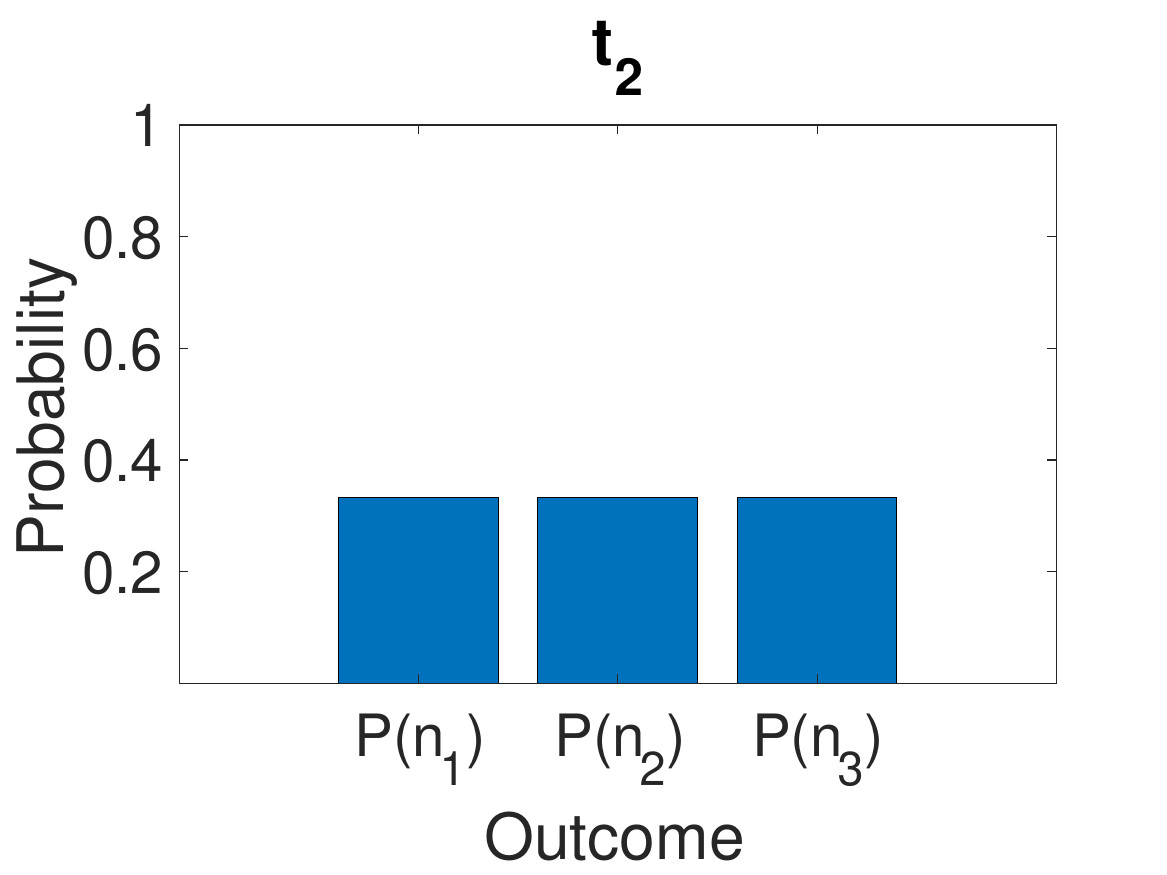}\includegraphics[width=0.5\columnwidth]{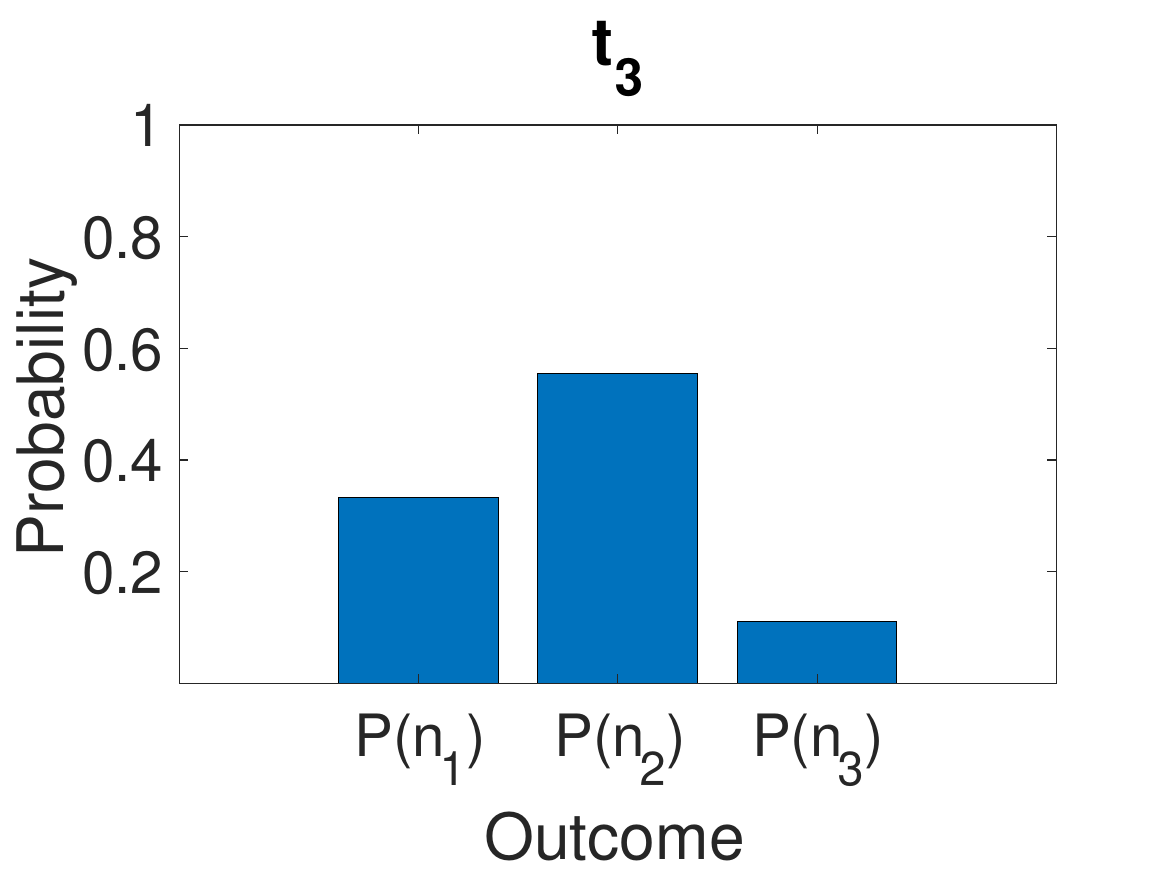}}
\par\end{centering}
\textcolor{black}{\caption{Measurable macroscopic differences occur for the probabilities of
outcomes at time $t_{3}$, depending on whether Bob makes a measurement
or not, but the differences emerge only \emph{after} the dynamics
of Alice's operations (shuffling). This gives consistency with weak
macroscopic realism. Plotted are the probabilities $P(n_{k})$ of
obtaining $N$ on measuring $\hat{n}$ of mode $k$ at the time $t$.
The top sequence shows the sequence of probabilities if Bob makes
no measurement. The lower sequence shows the probabilities if Bob
makes a measurement between times $t_{1}$ and $t_{2}$. The far right
plots show the probabilities at time $t_{3}$, after Alice's operation
$U_{f}$, which takes place between times $t_{2}$ and $t_{3}$.\textcolor{red}{}
\label{fig:compare-number-no-m}\textcolor{green}{}}
}
\end{figure*}

If it is assumed that \emph{macroscopic realism} holds, then how does
the three-box paradox occur? We solve for the dynamics as Bob and
Alice make their measurements, which includes the unitary transformations
$U$. At the times $t_{1}$, $t_{2}$ and $t_{3}$, \emph{after} the
unitary transformations corresponding to the shuffle, macroscopic
realism posits the system to have predetermined values for the final
measurement of $\hat{X}$ (or $\hat{n}$) at each mode i.e. relative
to the measurement basis. This is not inconsistent with the paradox,
because the predictions for the probabilities are indistinguishable
from those of a \emph{mixture}, for which there is a predetermination
of the outcome of $\hat{n}$. However,  the system is \emph{not}
at the time $t_{k}$ in any of the quantum eigenstates given by $|1\rangle$,
$|2\rangle$ or $|3\rangle$.  In between the times $t_{k}$, unitary
dynamics occurs which leads to different final states for the superposition
and the mixed state.

This is illustrated in Figure \ref{fig:compare-number-no-m}, where
we compare the dynamics if Bob does or does not make a measurement
at time $t_{2}$. The probabilities for the outcome of $\hat{n}$
for each mode immediately before and after Bob's measurements are
indistinguishable. However, \emph{after} Alice's unitary transformations,
macroscopic differences occur.\textcolor{black}{}

Now we comment on the subtle distinction between possible definitions
of \emph{macroscopic realism}. We specify in the definition of macroscopic
realism that the hidden variable $\lambda_{i}^{(k)}$ applies to the
system created at the times $t_{k}$\emph{ after} the system is prepared
in the measurement basis of $\hat{n}$ i.e. after any reversible ``shuffling''
corresponding to the unitary operations. Prior to $t_{k}$, before
the shuffling, the system is \emph{not} prepared for the measurement.
We also specify that \emph{macroscopic realism} for each mode $i$
includes the assumption that the value $\lambda_{i}^{(k)}$ is not
affected by any operations occurring at the other modes. Since other
definitions of macroscopic realism exist, where the unitary operations
associated with the preparation of the measurement setting are not
considered \cite{manushan-bell-cat-lg,legggarg}, we refer to the
definition used in this paper as \emph{weak macroscopic realism (wMR).}

We next illustrate that the predictions of weak macroscopic realism
are not negated by the paradox. Consider where Alice acts on the system
in the state
\begin{equation}
|\psi_{sup}\rangle=\frac{e^{i\varphi}}{\sqrt{3}}(|3\rangle+e^{i\varphi_{1}}(i|2\rangle-|1\rangle))\label{eq:sup321}
\end{equation}
We consider where the unitary parts $U_{21}$ and $U_{32}$ of Alice's
measurements are performed \emph{in sequence}. Her measurement is
\begin{equation}
U_{f}=U_{1f}^{-1}U_{2f}^{-1}\label{eq:trans}
\end{equation}
given by Eq. (\ref{eq:alice-uf}). Alice performs $U_{2f}^{-1}$ first,
which involves only systems $1$ and $2$, arising from $H_{21}$.
Then, weak macroscopic realism (wMR) posits that the hidden variables
$\lambda_{3}^{(k)}$ for mode $3$ \emph{should not be affected by
this transformation}.\textcolor{black}{}
\begin{figure*}[t]
\begin{centering}
\textcolor{black}{\includegraphics[width=0.5\columnwidth]{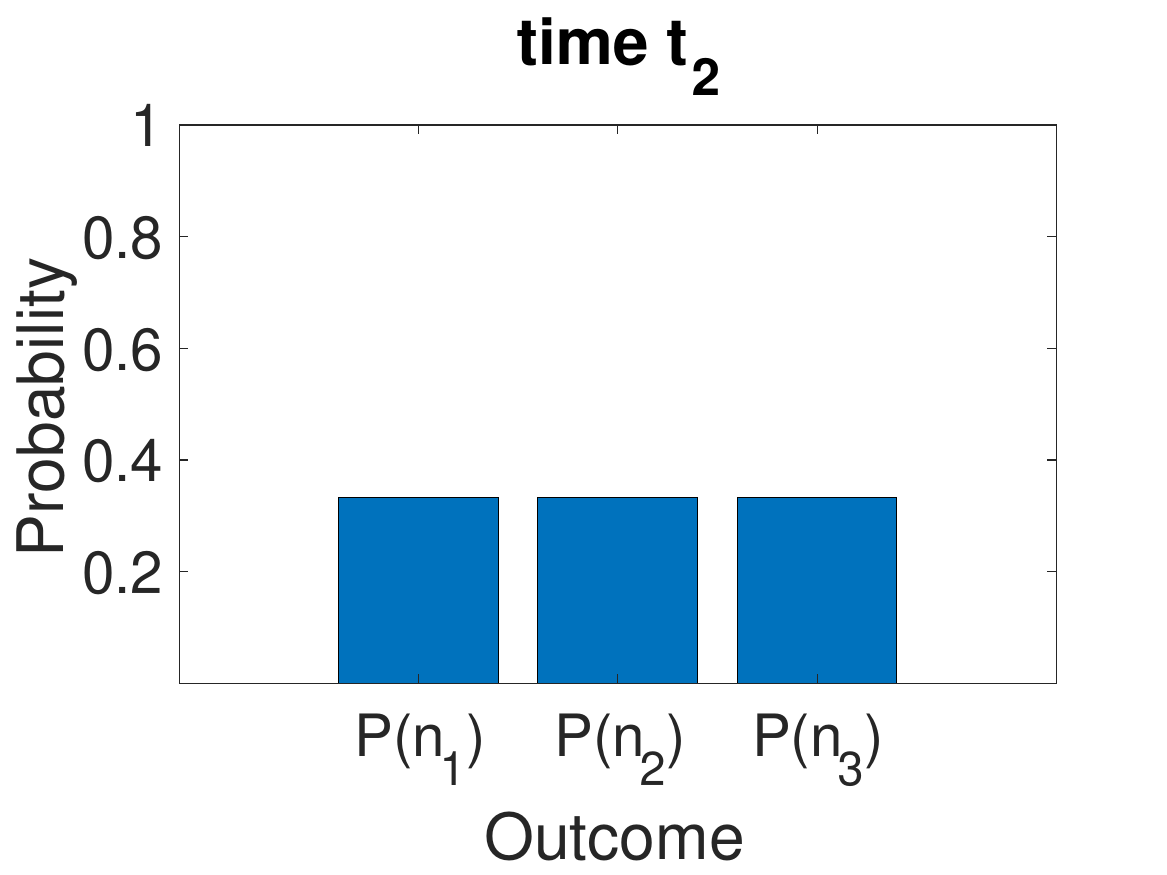}\includegraphics[width=0.5\columnwidth]{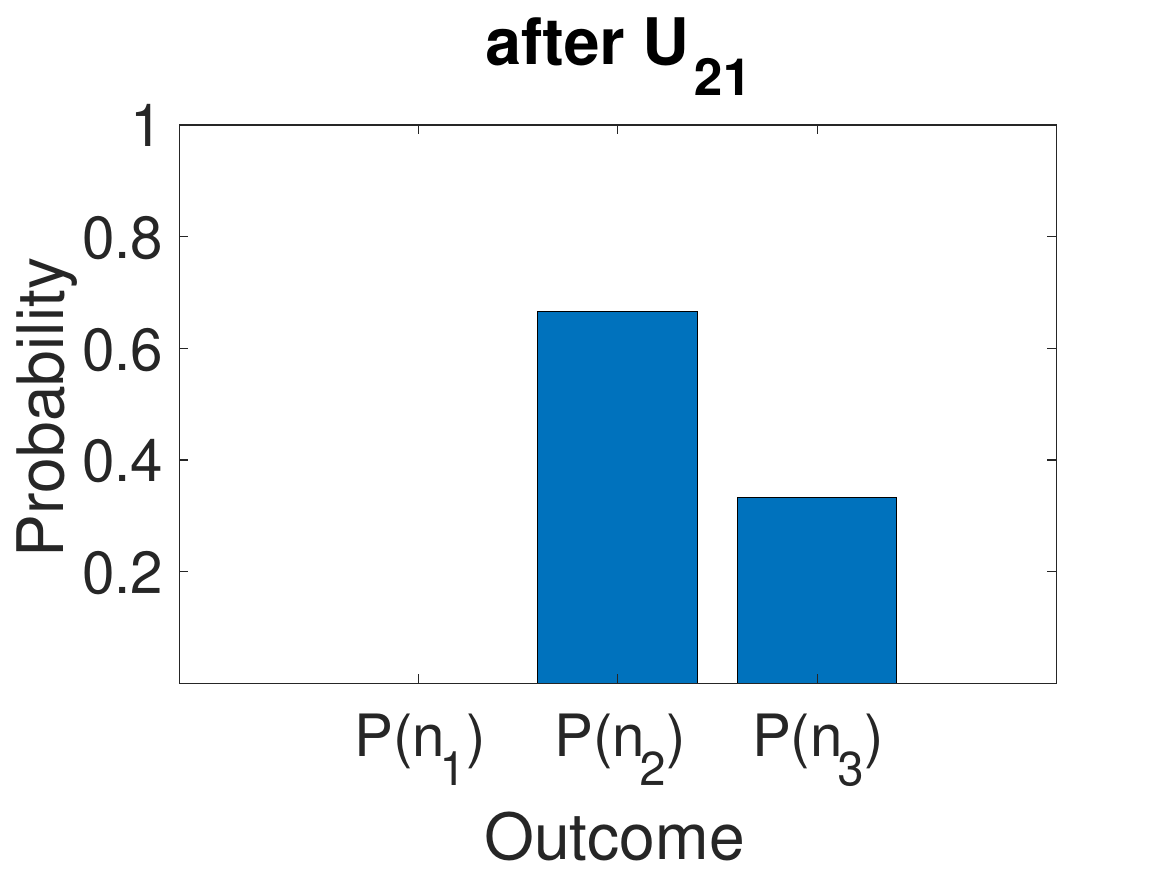}\includegraphics[width=0.5\columnwidth]{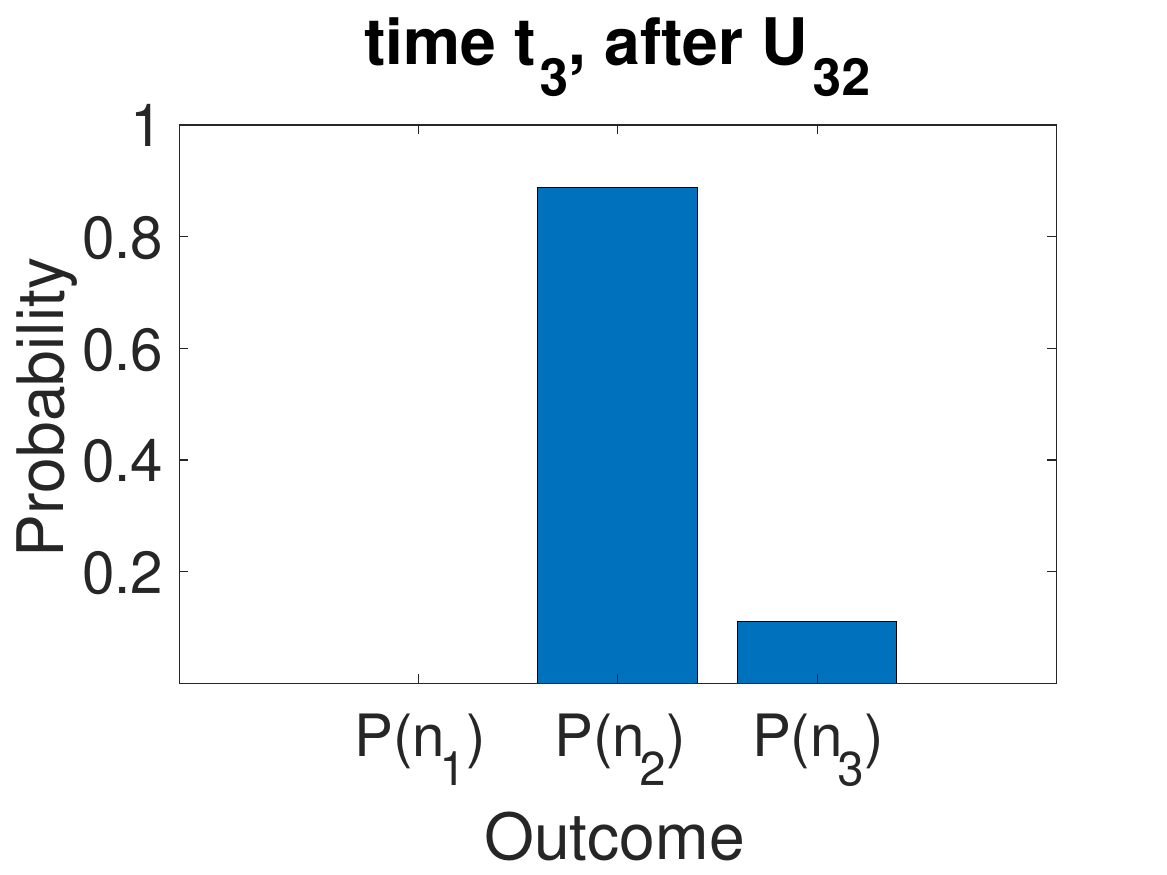}}
\par\end{centering}
\medskip{}

\begin{centering}
\textcolor{black}{\includegraphics[width=0.5\columnwidth]{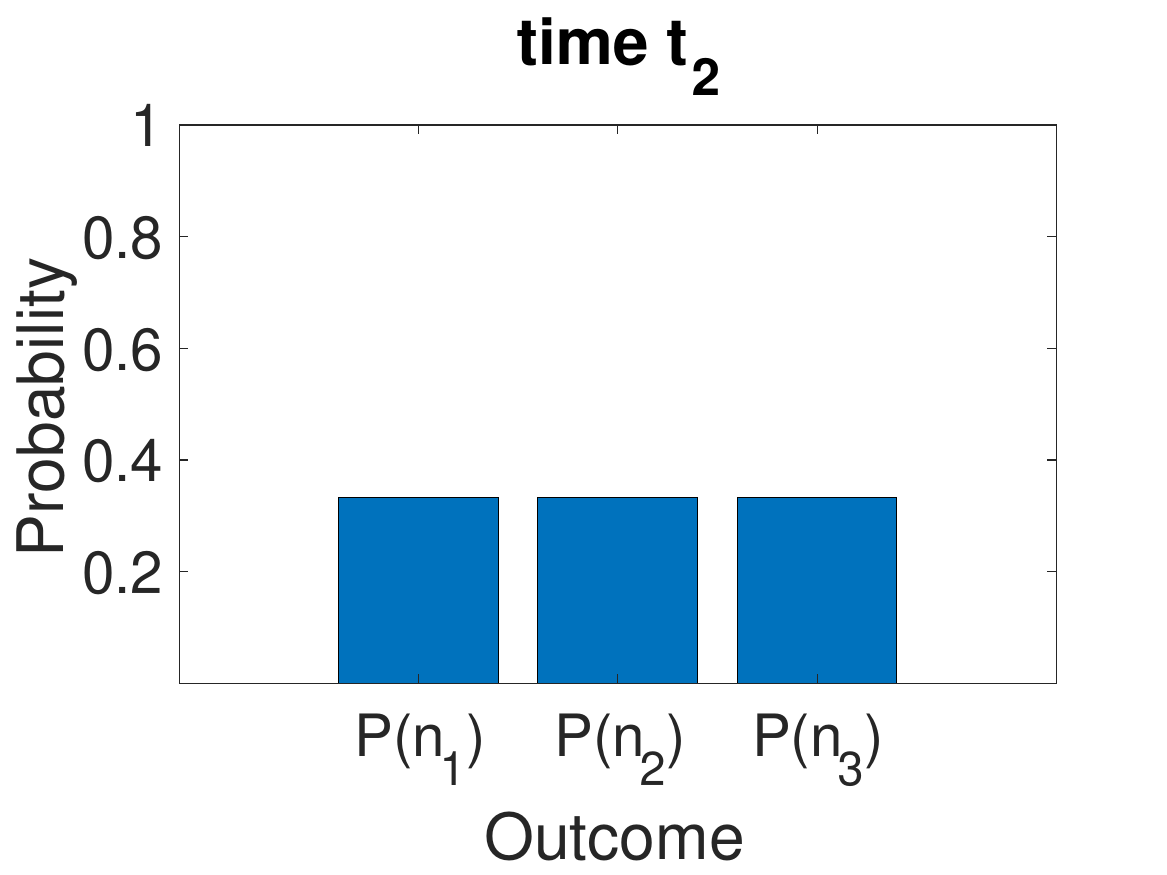}\includegraphics[width=0.5\columnwidth]{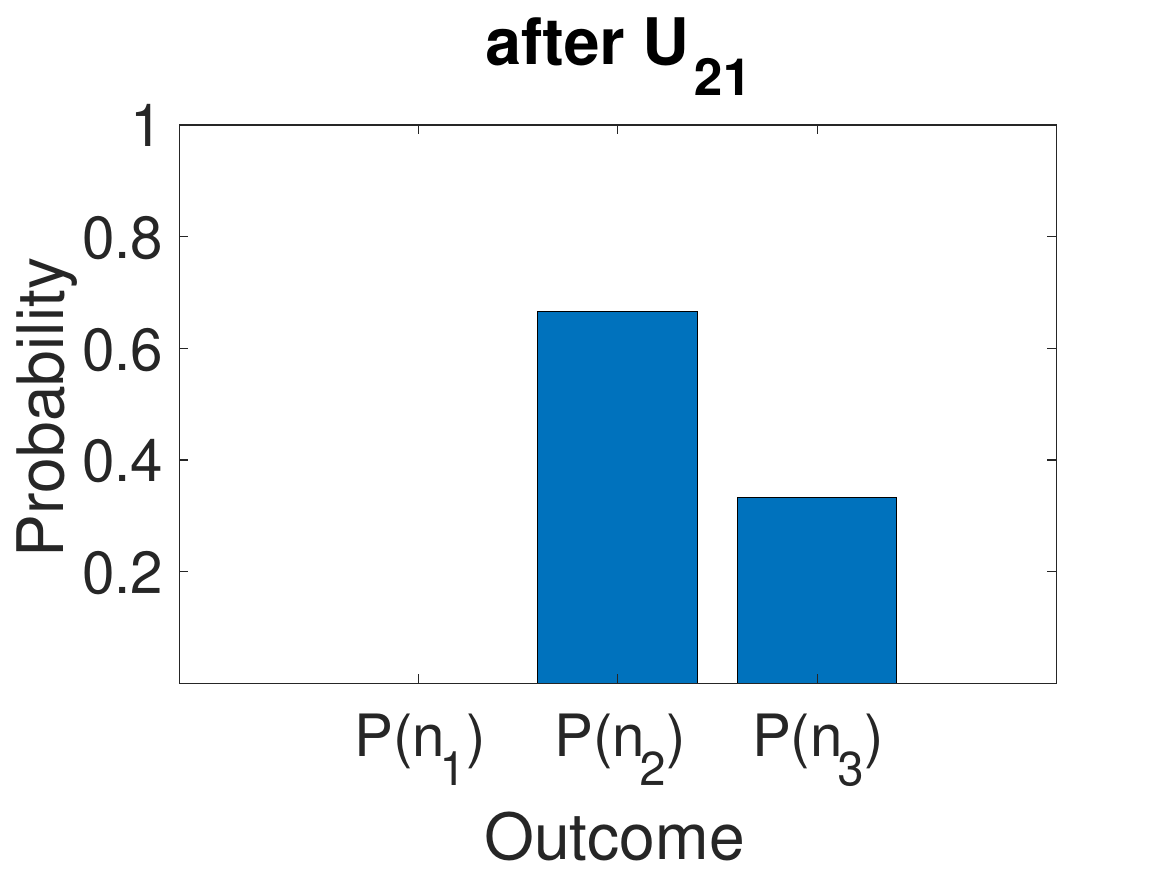}\includegraphics[width=0.5\columnwidth]{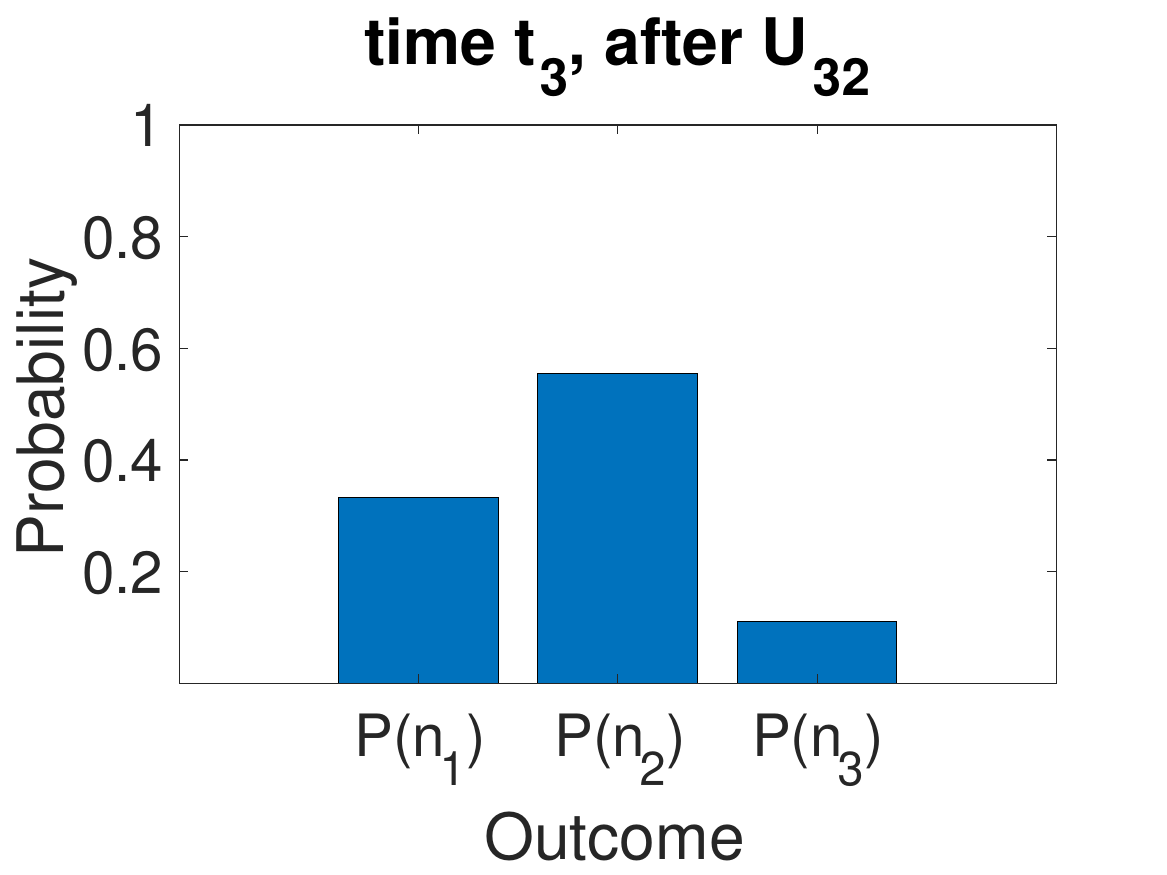}}
\par\end{centering}
\textcolor{black}{\caption{Validating the predictions of weak macroscopic realism: The sequence
as Alice makes the operations $U_{f}$ on the states (top) $|\psi_{sup}\rangle$
and (lower) $\rho_{3,mix}$, which is the state after Bob's measurement
detects the system in state $|3\rangle$ at time $t_{2}$. Alice first
transforms (shuffling between boxes 1 and 2) according to $U_{f2}^{-1}=U_{21}$,
leaving Box 3 untouched. Then she transforms (shuffling between boxes
3 and 2) by $U_{1f}^{-1}=U_{32}$. Far left are the plots of $P(n_{k})$
for the initial states (top) $|\psi_{sup}\rangle$ and (lower) $\rho_{mix}$.
Second from left are the states after Alice's transformation $U_{21}$.\textcolor{red}{{}
}The predictions for the $|\psi_{sup}\rangle$ and the mixture $\rho_{3,mix}$
are identical. Far right are the states after Alice's transformation
(shuffling between boxes 2 and 3) $U_{32}$, where the predictions
diverge.\textcolor{red}{} The results are in agreement with the predictions
of weak macroscopic realism, which posits that the system has a definite
property $\lambda_{3}$ for Box 3 (the ball is in the box or not)
at time $t_{2}$, and this property cannot be changed by any shuffling
$U_{21}$ on boxes 2 and 1. This is because throughout the shuffling
$U_{21}$, the predictions are identical to those of the mixture $\rho_{3,mix}$,
which is consistent with the definite value $\lambda_{3}$. \label{fig:compare-number-successive}\textcolor{green}{}}
}
\end{figure*}

We now confirm that the predictions of the quantum paradox are indeed
consistent with this hypothesis. We see that $U_{2f}=U_{2i}$. Hence,
the state after this transformation is
\begin{eqnarray}
U_{2f}^{-1}|\psi_{sup}\rangle & = & \frac{1}{\sqrt{2}}\left(\begin{array}{ccc}
ie^{-i\varphi_{1}} & e^{-i\varphi_{1}} & 0\\
-ie^{-i\varphi_{1}} & e^{-i\varphi_{1}} & 0\\
0 & 0 & \sqrt{2}
\end{array}\right)\frac{e^{-i\varphi}}{\sqrt{3}}\left(\begin{array}{c}
-e^{i\varphi_{1}}\\
ie^{i\varphi_{1}}\\
1
\end{array}\right)\nonumber \\
 & = & \frac{e^{-i\varphi}}{\sqrt{6}}\left(\begin{array}{c}
0\\
2i\\
\sqrt{2}
\end{array}\right)\label{eq:state-after-partial-trans-12}
\end{eqnarray}
which gives relative probabilities of $0$, $2/3$ and $1/3$, for
detection of the system in the states $|1\rangle$, $|2\rangle$ and
$|3\rangle$. We next compare the evolution under $U_{2f}^{-1}$
for the \emph{mixed} state 
\begin{eqnarray}
\rho_{3,mix} & = & \frac{1}{3}(ie^{i\varphi_{1}}|2\rangle-e^{i\varphi_{1}}|1\rangle)(-ie^{-i\varphi_{1}}\langle2|-e^{-i\varphi_{1}}\langle1|)\nonumber \\
 &  & +\frac{1}{3}|3\rangle\langle3|\label{eq:mix}
\end{eqnarray}
The system in a mixed state can be viewed as being in a state with
a \emph{definite} outcome for $\hat{n}$ on the mode $3$. Hence,
there is a hidden variable $\lambda_{3}$ for the system in this description.
Beginning with $\rho_{3,mix}$, the state after the transformation
$U_{2f}^{-1}$ is
\begin{eqnarray}
\rho_{3,mix,I} & = & \frac{2}{3}|2\rangle\langle2|+\frac{1}{3}|3\rangle\langle3|\label{eq:mix-state-after-partial-tarns}
\end{eqnarray}
which has identical relative probabilities of $0$, $2/3$ and $1/3$.
 There is consistency with wMR for mode $3$.

On the other hand, if we continue to evolve with $U_{1f}^{-1}$, then
the evolution of the two states $|\psi_{sup}\rangle$ and $\rho_{3,mix}$
diverges macroscopically. For $|\psi_{sup}\rangle$, we find 
\begin{eqnarray}
U_{1f}^{\dagger}U_{2f}^{\dagger} & = & \frac{1}{\sqrt{3}}\left(\begin{array}{ccc}
\sqrt{3} & 0 & 0\\
0 & ie^{-i\varphi} & -e^{-i\varphi}\sqrt{2}\\
0 & -ie^{-i\varphi}\sqrt{2} & -e^{-i\varphi}
\end{array}\right)\frac{e^{-i\varphi}}{\sqrt{6}}\left(\begin{array}{c}
0\\
2i\\
\sqrt{2}
\end{array}\right)\nonumber \\
 & = & \frac{e^{-i\varphi}}{\sqrt{18}}\left(\begin{array}{c}
0\\
-4e^{-i\varphi}\\
\sqrt{2}e^{-i\varphi}
\end{array}\right)\label{eq:final-state-from-sup-after-both-tarns}
\end{eqnarray}
There is zero probability of finding the system in state $|1\rangle$.
The relative probabilities are $0$, $8/9$, $1/9$. On the other
hand, for the system initially in $\rho_{3,mix}$, the final state
after both the transformations $U_{2f}^{-1}$ and then $U_{1f}^{-1}$
is
\begin{equation}
\rho_{3,mix,F}=\frac{2}{3}U_{1f}^{-1}|2\rangle\langle2|U_{1f}+\frac{1}{3}U_{1f}^{-1}|1\rangle\langle1|U_{1f}\label{eq:mix-final-state-after-both-trans}
\end{equation}
where
\begin{equation}
U_{1f}^{-1}|2\rangle=\frac{e^{-i\varphi}}{\sqrt{3}}\left(\begin{array}{c}
0\\
i\\
-i\sqrt{2}
\end{array}\right)\label{eq:sup1-1}
\end{equation}
and
\begin{equation}
U_{1f}^{-1}|1\rangle=\left(\begin{array}{c}
1\\
0\\
0
\end{array}\right)\label{eq:sup1-2}
\end{equation}
The probability of finding the system in state $|1\rangle$ is $1/3$.
The evolution of $|\psi_{sup}\rangle$ leads to the paradox, but does
not contradict the premise of weak macroscopic realism, because the
local unitary transformation $U_{32}$ acts on both mode $3$ and
$2$. The transformations are depicted in Figure \ref{fig:compare-number-successive}.

The paradox is seen to arise from the distinction between the state
before and after Bob's measurement, which is initially microscopic.
This gives the seemingly paradoxical situation whereby the measurement
disturbance from Bob vanishes, but the effect of it nonetheless can
be extracted by suitable dynamics.

\section{Macroscopic three-box paradox with cat states}

The example of Sections III and IV considered states distinct by
$N$ quanta. However, the calculations involved the interaction $H_{kl}$,
which was solved for $N\sim10$. One way to achieve a more macroscopic
realisation of the three-box paradox is to consider the coherent states
$|\alpha\rangle$ of a single-mode field. In this section, we propose
such a paradox, where the separation between the relevant coherent
states can be made arbitrarily large. In order to achieve a feasible
realisation, the macroscopic paradox is based on a modified version
of the original three-box paradox.

Superpositions of macroscopically distinct coherent sates are referred
to as ``cat states'' \cite{s-cat,cat-states,cat-states-review,yurke-stoler-1}.
Such states can be generated in optical cavities with dissipation
and /or using conditional measurements \foreignlanguage{australian}{\cite{cat-states,cat-det-map,cat-state-phil,collapse-revival-super-circuit-1,cat-states-super-cond,cat-bell-wang-1,cat-states-wc,transient-cat-states-leo,cats-hach,cat-even-odd-transient,cat-dynamics-ry,IBM-macrorealism-1,omran-cats}}.
Here, we take a simple model in which the cat states are created from
a nonlinear dispersive medium, where losses are assumed minimal \cite{yurke-stoler-1,wright-walls-gar-1}.
The unitary operations are solved analytically, and have been realised
experimentally, to create cat states for large $\alpha$ \cite{collapse-revival-super-circuit-1,collapse-revival-bec-2}.

\subsection{Coherent-state model: $k>2$}

We propose four states defined as
\begin{eqnarray}
|1\rangle & = & |-i\alpha_{0}\rangle\nonumber \\
|2\rangle & = & |i\alpha_{0}\rangle\nonumber \\
|3\rangle & = & |\alpha_{0}\rangle\nonumber \\
|4\rangle & = & |-\alpha_{0}\rangle\label{eq:coh-three-states}
\end{eqnarray}
where $|\alpha\rangle$ is a coherent state of a single-mode field.
As $\alpha_{0}\rightarrow\infty$, these states become orthogonal.
We note that for sufficiently large $\alpha_{0}$, the states can
be distinguished by simultaneous quadrature phase amplitude measurements
$\hat{X}=\hat{a}+\hat{a}^{\dagger}$ and $\hat{P}=(\hat{a}-\hat{a}^{\dagger})/i$,
where $\hat{a}$ and $\hat{a}^{\dagger}$ are the field boson destruction
and creation operators \cite{yurke-stoler-1}.

The system can be prepared at time $t_{1}$ from the state $|3\rangle$
in the superposition \cite{yurke-stoler-1,manushan-cat-lg} of type
Eq. (\ref{eq:sup1}), by applying a set of unitary transformations
based on nonlinear interactions. Following the work of Yurke and Stoler
\cite{yurke-stoler-1}, we consider the evolution of a single mode
system prepared in a coherent state under the influence of a nonlinear
Hamiltonian written in the Schrödinger picture as,
\begin{equation}
H_{NL}=\omega\hat{n}+\Omega\hat{n}^{k}\label{eq:H}
\end{equation}
where $\varOmega$ represents the strength of the nonlinear term and
$\hat{n}=\hat{a}^{\dagger}\hat{a}$ is the field number operator.
We take $\hbar=1$. Here, $k$ is a positive integer. After an interaction
time of $t=\frac{\pi}{2\Omega}$, the system with $k>2$ initially
prepared in a coherent state $|\alpha\rangle$ becomes 
\begin{eqnarray}
|\psi_{sup}\rangle & = & \frac{1}{2}(|1\rangle-|2\rangle+|3\rangle+|4\rangle)\nonumber \\
 & = & \frac{1}{2}(|-i\alpha_{0}\rangle-|i\alpha_{0}\rangle+|\alpha_{0}\rangle+|-\alpha_{0}\rangle)\label{eq:sup-coh}
\end{eqnarray}
which for large $\alpha_{0}$ is a a four-component cat state. Since
it is produced at time $t_{1}$, we also refer to the state $|\psi_{sup}\rangle$
as $|\psi_{1}\rangle$. We can interpret the creation of the state
as a transformation using a unitary operator $U_{i}=U_{\left(\frac{\pi}{2\Omega}\right)}$
given by
\begin{equation}
U_{i}|3\rangle=U_{i}|\alpha_{0}\rangle=|\psi_{sup}\rangle\label{eq:Ui1-coh}
\end{equation}
where $U_{\left(\frac{\pi}{2\Omega}\right)}=e^{-i\Omega\hat{n}^{k}t}$
with $\hbar=1$. For large $\alpha_{0}$ where the four states can
form a basis set $\{|1\rangle,$ $|2\rangle$, $|3\rangle$, $|4\rangle$\},
it is convenient to identify the transformation as a matrix
\begin{eqnarray}
U_{i} & = & \frac{1}{2}\left(\begin{array}{cccc}
1 & 1 & 1 & -1\\
1 & 1 & -1 & 1\\
-1 & 1 & 1 & 1\\
1 & -1 & 1 & 1
\end{array}\right)\label{eq:Ui-1-2}
\end{eqnarray}
where the basis states $|k\rangle$ correspond to column matrices
$(a_{j1})$ with coefficients given as $a_{j1}=\delta_{jk}$. It is
straightforward to verify that the operations $U_{i}|k\rangle$ realise
the correct final states.

To account for the four states, we consider a modified version of
the three-box paradox. Bob can consider to determine whether the system
is in one of the four states $|k\rangle$, analogous to four boxes.
The initial state $|\psi_{sup}\rangle\equiv|\psi_{1}\rangle$ which
Bob makes a measurement on is $|1\rangle-|2\rangle+|3\rangle+|4\rangle$.
Bob may make measurements to determine whether the system is in one
of the states $|2\rangle$ or $|4\rangle$, or not. If the system
can be determined to be in neither $|2\rangle$ nor $|4\rangle$,
then the reduced state for the system is $|1\rangle+|3\rangle$. Alternatively,
Bob may make measurements to determine whether the system is in one
of the states $|1\rangle$ or $|4\rangle$, or not. If the system
is determined to be in neither $|1\rangle$ or $|4\rangle$, then
the reduced state for the system is $-|2\rangle+|3\rangle$.\textcolor{black}{}
\begin{figure*}[t]
\textcolor{black}{}
\begin{centering}
\includegraphics[width=2\columnwidth]{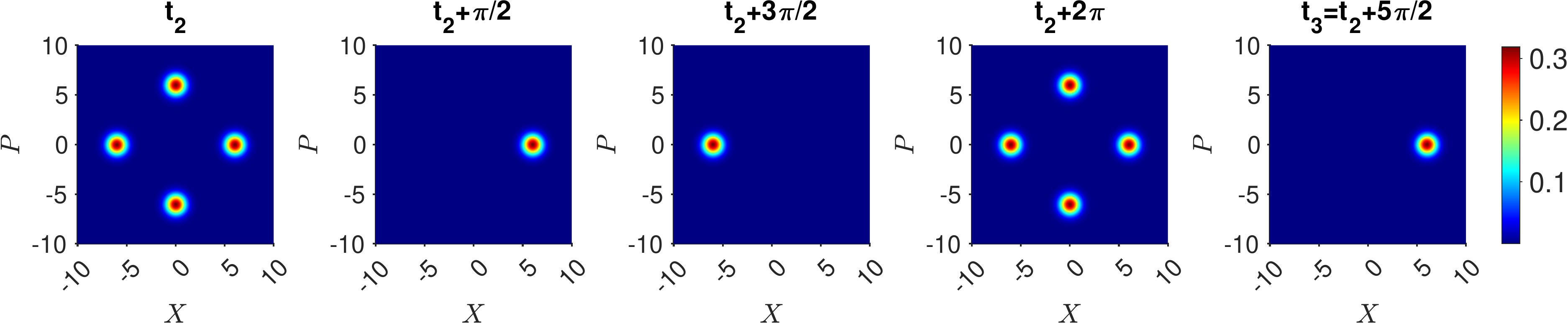}
\par\end{centering}
\textcolor{black}{\caption{Dynamics of Alice's transformation $U_{f}$ performed on the system
in state $|\psi_{f}\rangle=-|1\rangle+|2\rangle+|3\rangle+|4\rangle$
at time $t_{2}$. The operation $U_{f1}$ is carried out by evolving
under $H_{NL}$ for a time $t=3\pi/2\Omega$. Next, the operation
$U_{f2}$ is carried out by evolving under $H_{NL}$ for a further
time $t=\pi/\Omega$. At time $t_{3}$ after the combined operation
$U_{f}=U_{f2}U_{f1}$, the system is in state $|3\rangle$.\label{fig:test}\textcolor{green}{}}
}
\end{figure*}

Suppose Alice postselects for the state 
\begin{eqnarray}
|\psi_{f}\rangle & = & -|1\rangle+|2\rangle+|3\rangle+|4\rangle\label{eq:supf-coh}\\
 & = & \frac{1}{2}(|\alpha_{0}\rangle+|-\alpha_{0}\rangle-|-i\alpha_{0}\rangle+|i\alpha_{0}\rangle))
\end{eqnarray}
which is generated from $|\alpha_{0}\rangle=|3\rangle$ by a transformation
\begin{equation}
U_{f}^{-1}|3\rangle=|\psi_{f}\rangle\label{eq:test2}
\end{equation}
The dynamics for $U_{f}$ is shown in Figure \ref{fig:test}. The
inverse is done by first transforming $|\alpha_{0}\rangle\rightarrow$$|-\alpha_{0}\rangle$,
using the nonlinear interaction modelled by the Hamiltonian $H_{NL}$
of Eq. (\ref{eq:H}). Defining $U_{f2}=U_{\left(\frac{\pi}{\Omega}\right)}=e^{-i\Omega\hat{n}^{k}t}$
where the interaction time is $t=\pi/\Omega$, the state $|\alpha_{0}\rangle$
becomes $|-\alpha_{0}\rangle$, for all integers $k$. We identify
\begin{eqnarray}
U_{f2}^{-1} & = & \left(\begin{array}{cccc}
0 & 1 & 0 & 0\\
1 & 0 & 0 & 0\\
0 & 0 & 0 & 1\\
0 & 0 & 1 & 0
\end{array}\right)\label{eq:Ui-1-1-1-1}
\end{eqnarray}
and note that $U_{f2}^{-1}|\alpha_{0}\rangle=|-\alpha_{0}\rangle$.
We also see that $U_{f2}^{-1}=U_{f2}$. The second stage of the
generation is to transform $|4\rangle=|-\alpha_{0}\rangle$ by the
unitary transformation $U_{f1}^{-1}$ so that
\begin{equation}
U_{f1}^{-1}|4\rangle=|\psi_{f}\rangle\label{eq:Uf-12-coh}
\end{equation}
We find that $U_{f1}^{-1}=U_{\left(\frac{\pi}{2\Omega}\right)}=e^{-i\Omega\hat{n}^{k}t}$
where $t=\pi/2\Omega$, and note that $U_{f1}^{-1}=U_{i}$. It is
straightforward to verify that $U_{f1}^{-1}|4\rangle=|\psi_{f}\rangle$.
In fact, $U_{f1}=U_{i}^{-1}$. Since for any unitary interaction $U=e^{-iHt}$,
$U^{-1}=U^{\dagger}$, the inverse $U^{-1}$ corresponds to $U^{-1}=e^{iHt}$.
Since the evolution is periodic, with period $2\pi/\Omega$ \cite{yurke-stoler-1},
this corresponds to $U^{-1}=e^{iH_{NL}(t-2\pi/\Omega)}=e^{-iH_{NL}(2\pi/\Omega-t)}$
and we find that $U_{f1}=U_{i}^{-1}=U_{\left(\frac{3\pi}{2\Omega}\right)}$.
The $U_{f1}$ is realised by Alice evolving the system forward in
time by the amount $t=3\pi/2\Omega$. We may express this as
\begin{equation}
U_{f1}^{-1}U_{f2}^{-1}|3\rangle=|\psi_{f}\rangle\label{eq:trans2}
\end{equation}
which implies
\begin{equation}
U_{f}|\psi_{f}\rangle\equiv U_{f2}U_{f1}|\psi_{f}\rangle=|3\rangle\label{eq:trans5}
\end{equation}
The operations $U_{f}$ correspond to Alice first evolving the system
under $H$ with $k>2$ for a time $t=3\pi/2\Omega$, and then applying
the evolution $H$ for a time $t=\pi/\Omega$. We find that $U_{f}=U_{f2}U_{f1}=U_{f2}^{-1}U_{i}^{-1}$:
Hence, in matrix form
\begin{eqnarray}
U_{f} & = & \frac{1}{2}\left(\begin{array}{cccc}
1 & 1 & 1 & -1\\
1 & 1 & -1 & 1\\
-1 & 1 & 1 & 1\\
1 & -1 & 1 & 1
\end{array}\right)\label{eq:matrix-uf}
\end{eqnarray}
\textcolor{black}{Figure \ref{fig:test} shows the dynamics of Alice's
$U_{f}$, where the system is initially in the state $|\psi_{f}\rangle$.
The evolution confirms that the final state is indeed $|3\rangle$.}

\begin{figure}[t]
\textcolor{black}{}
\begin{centering}
\includegraphics[width=1\columnwidth]{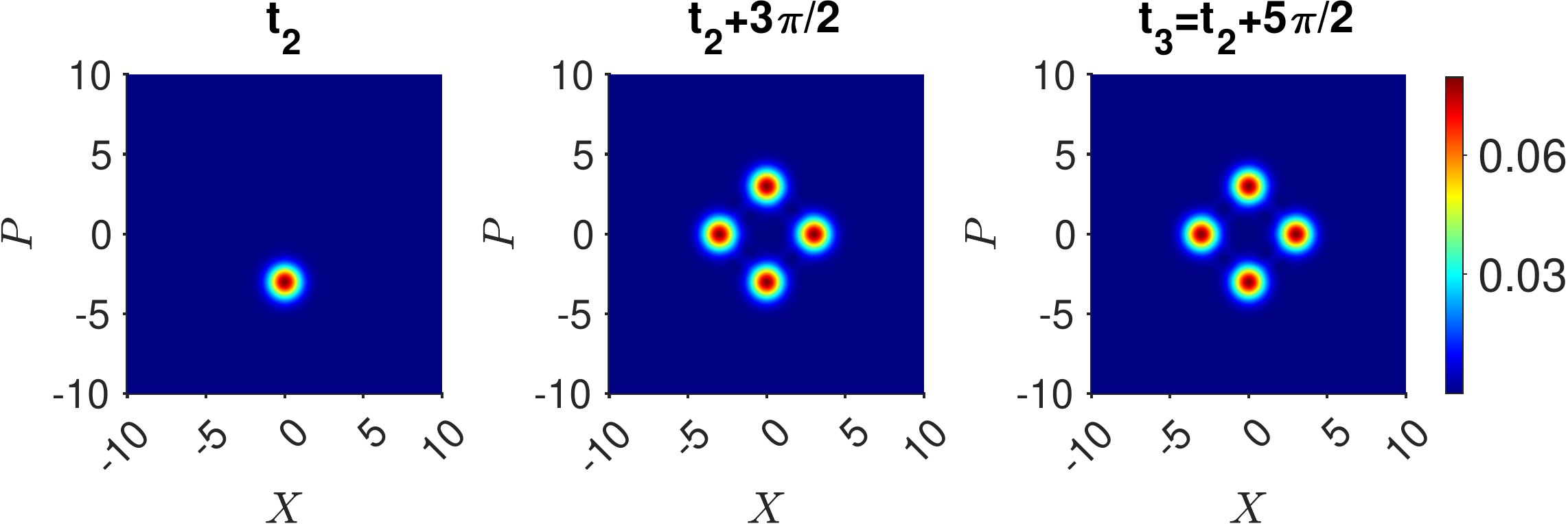}
\par\end{centering}
\medskip{}

\begin{centering}
\includegraphics[width=1\columnwidth]{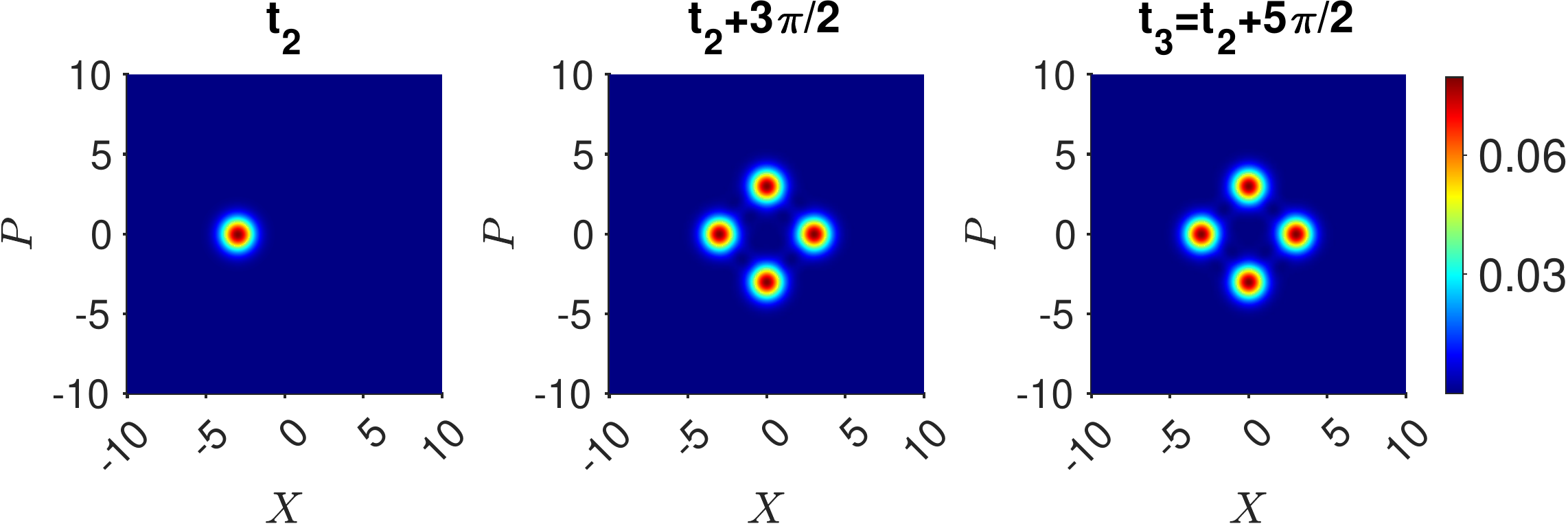}
\par\end{centering}
\textcolor{black}{\caption{Dynamics of Alice's transformation $U_{f}$ performed at time $t_{2}$
after Bob makes his measurement. Here, we take the case where Bob
obtains the result: (top) that the Ball is in Box $1$ (implying the
system is in state $|1\rangle$ at time $t_{2}$); or (lower) that
the Ball is in Box 4 (implying the system is in state $|4\rangle$
at time $t_{2}$). The figures show contour plots of $Q(\alpha)$
\textcolor{black}{for $\alpha_{0}=3$ and $k=3$.}\textcolor{red}{{}
}\textcolor{black}{Here, $U_{f}=U_{f2}U_{f1}$. The plots show the
state after the first transformation $U_{f1}$ at time $t_{2}+3\pi/2$.
} In both cases, there is a finite probability of Alice finding
the Ball in Box $3$. \label{fig:q-for-bob-detecting-ball-in1-then-evolve}\textcolor{green}{}}
}
\end{figure}

\begin{figure}[t]
\textcolor{black}{}
\begin{centering}
\includegraphics[width=1\columnwidth]{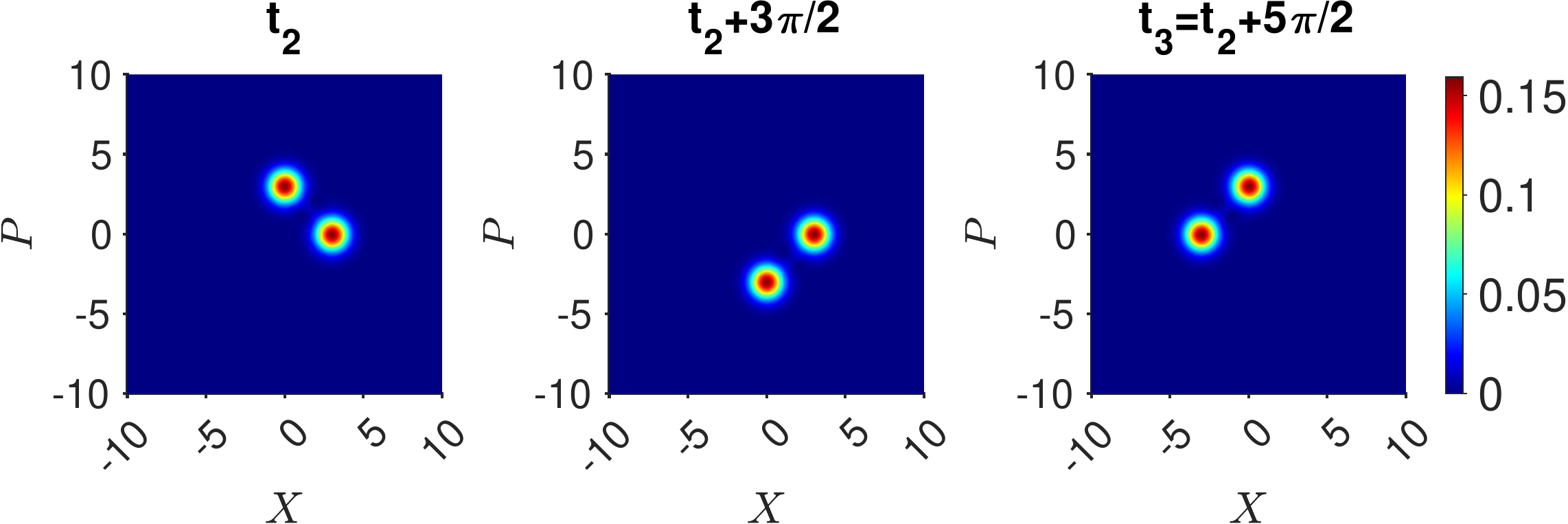}
\par\end{centering}
\medskip{}

\begin{centering}
\includegraphics[width=1\columnwidth]{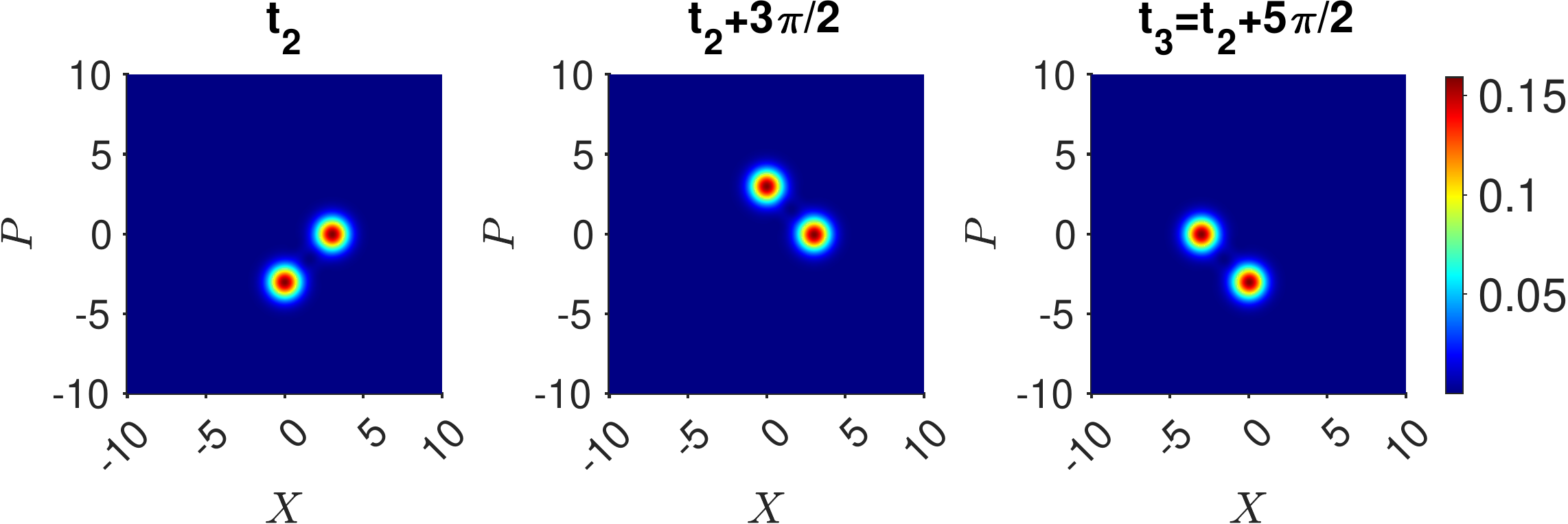}
\par\end{centering}
\textcolor{black}{\caption{Dynamics of Alice's transformation $U_{f}$ performed at time $t_{2}$
after Bob makes his measurement. Here, we take the case where Bob
obtains the result: (top) that the Ball is not in Box 1 or 4 (implying
the system in state $-|2\rangle+|3\rangle$ at time $t_{2}$); or
(lower) that the Ball is not in Box 2 or 4 (so that the system in
state $|1\rangle+|3\rangle$ at time $t_{2}$). The figures show contour
plots of $Q(\alpha)$ \textcolor{black}{for $\alpha_{0}=3$ and $k=3$.}
\label{fig:q-plots-for-Bob-notdetecting-ball-in-1-or4} In both
cases, there is zero probability of Alice finding the ball in Box
3.}
}
\end{figure}

The paradox is realised when Alice performs the measurement given
by the transformation $U_{f}$ and detects whether the system is
in state $|3\rangle$. We check for the paradox as follows. If Bob
detects the system to be in $|2\rangle$, then the evolution by Alice
gives 
\begin{eqnarray}
U_{f}|2\rangle & = & \frac{1}{2}\left(\begin{array}{c}
1\\
1\\
1\\
-1
\end{array}\right)\label{eq:ma1}
\end{eqnarray}
If Bob detects that the system is in state $|4\rangle$, then the
evolution by Alice gives
\begin{eqnarray}
U_{f}|4\rangle & = & \frac{1}{2}\left(\begin{array}{c}
-1\\
1\\
1\\
1
\end{array}\right)\label{eq:ma2}
\end{eqnarray}
If Bob detects that the system is in state $|1\rangle$, then the
evolution by Alice gives
\begin{eqnarray}
U_{f}|1\rangle & = & \frac{1}{2}\left(\begin{array}{c}
1\\
1\\
-1\\
1
\end{array}\right)\label{eq:ma3}
\end{eqnarray}
 If Bob detected that the system is not in $|2\rangle$ or $|4\rangle$,
then the system is in reduced state $|1\rangle+|3\rangle$, and the
evolution by Alice gives the final state
\begin{eqnarray}
U_{f}(|1\rangle+|3\rangle)/\sqrt{2} & = & \frac{1}{\sqrt{2}}\left(\begin{array}{c}
1\\
0\\
0\\
1
\end{array}\right)\label{eq:mqa5}
\end{eqnarray}
\textcolor{red}{}If Bob detected that the system is not in $|1\rangle$
or $|4\rangle$, then the system is in reduced state $-|2\rangle+|3\rangle$,
and the evolution by Alice gives the final state
\begin{eqnarray}
U_{f}(-|2\rangle+|3\rangle)/\sqrt{2} & = & \frac{1}{\sqrt{2}}\left(\begin{array}{c}
0\\
-1\\
0\\
1
\end{array}\right)\label{eq:ma6}
\end{eqnarray}
In both cases, there is zero probability of Alice measuring the system
to be in state $|3\rangle=|\alpha\rangle$. Hence, if Alice detects
the system to be in $|\alpha_{0}\rangle$ at time $t_{3}$, she knows
that Bob detected the ball to be in one of the boxes he opened. The
dynamics is confirmed in Figures \ref{fig:q-for-bob-detecting-ball-in1-then-evolve}
and \ref{fig:q-plots-for-Bob-notdetecting-ball-in-1-or4}. The figures
plot the dynamical sequences where Bob detects the Ball, and where
Bob detects no Ball in Boxes 1 or 4. We confirm for the latter, there
is zero probability of the system being found by Alice in state $|3\rangle.$\textcolor{black}{}

\textcolor{black}{}

The calculations can be summarised. If Bob opens Box 1 ($B1$) , then
the probability of him detecting the ball is $1/4$. Following \cite{Maroney},
we write this as $P_{B1}(1_{2})=1/4$. We see also that $P_{B1}(3_{3}|1_{2})=1/4$
and hence $P_{B1}(1_{2},3_{3})=P_{B1}(3_{3}|1_{2})P_{B1}(1_{2})=1/16$.
Similarly, denoting the probability of detecting the ball in either
Box 1 or 4 if both boxes are opened as $P_{B1,B4}(\{1_{2},4_{2}\})$,
we find $P_{B1,B2}(\{1_{2},4_{2}\})=1/2$. From above, $P_{B1}(3_{3}|1_{2})=1/4$,
and similarly, $P_{B1}(3_{3}|1_{2})=1/4$, but only one box can be
detected with a ball. Hence, the probability of detecting a ball in
Box 3 at time $t_{3}$, given Bob detected a ball in either Box 1
or 4 at time $t_{2}$, is $P_{B1,B4}(3_{3}|\{1_{2},4_{2}\})=1/4$.
Hence
\begin{eqnarray}
P_{B1,B4}(\{1_{2},4_{2}\},3_{3}) & = & P_{B1,B4}(3_{3}|\{1_{2},4_{2}\})\nonumber \\
 &  & \times P_{B1,B4}(\{1_{2},4_{2}\})\nonumber \\
 & = & 1/8\label{eq:probtwobox}
\end{eqnarray}
The probability the ball is detected in Box 3, if Bob opens Boxes
1 and 4, is $P_{B1,B4}(3_{3})=\frac{1}{2}\{\frac{1}{4}\}+\frac{1}{2}\{0\}=1/8$.
Hence, we obtain the key result:
\begin{equation}
P_{B1,B4}(\{1_{2},4_{2}\}|3_{3})=1\label{eq:prob-answer}
\end{equation}
Similarly, 
\begin{equation}
P_{B2,B4}(\{2_{2},4_{2}\}|3_{3})=1\label{eq:answer2}
\end{equation}
We also note that
\begin{eqnarray*}
P_{B1,B4}(4_{2},3_{3}) & = & P_{B1,B4}(3_{3}|4_{2})P_{B1,B4}(4_{2})\\
 & = & 1/16
\end{eqnarray*}
which implies, since $P_{B1,B4}(3_{3})=1/8$,
\[
P_{B1,B4}(4_{2}|3_{3})=1/2
\]
Hence, Alice knows that, if she detects the ball in Box 3, the ball
was detected in Box 4 only $50\%$ of the time. Yet, we see that the
ball was detected with certainty in the set of boxes (either $1$
and $4$, or $2$ and $4$) that Bob opened. The paradox is that it
seems as though for 50\% of the time, the ball would be have to be
detected in box 2 or 1, had that box been opened. The paradox is
expressed as a violation of a Leggett-Garg inequality, in Section
VI.

The effect of measurement disturbance is made apparent in this example.
If Bob makes \emph{no} measurement at time $t_{2}$, then the final
state at time $t_{3}$ is $U_{f}|\psi_{sup}\rangle$: 
\begin{eqnarray}
U_{f}|\psi_{sup}\rangle & = & \left(\begin{array}{c}
0\\
0\\
0\\
1
\end{array}\right)\label{eq:no-m}
\end{eqnarray}
There is zero probability that Alice detects a ball in Box 3: $P_{N}(3_{3})=0$.
Yet, from the results above, if Bob opens Boxes 1 and 4, the probability
of Alice observing the ball in Box 3 is $1/4$. Unlike the standard
paradox, $P_{N}(3_{3})\neq P_{B1,B4}(3_{3})$. This means that Condition
1 of Ref. \cite{Maroney} is not satisfied. We note however that the
probability of Alice's detecting a ball in Box 3, given Bob makes
a measurement, is not affected by which pair of Boxes he opens: $P_{B1,B4}(3^{3})=1/4$
and $P_{B2,B4}(3^{3})=1/4$. This means part of Condition 1 of \cite{Maroney}
is satisfied. The transformation $U_{f}$, equivalent to a shuffle,
does not give a relative enhancement of the probability of Alice detecting
a Ball in Box 3 depending on which Boxes Bob opened.

\subsection{Coherent-state model: $k=2$}

The experimental realisation of $k>2$ may be challenging. However,
the unitary interactions described in Yurke and Stoler have been experimentally
verified for $k=2$ \cite{collapse-revival-super-circuit-1,collapse-revival-bec-2}.
We can use
\begin{eqnarray}
|1\rangle & = & |-i\alpha_{0}\rangle\nonumber \\
|2\rangle & = & |i\alpha_{0}\rangle\nonumber \\
|3\rangle & = & e^{-i\pi/4}|\alpha_{0}\rangle\nonumber \\
|4\rangle & = & e^{-i\pi/4}|-\alpha_{0}\rangle\label{eq:coh-three-states-1}
\end{eqnarray}
We consider
\[
|\psi_{sup}\rangle\equiv|\psi_{1}\rangle=\frac{1}{2}\{|1\rangle+|2\rangle+|3\rangle-|4\rangle\}
\]
This state \cite{yurke-stoler-1,manushan-cat-lg}
\begin{equation}
\begin{array}{c}
|\psi_{sup}\rangle=\frac{1}{2}\{|-i\alpha_{0}\rangle+|i\alpha_{0}\rangle+e^{-i\pi/4}|\alpha_{0}\rangle-e^{-i\pi/4}|-\alpha_{0}\rangle\}\\
\\
\end{array}\label{eq:sup3-coh2}
\end{equation}
is formed at time \textcolor{black}{$t=\pi/4\Omega$} from $|\alpha_{0}\rangle$,
using the evolution given by $H_{NL}$ of Eq. (\ref{eq:H}) with $k=2$.
We can interpret the creation of the state as a transformation using
a unitary operator $U_{i}=U_{\left(\frac{\pi}{4\Omega}\right)}$ given
by
\begin{equation}
U_{i}|3\rangle=U_{i}|\alpha_{0}\rangle=|\psi_{sup}\rangle\label{eq:Ui1-coh-1}
\end{equation}
where $U_{\left(\frac{\pi}{4\Omega}\right)}=e^{\frac{-i\Omega\hat{n}^{k}t}{\hbar}}$
with $\hbar=1$.

Bob measures whether the system is in state $|1\rangle$ or $|4\rangle$,
or not. If the system can be determined to be in neither $|1\rangle$
nor $|4\rangle$, then the reduced state for the system is $|2\rangle+|3\rangle$.
Alternatively, Bob may make measurements to determine whether the
system is in one of the states $|2\rangle$ or $|4\rangle$, or not.
If the system is determined to be in neither $|2\rangle$ or $|4\rangle$,
then the reduced state for the system is $|1\rangle+|3\rangle$.

Alice postselects for the state
\begin{eqnarray}
|\psi_{f}\rangle & = & \frac{1}{2}\{|1\rangle+|2\rangle-|3\rangle+|4\rangle\}\nonumber \\
 & = & \frac{1}{2}\{|-i\alpha_{0}\rangle+|i\alpha_{0}\rangle\nonumber \\
 &  & -e^{-i\pi/4}|\alpha_{0}\rangle+e^{-i\pi/4}|-\alpha_{0}\rangle\}
\end{eqnarray}
which is formed at time \textcolor{black}{$t=\pi/4\Omega$} from $|-\alpha_{0}\rangle$,
using $k=2$. We aim to find $U_{f}$ such that
\begin{equation}
U_{f}^{-1}|3\rangle=|\psi_{f}\rangle\label{eq:Uf-1-2-coh2}
\end{equation}
As above, we first transform to $|4\rangle$. Defining $U_{f2}=U_{\left(\frac{\pi}{\Omega}\right)}=e^{\frac{-i\Omega\hat{n}^{k}t}{\hbar}}$
where the interaction time is $t=\pi/\Omega$, the state $|\alpha_{0}\rangle$
becomes $|-\alpha_{0}\rangle$, for $k=2$. We see that $U_{f2}=U_{f2}^{-1}$.
We then apply
\begin{equation}
U_{f3}^{-1}U_{f2}^{-1}|3\rangle=|\psi_{f}\rangle\label{eq:trans2-1}
\end{equation}
where we define $U_{f3}^{-1}=U_{\left(\frac{\pi}{4\Omega}\right)}=e^{-i\Omega\hat{n}^{k}t}$
with interaction time is $t=\pi/4\Omega$. We note that $U_{f3}=U_{\left(\frac{7\pi}{4\Omega}\right)}=e^{-i\Omega\hat{n}^{k}t}$
which corresponds to the interaction time $t=7\pi/4\Omega$. We find
that $U_{f}=U_{f2}U_{f3}$.%
We find that $U_{f}=U_{f2}U_{f3}$.

Hence, Alice transforms the system according to $U_{f}$ and then
detects whether the system is in state $|3\rangle$. The operations
$U_{f}$ correspond to Alice first evolving the system under $H_{NL}$
with $k=2$ for a time $t=7\pi/4\Omega$, and then applying the evolution
$H$ for a time $t=\pi/\Omega$. For the system in either $|2\rangle+|3\rangle$
or $|1\rangle+|3\rangle$ at time $t_{2}$, the probability for Alice
obtaining a result $|3\rangle$ is zero, which leads to the paradox.

Realisation of the interactions $H_{NL}$ of Eq. (\ref{eq:H}) with
$k=2$ have been achieved in the experiments of Kirchmair et al \cite{collapse-revival-super-circuit-1}.
Bob's projective measurements can be performed in principle by measuring
the quadrature phase amplitude, along a chosen direction. For example,
determining whether the system is in $|3\rangle$ or $|4\rangle$
can be determined by measuring $\hat{X}$. This will distinguish $|3\rangle$
and $|4\rangle$, but the outcome of zero gives no information about
whether the system is in $|1\rangle$ or $|2\rangle$. Similarly,
whether the system is in $|2\rangle$ or $|4\rangle$ (or $|1\rangle$
or $|4\rangle$) can be determined by measuring whether $\hat{X}_{\theta}$
is zero or not, for the right choice of rotated axis.

\textcolor{black}{}
\begin{figure}[h]
\begin{centering}
\textcolor{black}{\includegraphics[width=0.47\columnwidth]{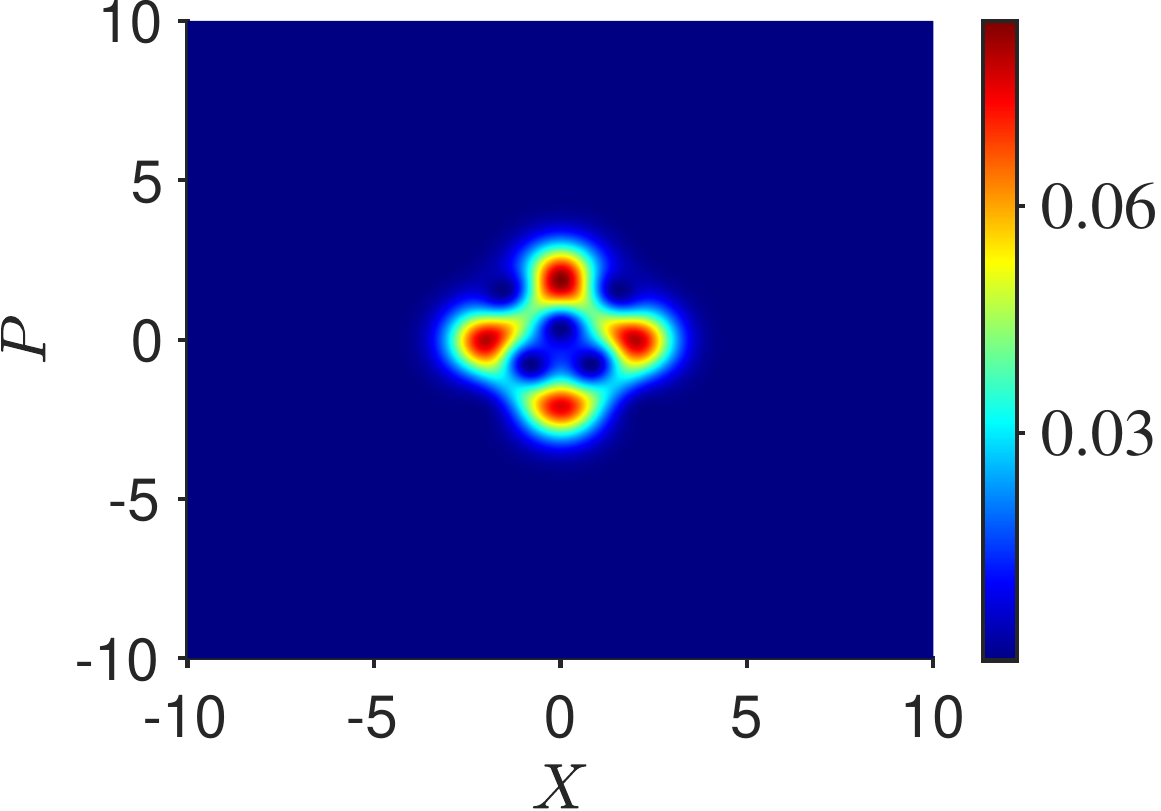}\hspace{0.03\columnwidth}\includegraphics[width=0.47\columnwidth]{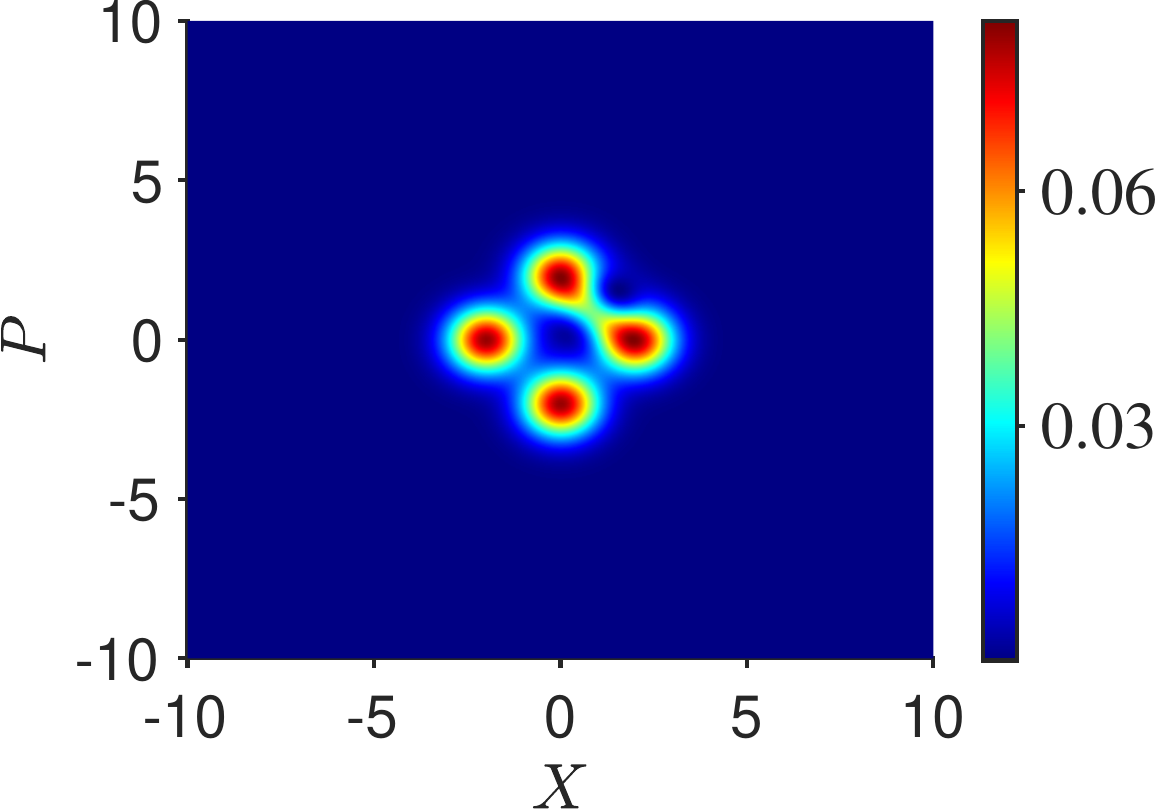}}
\par\end{centering}
\medskip{}

\begin{centering}
\textcolor{black}{\includegraphics[width=0.47\columnwidth]{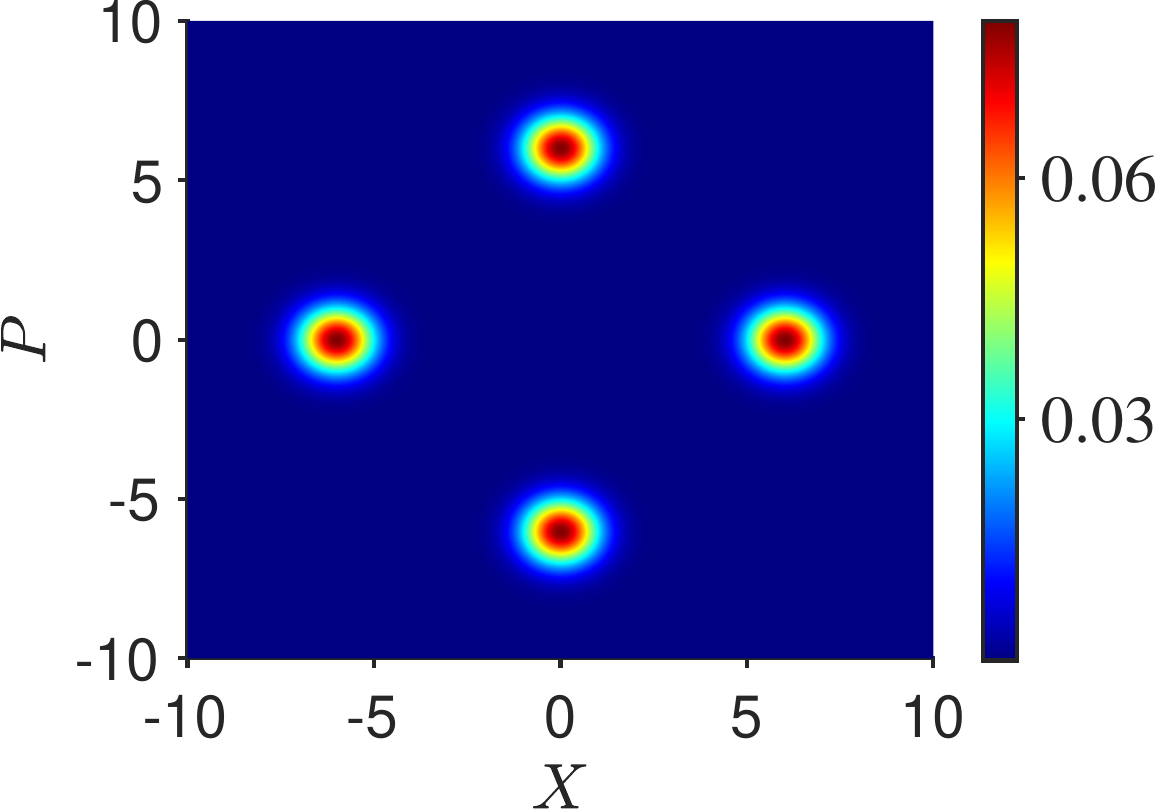}\hspace{0.03\columnwidth}\includegraphics[width=0.47\columnwidth]{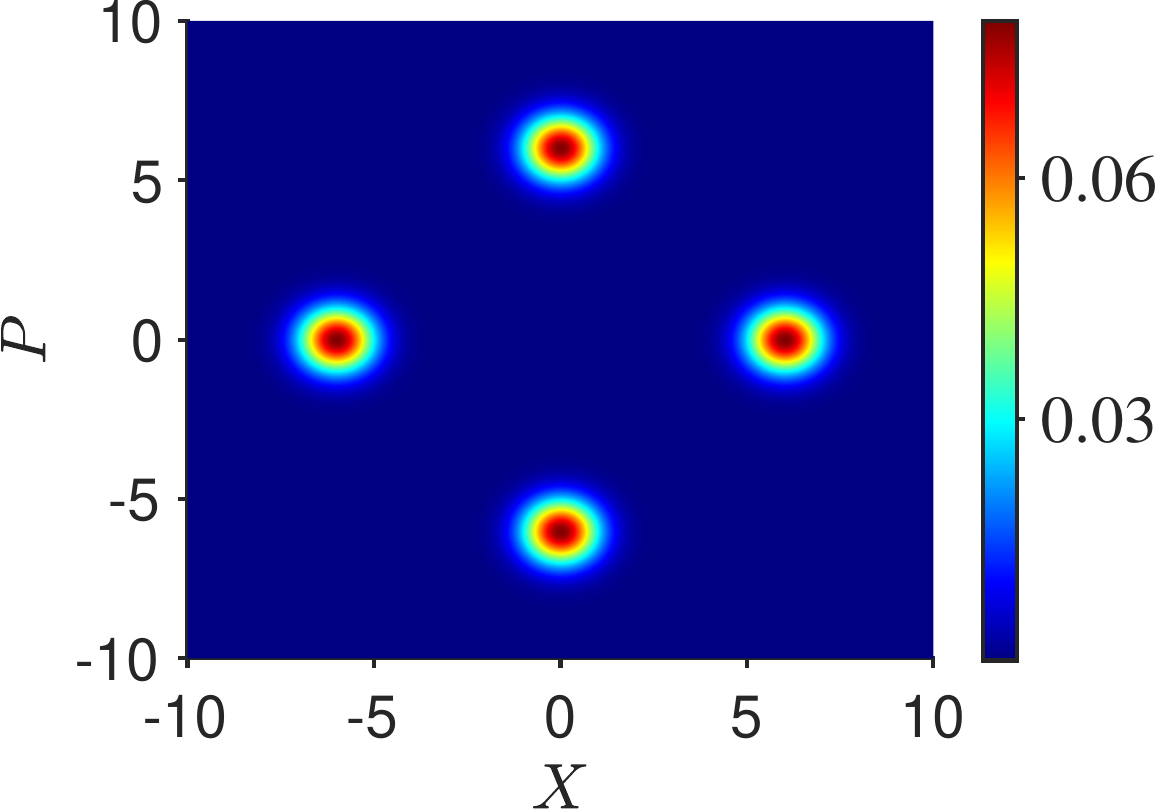}}
\par\end{centering}
\textcolor{black}{\caption{The $Q$ functions for the superposition $|\psi_{sup}\rangle$ \textcolor{black}{(Eq.
\ref{eq:sup-coh}) }(left) and the mixture $\rho_{mix,14}$ \textcolor{black}{(Eq.
\ref{eq:mix-coh})} (right) become indistinguishable for large $\alpha_{0}$\textcolor{black}{.
}The top and lower pairs are for $\alpha_{0}=2$ and $\alpha_{0}=6$
respectively.\textcolor{green}{}\textcolor{red}{{} \label{fig:Contour-plots-of-q-sup-and-mix}}}
}
\end{figure}

\textcolor{black}{}
\begin{figure*}[t]
\textcolor{black}{}
\begin{centering}
\includegraphics[width=2\columnwidth]{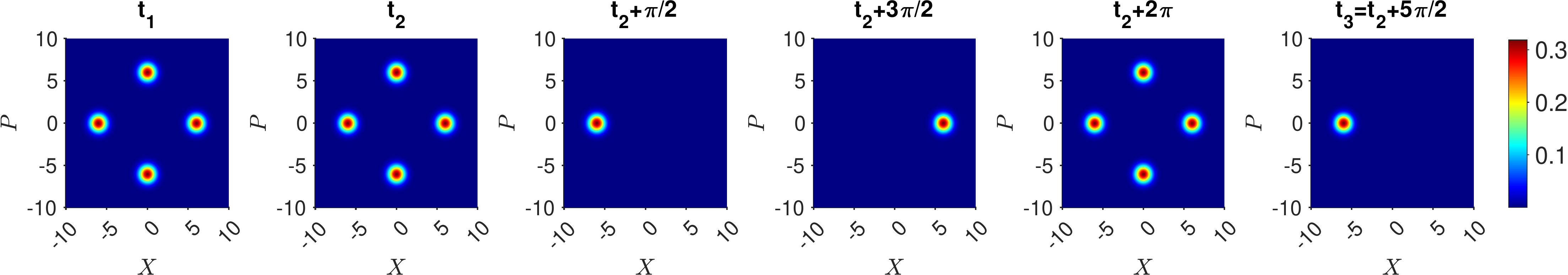}\medskip{}
\par\end{centering}
\begin{centering}
\includegraphics[width=2\columnwidth]{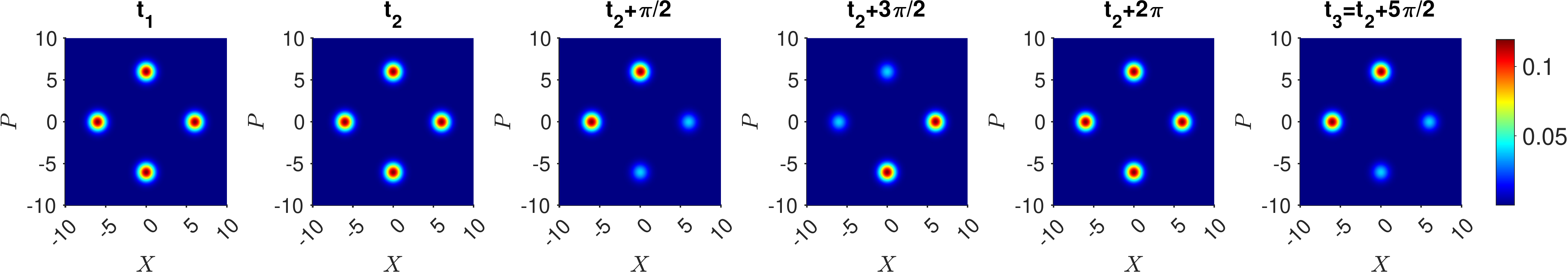}
\par\end{centering}
\textcolor{black}{\caption{The dynamics of the macroscopic three-box paradox. The dynamics induced
by Alice's transformations $U_{f}$ is macroscopically sensitive to
whether or not Bob has made a prior measurement on the system at time
$t_{1}$, despite Bob's measurement being seemingly non-invasive.
The figure shows contour plots of $Q(\alpha)$ \textcolor{black}{for
the system prepared at time $t_{1}$ in the }superposition state $|\psi_{sup}\rangle$
(Eq. (\ref{eq:sup-coh}))\textcolor{black}{{} (far left), as it evolves
under the action of Bob and Alice's operations. The top sequence shows
the dynamics if }there is no measurement made by Bob. Alice makes
her transformation $U_{f}$ on the system starting at time $t_{2}$
(second from left).\textcolor{black}{{} The dynamics of the transformation
$U_{f}$ is completed at time $t_{3}$. }The lower sequence shows
the evolution if Bob makes a measurement on the system at time $t_{1}$
(the outcome revealing whether the system is in the state $|1\rangle$
or $|4\rangle$, or not). At time $t_{2}$, after Bob's measurement,
the system is in the mixed state $\rho_{mix,14}$. While the $Q$
function for $\rho_{mix,4}$ (lower second from left) is indistinguishable
from that of $|\psi_{sup}\rangle$ (top second from left) as $\alpha\rightarrow\infty$,
indicating a non-invasive measurement, there is a macroscopic difference
between the final probabilities (top far right and lower far right)
after the dynamics of Alice's transformations, at time $t_{3}$. Here,\textcolor{black}{{}
}\textcolor{black}{$\alpha_{0}=6$ and }$k=3$.\textcolor{red}{}\textcolor{green}{}\textcolor{red}{{}
\label{fig:q-plots-dynamics-bob-measure-or-not}}\textcolor{green}{}}
}
\end{figure*}

\subsection{Consistency with macroscopic realism}

We now analyse the dynamics of the three-box paradox for the macroscopic
example using coherent states. The paradox can be modelled using the
$Q$ function, defined by \cite{Husimi-Q-1}
\begin{equation}
Q(\alpha)=\frac{\langle\alpha|\rho|\alpha\rangle}{\pi}\label{eq:Qfun}
\end{equation}
where $\rho$ is the system density operator. We examine the case
of $k>2$. The case of $k=2$ will show similar behaviour. The system
at time $t_{1}$ is in the superposition $|\psi_{sup}\rangle$ (Eq.
(\ref{eq:sup-coh})). The Q function is
\begin{eqnarray}
Q(x,p) & = & e^{-\frac{(|\alpha|^{2}+|\alpha_{0}|^{2})}{2\pi}}\Bigr(2\cos\left(x\alpha_{0}-p\alpha_{0}\right)\sinh\left(x\alpha_{0}-p\alpha_{0}\right)\nonumber \\
 &  & -2\cos\left(x\alpha_{0}+p\alpha_{0}\right)\sinh\left(x\alpha_{0}+p\alpha_{0}\right)\nonumber \\
 &  & +\cosh\left(2x\alpha_{0}\right)+\cosh\left(2p\alpha_{0}\right)\nonumber \\
 &  & +\cos\left(2p\alpha_{0}\right)-\cos\left(2x\alpha_{0}\right)\Bigl)\label{eq:q-sup}
\end{eqnarray}
\textcolor{red}{}where $\alpha=x+ip$ and $\alpha_{0}$ is taken
to be real. The function is depicted by the contour graph, in Figure
\ref{fig:Contour-plots-of-q-sup-and-mix}.

Suppose Bob looks at Boxes 1 and 4. After the projective measurement
by Bob to determine whether the system is in state $|1\rangle$ or
$|4\rangle$, or not ($i=1,2$), the system is in the mixed state
given by density operator
\begin{equation}
\rho_{mix,14}=\frac{1}{4}|1\rangle\langle1|+\frac{1}{4}|4\rangle\langle4|+\frac{1}{4}(-|2\rangle+|3\rangle)(-\langle2|+\langle3|)\label{eq:mix-coh}
\end{equation}
\textcolor{green}{}The Q function of the mixed state created at
time $t_{2}$ is
\begin{eqnarray}
Q_{mix,14}(x,p) & = & e^{-\frac{(|\alpha|^{2}+|\alpha_{0}|^{2})}{2\pi}}\Bigr(\cosh\left(2x\alpha_{0}\right)+\cosh\left(2p\alpha_{0}\right)\nonumber \\
 &  & -\exp\bigl(x\alpha_{0}+p\alpha_{0}\bigr)\cos\bigl(x\alpha_{0}+p\alpha_{0}\bigr)\Bigr)\label{eq:qmix}
\end{eqnarray}
plotted in Figure \ref{fig:Contour-plots-of-q-sup-and-mix}. Comparison
of $Q(x,p)$ with $Q(x,p)_{mix,14}$ shows that as $\alpha_{0}\rightarrow\infty$,
$Q(x,p)\rightarrow Q_{mix,14}(x,p)$. The terms contributing from
the superposition are damped by a term exponential in $\alpha_{0}$.\textcolor{red}{}

Now Alice performs the transformations. The system evolves according
to $U_{f}$. The final state if Bob made no measurement is given by
Eq. (\ref{eq:no-m}), since the system remains in the superposition
$|\psi_{sup}\rangle$ (Figure \ref{fig:q-plots-dynamics-bob-measure-or-not}).
On the other hand, the result if Bob opened Boxes 1 and 4 is different.
The system is $\rho_{mix,14}$ at time $t_{2}$. After the transformations
$U_{f}$, the system is in the mixed state
\begin{eqnarray}
\rho_{mix,14}(t_{3}) & = & \frac{1}{4}[U_{f}|1\rangle\langle1|U_{f}^{\dagger}]+\frac{1}{4}[U_{f}|4\rangle\langle4|U_{f}^{\dagger}]\nonumber \\
 &  & +\frac{1}{2}U_{f}(\frac{-|2\rangle+|3\rangle}{\sqrt{2}})(\frac{-\langle2|+\langle3|}{\sqrt{2}})U_{f}^{\dagger}\nonumber \\
\label{eq:mix14trans}
\end{eqnarray}
The states $U_{f}|1\rangle$, $U_{f}|4\rangle$ and $U_{f}(-|2\rangle+|3\rangle)/\sqrt{2}$
are given by Eqs. (\ref{eq:ma3}), (\ref{eq:ma2}) and (\ref{eq:ma6}).
Figure \ref{fig:q-plots-dynamics-bob-measure-or-not} shows the final
state after Alice's operations $U_{f}$, for the two initial states
$|\psi_{sup}\rangle$ and $\rho_{mix,14}$ at time $t_{2}$. While
the $Q$ functions are initially indistinguishable, after the $U_{f}$,
a macroscopic difference between the final states emerges. This is
consistent with a model in which weak macroscopic realism holds, but
where there is measurement disturbance due to Bob's interactions.

\section{Leggett-Garg test of macrorealism}

The Leggett-Garg inequality can be violated for systems which do not
jointly satisfy the combined premise of macroscopic realism and no
measurement-disturbance. Leggett and Garg considered a system which
at three times $T_{1}<T_{2}<T_{3}$ can be found to be in one of two
macroscopically distinguishable states. If macroscopic realism holds,
then the system is at each time always in one or other of the states,
and can be ascribed the variable $\lambda_{k}$ at time $T_{k}$,
where $\lambda_{k}=\pm1$. If noninvasive measurability holds, there
is no disturbance from a measurement made on the system to determine
the value of $\lambda_{k}$. In this case, \emph{macrorealism} is
said to hold, and inequalities follow.

The application of a Leggett-Garg test to the three-box paradox elucidates
the origin of the paradox \cite{Maroney}. Maroney derived new versions
of the Leggett-Garg inequality that applied to the paradox. The Condition
1 by Bob ensures that 
\begin{equation}
P_{N}(3_{3})=P_{B1}(3_{3})=P_{B2}(3_{3})\label{eq:operation-nd}
\end{equation}
This means that Alice observes no change in the probability of her
detecting a Ball in Box 3, due to whether Bob opens one of the Boxes
or not. Maroney referred to Bob's measurement, which satisfies this
condition, as operationally nondisturbing. Maroney pointed out that
this condition is satisfied in the three-box paradox, but has not
been satisfied in other tests of Leggett-Garg inequalities. The condition
gives a particularly strong test of macrorealism. Here, we illustrate
that the mesoscopic paradox of Section IV will satisfy the strict
Maroney-Leggett-Garg test of macrorealism, and that violation of a
Leggett-Garg inequality is also possible for the macroscopic system
of Section III.

Following Maroney, we let $\lambda_{k}=-1$ if the system is found
in $|1\rangle$ or $|2\rangle$, and $\lambda_{k}=1$ if the system
is found in Box 3. At time $T_{1}$, the ball is in Box 3, and $\lambda_{1}=1$.
The premise of macrorealism implies the Leggett-Garg inequality $-1\leq Q\leq3$,
where 
\begin{equation}
Q=\langle\lambda_{1}\lambda_{2}\rangle+\langle\lambda_{2}\lambda_{3}\rangle+\langle\lambda_{1}\lambda_{3}\rangle\label{eq:lg}
\end{equation}
Consideration that macrorealism holds leads to constraints on the
relations between the probabilities. We summarise the work of Maroney,
since this will apply directly to the mesoscopic example. Using the
initial condition, we find that $Q=\langle\lambda_{2}\rangle+\langle\lambda_{3}\rangle+\langle\lambda_{2}\lambda_{3}\rangle$
where $\langle\lambda_{2}\rangle=-P(1_{2})-P(2_{2})+P(3_{2})=2P(3_{2})-1$,
and

$\langle\lambda_{3}\rangle=2P(3_{3})-1$. Macrorealism posits the
ball to be in one of the boxes, which implies\textcolor{green}{}
\begin{equation}
P(3_{2})=P(3_{2},1_{3})+P(3_{2},2_{3})+P(3_{2},3_{3})\label{eq:probs}
\end{equation}
and similarly, 
\begin{equation}
P(3_{3})=P(1_{2},3_{3})+P(2_{2},3_{3})+P(3_{2},3_{3})\label{eq:probs2}
\end{equation}
Macrorealism also implies relations between joint probabilities and
marginals, for example

\begin{eqnarray}
P(1_{2}+2_{2},1_{3}+2_{3}+3_{3}) & = & \sum_{I_{2}=1,2}\sum_{J_{3}=1,2,3}P(I_{2},J_{3})\nonumber \\
 & = & P(1_{2}+2_{2})\nonumber \\
 & = & 1-P(3_{2})\label{eq:prob7}
\end{eqnarray}
Hence, macrorealism implies
\begin{eqnarray}
\langle\lambda_{2}\lambda_{3}\rangle & = & P(3_{2},3_{3})+P(2_{2},2_{3})+P(1_{2},1_{3})+P(1_{2},2_{3})\nonumber \\
 &  & +P(2_{2},1_{3})-P(3_{2},1_{3})-P(3_{2},2_{3})\nonumber \\
 &  & {\color{red}{\normalcolor -P(1_{2},3_{3})-P(2_{2},3_{3})}}\nonumber \\
 & = & P(3_{2},3_{3})-(P(3_{2})-P(3_{2},3_{3}))\nonumber \\
 &  & -(P(3_{3})-P(3_{2},3_{3}))\nonumber \\
 &  & +1-P(3_{2})-(P(3_{3})-P(3_{2},3_{3}))\label{eq:probaverage}
\end{eqnarray}
This leads to 
\begin{eqnarray}
Q & = & 4P(3_{2},3_{3})-1\nonumber \\
 & = & 4(P(3_{3})-P(1_{2},3_{3})-P(2_{2},3_{3}))-1\nonumber \\
 & = & 4P(3_{3})(1-P(1_{2}|3_{3})-P(2_{2}|3_{3}))-1\label{eq:qfinal}
\end{eqnarray}

Considering the paradox of Section III, the probability of Alice detecting
a Ball in box 3 at time $T_{3}$, regardless of a measurement by Bob,
is $1/9$. The joint probability $P(1_{2},3_{3})$ is measurable by
Alice and Bob, and calculable as $P(1_{2})P(3_{3}|1_{2})$. Here,
$P(1_{2})=1/3$ and $P(3_{3}|1_{2})=1/3$, from the earlier sections.
Alternatively, $P(1_{2},3_{3})=P(3_{3})P(1_{2}|3_{3})=P(3_{3})=1/9$,
since $P(1_{2}|3_{3})=1$. The results for $P(2_{2},3_{3})$ are identical,
giving $Q=-13/9$, which is a violation of the Leggett-Garg inequality,
implying a negation of macrorealism. The test can be performed
using the mesoscopic version of Section III, since the predictions
are identical to the original paradox.

The calculation of the prediction $P(3_{3}|1_{2})$ assumes the system
at time $T_{2}$ after Bob's measurement is $|1\rangle$ which is
the eigenstate. We have seen how the $Q$ function for the superposition
differs from that of a mixed state. Prior to Bob's measurement, the
system can be viewed to be in a state with a definite outcome for
the detection of the ball in the box, but the ``state'' the system
is in is \emph{not} the corresponding eigenstate.

The Leggett-Garg test is also applicable to the macroscopic set-up
proposed in Section IV. We we let $\lambda_{k}=-1$ if the system
is found in $|1\rangle$ or $|4\rangle$, $|2\rangle$ or $|4\rangle$,
and $\lambda_{k}=1$ if the system is found in Box 3. At time $T_{1}$,
the ball is in Box 3, and $\lambda_{1}=1$. We find $\langle\lambda_{2}\rangle=2P(3_{2})-1$
and $\langle\lambda_{3}\rangle=2P(3_{3})-1$. To derive $\langle\lambda_{2}\lambda_{3}\rangle$,
extending the logic to apply to four possible states is straightforward
e.g. macrorealism implies
\begin{eqnarray}
P(3_{2}) & = & P(3_{2},1_{3})+P(3_{2},2_{3})+P(3_{2},3_{3})+P(3_{2},4_{3})\nonumber \\
\label{eq:prob19}
\end{eqnarray}
This leads to the expression above, and 
\begin{eqnarray}
Q & = & 4P(3_{2},3_{3})-1\nonumber \\
 & = & 4(P(3_{3})-P(1_{2},3_{3})-P(2_{2},3_{3})-P(4_{2},3_{3}))-1\nonumber \\
\label{eq:Qterms}
\end{eqnarray}
Now, we know that since the ball can only be in one box, $P(\{1_{2},4_{2}\},3_{3})=P(1_{2},3_{3})+P(4_{2},3_{3})$.
Hence, 
\begin{eqnarray}
Q & = & 4(P(3_{3})-P(1_{2},3_{3})-P(2_{2},3_{3})-P(4_{2},3_{3}))-1\nonumber \\
 & = & 4(P(3_{3})-P(\{1_{2},4_{2}\},3_{3})-P(\{2_{2},4_{2}\},3_{3})\nonumber \\
 &  & +P(4_{2},3_{3}))-1\nonumber \\
 & = & 4(P(3_{3})-P_{B1,B4}(\{1_{2},4_{2}\},3_{3})-P_{B2,B4}(\{2_{2},4_{2}\},3_{3})\nonumber \\
 &  & +P_{B1,B4}(4_{2},3_{3}))-1\nonumber \\
 & = & 4P_{B1,B4}(3_{3})\{1-\{P(\{1_{2},4_{2}\}|3_{3})-P(\{2_{2},4_{2}\}|3_{3})\nonumber \\
 &  & +P_{B4}(4_{2}|3_{3})\}-1\label{eq:qfinalcoh}\\
\nonumber 
\end{eqnarray}
Using the solutions $P_{B1,B4}(3_{3})=P_{B1,B2}(3_{3})=1/8$, $P_{B4}(4_{2}|3_{3})=1/2$
and $P(1_{2},4_{2}|3_{3})=1$ and $P(2_{2},4_{2}|3_{3})=1$, we find
$Q=-5/4$, giving a violation of the Leggett-Garg inequality and hence
a negation of macrorealism.

\section{Conclusion}

This paper gives proposals for mesoscopic and macroscopic quantum
three-box paradoxes. The unitary operations (shuffling) required for
the three-box paradox are realised by nonlinear interactions which
we model by specific Hamiltonians. The motivation for considering
the macroscopic versions is to argue the case for realism: The paradox
may be explained as a failure of realism, or else explained by measurement
disturbance.

We show how macroscopic realism can be upheld consistently with the
paradox. Macroscopic realism asserts that the system found to be in
two macroscopically distinct states has a predetermined value for
the outcome of a measurement that distinguishes those states. In order
to achieve consistency with macroscopic realism, the definition of
macroscopic realism is refined, so that it applies to the system created
at the time $t_{i}$ \emph{after} the unitary operations that determine
the local measurement basis. Moreover, the predetermined value cannot
be changed by any operations or measurements on spatially separated
systems. We refer to this restricted definition as \emph{weak macroscopic
realism} (wMR). Weak macroscopic realism has been shown consistent
with violations of macroscopic Bell inequalities \cite{manushan-bell-cat-lg}.

Following Maroney \cite{Maroney}, we have shown that the realisation
of the paradox corresponds to a violation of a Leggett-Garg inequality.
Hence, the combined assumptions of macroscopic realism and noninvasive
measurability (``macrorealism'') are negated by the paradox. Our
proposals however are macroscopic, and have the advantage that macrorealism
is tested in the spirit of the Leggett-Garg paper \cite{legggarg},
applying to a system where macroscopic realism can be genuinely applied.
We illustrate how the Leggett-Garg inequality is violated, and yet
macroscopic realism upheld, the violation occurring due to a failure
of noninvasive measurability.

Further, in this paper we illustrate the paradoxical features of the
measurement disturbance, by manipulating the parameter that determines
the size of the system. The disturbance becomes minimal with increasing
size, yet the probabilities after Alice's unitary operations remain
macroscopically distinguishable, depending on whether a measurement
occurred or not. This effect is similar to a quantum revival. We expect
the origin is non-classical, arising from future boundary conditions
based on the Q function \cite{DrummondReid2020}.

The definition of macroscopic realism is \emph{required} to be minimal.
Macroscopic realism posits that there is a predetermined value for
the outcome of the macroscopic measurement: This means that the ball
is either in the Box, or not, prior to Alice or Bob opening the Box.
However,  it can be shown that if the system is viewed as being
in a 'state' with the predetermined outcome $+$ or $-$, then that
'state' cannot be given as a quantum state $|\psi_{+}\rangle$ or
$|\psi_{-}\rangle$, prior to measurement \cite{manushan-bell-cat-lg}.
This points to an inconsistency between wMR and (the standard interpretation
of) quantum mechanics, as in Schrödinger's argument \cite{s-cat}.
The acceptance of wMR as part of the explanation of the paradox may
raise other open questions.

Finally, we consider the possibility of an experiment. The unitary
dynamics required for the proposal with coherent states has been realised
in experiments \cite{collapse-revival-super-circuit-1}. The mesoscopic
system may be realised for $N=2$ using the Hong-Ou-Mandel effect.
We note that similar mesoscopic interactions may also be realisable
for moderate $N$, by applying the CNOT gates of the IBM computer
\cite{IBM-macrorealism-1}.

\section*{Acknowledgements}

\textcolor{black}{This research has been supported by the Australian
Research Council Discovery Project Grants schemes under Grant DP180102470
and }DP190101480. The authors also wish to thank NTT Research for
their financial and technical support.

\end{document}